\documentclass[12pt,preprint]{aastex}
\usepackage{amsmath}

\usepackage{epsfig}
\usepackage{xcolor}

\usepackage{lineno}
\newcommand{\nustar}{\textit{NuSTAR}}

\def\greenphoton{\dot N_\epsilon^{\rm G}}

\def\greencolumn{\Phi_\epsilon^{\rm G}}

\def\columntotal{\Phi_\epsilon^{\rm tot}}

\newcommand{\chimin}{\chi_{\rm min}}
\newcommand{\chimax}{\chi_{\rm max}}

\newcommand{\vel}{\upsilon}

\def\green{f_{_{\rm G}}}
\def\columntotal{\Phi_\epsilon^{\rm tot}}
\newcommand{\gapprox}{\lower.4ex\hbox{$\;\buildrel >\over{\scriptstyle\sim}\;$}}
\newcommand{\lapprox}{\lower.4ex\hbox{$\;\buildrel <\over{\scriptstyle\sim}\;$}}
\newcommand{\begeq}{\begin{equation}}
\newcommand{\fineq}{\end{equation}}
\newcommand{\msun}{M_\odot} 
\newcommand{\epsmin}{\epsilon_{\rm min}}
\newcommand{\epsmax}{\epsilon_{\rm max}}
\newcommand{\Tmound}{T_{\rm th}}
\newcommand{\rhomound}{\rho_{\rm th}}
\newcommand{\vmound}{\vel_{\rm th}}

\newcommand{\taumound}{\tau_{\rm th}}
\newcommand{\epsilonabs}{\epsilon_{\rm abs}}
\newcommand{\chiabs}{\chi_{\rm abs}}

\def\colrad{r_\perp}

\def\Msun{M_\odot}

\def\tauperp{\tau_\perp}

\newcommand{\deltapar}{\psi}

\newcommand{\sigmaT}{\sigma_{_{\rm T}}}
\def\sig{\sigma_{_{\rm T}}}
\def\sigpar{\sigma_{_{||}}}
\def\sigperp{\sigma_\perp}
\def\sigbar{\overline\sigma}

\def\green{f_{_{\rm G}}}

\def\sigmapar{\sigma_{||}}
\slugcomment{Draft: August 29, 2022 at 11:00 PM; Accepted by ApJ}

\shorttitle{COMPTONIZATION IN X-RAY PULSARS}
\shortauthors{Becker \& Wolff}

\begin{document}

\title{A GENERALIZED ANALYTICAL MODEL FOR THERMAL AND BULK COMPTONIZATION IN ACCRETION-POWERED X-RAY PULSARS}

\author{Peter A. Becker\altaffilmark{1}}

\affil{Department of Physics and Astronomy,
George Mason University,
Fairfax, VA 22030-4444, USA}

\and

\author{Michael T. Wolff\altaffilmark{2}}

\affil{Space Science Division, Naval Research Laboratory, Washington, DC 20375-5352, USA}

\vfil

\altaffiltext{1}{pbecker@gmu.edu}
\altaffiltext{2}{michael.wolff@nrl.navy.mil}

\begin{abstract}
We develop a new theoretical model describing the formation of the radiation spectrum in accretion-powered X-ray pulsars as a result of bulk and thermal Comptonization of photons in the accretion column.
The new model extends the previous model developed by the authors in four ways:
(1) we utilize a conical rather than cylindrical geometry;
(2) the radiation components emitted from the column wall and the column top are computed separately;
(3) the model allows for a non-zero impact velocity at the stellar surface; and
(4) the velocity profile of the gas merges with Newtonian free-fall far from the star.
We show that these extensions allow the new model to simulate sources over a wide range of accretion rates.
The model is based on a rigorous mathematical approach in which we obtain an exact series solution for the Green's function describing the reprocessing of monochromatic seed photons. Emergent spectra are then computed by convolving the Green's function with bremsstrahlung, cyclotron, and blackbody photon sources. The range of the new model is demonstrated via applications to the high-luminosity source Her X-1, and the low-luminosity source X Per.
The new model suggests that the observed increase in spectral hardness associated with increasing luminosity in Her X-1 may be due to a decrease in the surface impact velocity, which increases the $P$d$V$ work done on the radiation field by the gas.

\end{abstract}

% The different journals have different requirements for keywords.  The
% keywords.apj file, found on aas.org in the pubs/aastex-misc directory, 
% contains a list of keywords used with the ApJ and Letters.  These are 
% usually assigned by the editor, but authors may include them in their 
% manuscripts if they wish. 


\section{INTRODUCTION}
\label{sec:intro}

The X-ray emission from binary pulsars such as Her X-1, LMC X-4, Cen X-3, GX 304-1, and X Per is powered by the accretion of material from the ``normal'' companion onto the neutron star. The inflowing material is channeled by the strong magnetic field onto one or both of the neutron star's magnetic poles. The X-ray luminosity, $L_X$, of accretion-powered X-ray pulsars extends over a very wide range, from $L_X \sim 10^{34}\,{\rm ergs\,s}^{-1}$ for X Per up to $L_X \sim 10^{38}\,{\rm ergs\,s}^{-1}$ for LMC X-1.
Modeling the accretion of material from the binary stellar companion onto the surface of the neutron star,
and the formation of the associated X-ray spectrum, is a challenging physical problem that has not yet been fully solved. 
Physical simulation of these systems requires consideration of high-energy plasma processes in
strong magnetic fields, strong shock physics involving both gas-dominated and 
radiation-dominated shocks, general relativistic light bending in the gravitational 
potential well of the neutron star, and complicated multi-dimensional 
radiative transfer.
Many attempts have been made to relate the formation of the spectrum in X-ray pulsars 
to the physics of the accretion dynamics, based on either gas-mediated 
shocks \citep{LangerandRappaport1982}, or Coulomb collisional stopping
\citep[e.g.,][]{MillerWassermanandSalpeter1989}, but these have not demonstrated good 
agreement with X-ray pulsar spectral data \citep[e.g.,][]{MeszarosandNagel1985a,MeszarosandNagel1985b}.

More recently, the development of a new class of models that focuses on the effects of Comptonization in
the accretion columns has significantly improved the agreement between the theoretical predictions and the observed phase-averaged
X-ray spectra for a number of sources.
The most comprehensive physical model currently available for the formation of the
spectra in accretion-powered X-ray pulsars was developed 
by \citet[][hereafter BW07]{BeckerandWolff2007},
who studied the reprocessing of cyclotron, bremsstrahlung, and blackbody seed photons 
injected into a cylindrical accretion column located over one of the neutron star's magnetic poles. 
The model is based on a detailed treatment of Comptonization, which is the energization of photons via 
electron scattering. 
This process can be broken into two components, namely thermal Comptonization (due to the stochastic 
velocity of the electrons), and bulk Comptonization (due to the accretion velocity of the electrons). 
Photons injected with a monochromatic energy distribution subsequently propagate throughout
the energy space as a result of Comptonization, and they also diffuse through the 
physical space as they execute a random walk by scattering off the electrons, 
until they eventually escape through the walls or the top of the accretion column.
Although the process is stochastic, in an average scattering event, net energy is transferred 
from the electrons to the photons, and the escape of the upscattered seed photons through the 
walls or top of the column carries away the kinetic energy of the accreting material in the form of X-rays, 
thereby allowing the gas to settle onto the star.
This mechanism characteristically produces a power-law continuum in the energy range $\sim 1-30\,{\rm keV}$,
with a quasi-exponential cutoff (due to electron recoil) at higher energies. 
BW07 successfully applied their model to the interpretation of the phase-averaged X-ray spectra observed 
from Her X-1, Cen X-3, and LMC X-4, hence providing for the first time a firm 
theoretical foundation for the formation of the X-ray spectra emitted by luminous X-ray pulsars. 
Related work, reaching similar conclusions, was carried out by \citet{Farinelli_etal2016}, \citet{Wolff_etal2016}, \citet{West_etal2017a,West_etal2017b}, and \cite{Thalhammer_etal2021}.

The BW07 model has demonstrated success in applications to high-luminosity accretion-powered X-ray pulsars, with relatively flat power-law spectra, but it has proven more difficult to apply the model to low-luminosity sources such as X Per, which display steeper spectra. The problems with the low-luminosity sources stem from the fact that in the BW07 model, the accretion velocity is assumed to approach zero at the neutron star 
surface as a consequence of the strong deceleration occurring in an extended, smooth, radiation-dominated
shock wave. This type of velocity profile is expected in high-luminosity ``supercritical'' sources such
as Her X-1 \citep{Becker_etal2012}, 
in which the radiation luminosity exceeds the effective Eddington limit for the gas, which is usually 
referred to as the critical luminosity, given by \citep[e.g.,][]{Burnard_etal1991}
\begin{equation}
L_{\rm crit} = L_{\rm E}
\, \frac{\pi r_0^2}{4 \pi R_*^2} \, \frac{\sigmaT}{\sigmapar} \ ,
\label{eq1.1}
\end{equation}
where $r_0$ denotes the radius of the cylindrical accretion column, $R_*$ is the stellar radius, $\sigmaT$ is the Thomson cross section, $\sigpar$ denotes the electron scattering cross section for photons propagating parallel to the axis of the accretion column (which is also the magnetic field axis), and $L_{\rm E}$ represents the standard spherical Eddington limit for pure, fully-ionized hydrogen, defined by
\begin{equation}
L_{\rm E} \equiv \frac{4 \pi G M_* m_p c}{\sigmaT} =1.26 \times 10^{38} \left(\frac{M_*}{\msun}\right) \, {\rm ergs \ sec}^{-1} \ ,
\label{eq1.1b}
\end{equation}
with $m_p$ denoting the proton mass, $c$ the speed of light, and $M_*$ the stellar mass. The second factor on the right-hand side of Equation~(\ref{eq1.1}) represents the reduction in the critical luminosity due to the area of the accretion column, compared with the stellar surface area, and the third factor represents the increase in the critical luminosity, resulting from the fact that $\sigmapar\lesssim\sigmaT$ in a strong magnetic field, for energies significantly below the cyclotron energy
\citep[e.g.,][]{Ventura1979,MeszarosandVentura1979}.

The accretion dynamics for sources with X-ray luminosity $L_X \gtrsim L_{\rm crit}$, such as Her X-1, are expected to be dominated by radiation pressure, with the gas decelerating essentially to rest at the stellar surface, after passing through a radiative, radiation-dominated standing shock \citep{BeckerandWolff2005a,BeckerandWolff2005b}. Conversely, for sources with $L_X \ll L_{\rm crit}$, such as X Per, the role of radiation pressure is greatly reduced, and the gas may collide with the stellar surface with a substantial fraction of the local free-fall velocity. For sources with luminosity $L_X \sim L_{\rm crit}$, the situation is more complex, and the accretion dynamics will be affected by additional details, such as the dependence of the electron scattering cross section on the photon energy and propagation direction \citep{Becker_etal2012,Mushtukov_etal2015a}.

An interesting trend emerging from observations of accretion-powered X-ray pulsars over the
past decades suggests that the index of the power-law continuum tends to 
reduce (i.e., the spectrum becomes flatter and harder) with increasing luminosity. 
In the case of Her X-1, this same trend is observed both in the long-term variability of the 
source luminosity,
and also in the pulse-to-pulse variability \citep{Klochkov_etal2011}. 
In the context of the BW07 model, this type of spectral variability with 
changes in the luminosity could be a natural consequence of changes in the 
magnitude of the radiation pressure, which ultimately controls the impact 
velocity of the material accreting onto the surface of the neutron star, and 
therefore determines the amount of $P$d$V$ work done on the radiation field by the compressing gas.
This effect is stronger in the high-luminosity (supercritical) sources, because of the enhanced compression,
resulting in a flatter continuum. On the other hand, in the subcritical sources, the compression is weaker and the resulting X-ray spectrum is steeper and softer.
We discuss this idea further in Section~\ref{sec:concl}.

Motivated by the successes and limitations of the BW07 model, our goal in this paper is to develop a new, generalized model
that expands upon the BW07 model to include a variety of new enhancements.
Namely, the new model:
(1) utilizes a conical geometry rather than the cylindrical geometry employed by BW07;
(2) allows the computation of separate radiation components emitted through the walls and top of the column;
(3) includes a new boundary condition at the stellar surface that allows for any impact velocity between free-fall and zero velocity;
(4) incorporates a new velocity profile that smoothly merges with Newtonian free-fall far from the star;
(5) utilizes a proper free-streaming boundary condition at the top of the accretion column;
and (6) allows for the possibility of a standing shock at the column top.

We solve the steady-state radiation transport equation in a conical geometry, including the effects of thermal and bulk Comptonization, to obtain the analytical solution for the Green's function, which represents the contribution to the observed steady-state photon spectrum resulting from the continual injection of monochromatic seed photons. By exploiting the linear structure of the mathematical problem, we show how the Green's function can be convolved with an arbitrary source distribution to obtain the particular solution for the steady-state spectrum resulting from the continual injection of photons from any physical source mechanism. Examples include bremsstrahlung, blackbody, and cyclotron emission. The formalism also allows us to the calculate the separate radiation components emitted through the walls and top of the accretion column, which facilitates the computation of phase-dependent X-ray spectra, although we do not perform such calculations here.

We demonstrate that our new theoretical model is able to successfully reproduce the observed spectra for X-ray pulsars across five orders of magnitude in luminosity, from low-luminosity (subcritical) sources, up to relatively high-luminosity (supercritical) sources. We find that the spectra of the high-luminosity sources is dominated by Comptonized bremsstrahlung emission, and the spectra of the low-luminosity sources is dominated by Comptonized blackbody emission, powered by the residual kinetic energy of the flow at the stellar surface, as first suggested by \citet{BeckerandWolff2005b}. We illustrate the range of application of the new model by using it to qualitatively fit the X-ray continuum spectra for the supercritical source Her X-1 and the subcritical source X Per.

The remainder of the paper is organized as follows. In Section~\ref{sec:radproc} we briefly review the nature of the primary radiation transport mechanisms in the accretion column with a focus on dynamical, thermal, and magnetic effects. The transport equation governing the formation of the radiation spectrum is introduced and analyzed in Section~\ref{sec:radxfer}, and in Section~\ref{sec:solgrnfun} the exact analytical solution for the Green's function describing the radiation distribution inside the accretion column is derived. The spectrum of the radiation escaping through the walls and top of the accretion column is developed in Section~\ref{sec:emradspec}, and the physical constraints for the various model parameters are considered in Section~\ref{sec:modparcon}. The nature of the source terms describing the injection of blackbody, cyclotron, and bremsstrahlung seed photons into the accretion column is discussed in Section~\ref{sec:phsources}. Emergent X-ray spectra are computed in Section~\ref{sec:astroapps}, and the results are compared with the observational data for Her X-1 and X Per, which have widely differing luminosities. The self-consistency of the model is investigated in Section~\ref{sec:modselfcon}, and the implications of our work for the production of X-ray spectra in accretion-powered X-ray pulsars are discussed in Section~\ref{sec:concl}.

\section{RADIATIVE PROCESSES}
\label{sec:radproc}

The formation of the emergent X-ray continuum spectra in accretion-powered X-ray pulsars is a complex process that is powered fundamentally by the conversion of gravitational potential energy into kinetic energy of the accreting gas, which is transferred to the radiation field via electron scattering, and ultimately carried away by the energy of the escaping radiation. Accretion onto the surface of a neutron star differs qualitatively from black-hole accretion in the sense that the star obviously possesses a hard surface, and the accreting gas must therefore eventually decelerate to merge with the stellar crust. Conversely, in the case of black-hole accretion, no such solid solid barrier exists, although a centrifugal barrier may still develop in the flow due to a balance between centripetal and gravitational forces \citep[e.g.,][]{LeandBecker2004}. When an obstacle exists in the flow, whether due to a solid surface or due to a centrifugal barrier, shocks may form. In the case of neutron star accretion, the nature of the shock depends primarily on the accretion rate, which determines the luminosity of the escaping radiation field. In low-luminosity sources, the shock is expected to take the form of a discontinuous gas-mediated shock \citep{LangerandRappaport1982}, and in high-luminosity sources, the shock is likely to be radiation-dominated, smooth, and radiative, meaning that the kinetic energy of the inflowing gas is radiated away in the shock itself.

\begin{figure}[t]
\begin{center}
\hskip-0.4truein
\epsfig{file=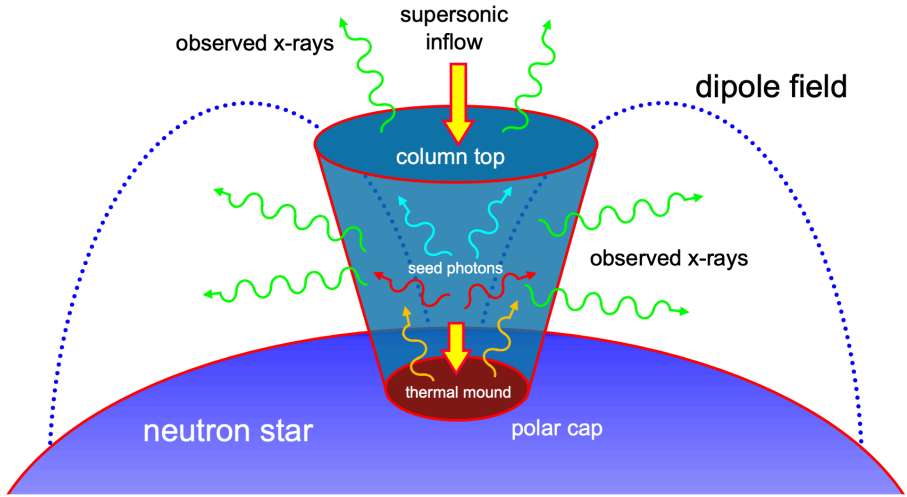,height=12.0cm,angle=-90}
\end{center}
\vskip-0.5truein
\caption{Schematic depiction of gas in a conical accretion column accreting onto the magnetic pole of a neutron star. Seed photons created via blackbody ({\it gold lines}), cyclotron ({\it red lines}), and bremsstrahlung ({\it cyan lines}) emission are reprocessed via electron scattering and eventually escape through the walls and top of the column to form the observed X-ray spectrum ({\it green lines}).
\label{fig:schematic}}
\end{figure}

The accretion flow in an X-ray pulsar is illustrated schematically in Figure~\ref{fig:schematic}. Physically, the accretion scenario corresponds to the flow of a mixture of gas and radiation inside a dipole-shaped magnetic ``pipe'' in which the fully-ionized gas is trapped by the strong magnetic field, but from which the radiation can escape via a three-dimensional random walk mediated by electron scattering. The primary mechanisms for the production of seed photons in X-ray pulsar accretion columns are cyclotron, bremsstrahlung, and blackbody emission. The seed photons are subsequently Compton scattered in energy due to collisions with hot electrons in the accreting gas, before escaping through the walls or top of the column. Bremsstrahlung emission provides a broad-band source of seed photons that are generated throughout the accretion column. Cyclotron emission also generates seed photons throughout the column, but rather than being a broad-band source, this process generates photons with an energy equal to the difference between the energy of the first excited Landau level and the ground state. Finally, blackbody seed photons are emitted from the surface of the ``thermal mound'' located close to the stellar surface, where the accreting gas gets sufficiently dense that thermodynamic equilibrium is achieved. Above the thermal mound, the opacity is dominated by electron scattering.

\subsection{Radiative Transfer in Magnetized Media}

The strong ($B \sim 10^{12}\,$G) magnetic field channels the flow of gas from the accretion disk surrounding the neutron star onto one (or both) of the star's magnetic poles. The presence of the strong magnetic field in X-ray pulsars has profound implications not only for the dynamics of the accretion flow, but also for the nature of the photon propagation through the accreting plasma. For example, vacuum polarization leads to birefringent behavior that produces two linearly polarized normal modes \citep{Ventura1979,Nagel1980,ChananNovickandSilver1979}. The electric field vector for the radiation is located in the plane formed by the photon propagation direction and the neutron star's magnetic field for the ordinary mode. Conversely, for the extraordinary mode, the electric field vector of the radiation is pointed perpendicular to this plane. The nature of the photon-electron scattering process differs qualitatively for the two polarization modes, and it also depends critically on whether the photon energy, $\epsilon$, exceeds the cyclotron energy, $\epsilon_c$, defined by
\begin{equation}
\epsilon_c \equiv \frac{e B h}{2 \pi m_e c} = 11.57 \left(\frac{B}{10^{12}\,{\rm G}}\right) \ {\rm keV}
\ ,
\label{eq2.1}
\end{equation}
where $c$, $h$, $m_e$, and $e$ represent the speed of light, Planck's constant, and the electron mass and charge, respectively.

\subsubsection{Electron Scattering Cross Sections in Magnetized Plasma}

\citet{Ventura1979} derived expressions for the electron scattering cross sections for extraordinary and ordinary mode photons propagating in a magnetized plasma, neglecting the effects of vacuum polarization. The results he obtained for the plasma-only cross sections are valid provided the photon energy, $\epsilon$, greatly exceeds the plasma energy, $\epsilon_{p}$, given by
\begin{equation}
\epsilon_{p} = \frac{h \omega_{p}}{2 \pi } = 0.371\,\left(\frac{n_e}{10^{26}\,{\rm cm}^{-3}}\right)^{1/2} \, {\rm keV}
\ ,
\label{eqElecPlasma1}
\end{equation}
where $n_e$ denotes the electron number density, and the electron plasma frequency, $\omega_{p}$, is defined by
\begin{equation}
\omega_{p} \equiv \left(\frac{4 \pi n_e e^2}{m_e}\right)^{1/2}
\ .
\label{eqElecPlasma2}
\end{equation}
Our primary focus here is on the pulsars X Per and Her X-1, in which case the electron number density, $n_e$, at the base of the accretion column is in the range $10^{18}\,{\rm cm}^{-3} \lesssim n_e \lesssim 10^{24}\,{\rm cm}^{-3}$, with the lower limit corresponding to X Per and the upper limit to Her X-1. Substituting this range of number densities into Equation~(\ref{eqElecPlasma1}) leads to the conclusion that $\epsilon_p \ll 1\,$keV for both of the sources. Hence $\epsilon \gg \epsilon_{p}$ in the X-ray pulsar application, and therefore the expressions derived by \citet{Ventura1979} are applicable, provided the effects of vacuum polarization are negligible, which we discuss further in Section~\ref{sec:EVP}. The primary results derived by \citet{Ventura1979} are summarized below.

\begin{figure}[t]
\begin{center}
\epsfig{file=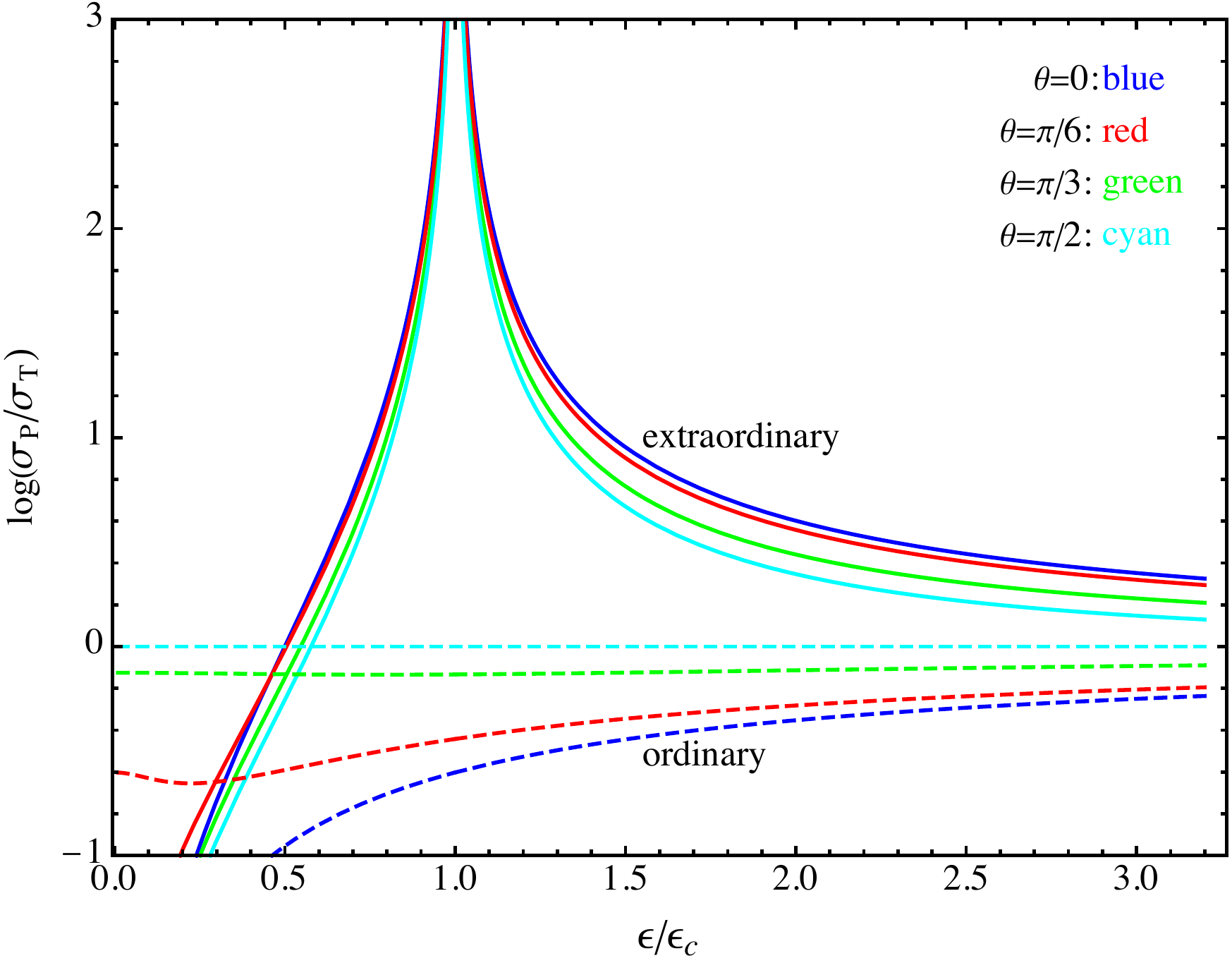,height=8.0cm}
\end{center}
\vskip-0.2truein
\caption{Electron scattering cross sections in a magnetized plasma, $\sigma_{_{P1}}$ and $\sigma_{_{P2}}$, plotted as functions of the energy ratio $\epsilon/\epsilon_c$ relative to the Thomson value $\sigma_{_{\rm T}}$ for extraordinary mode ({\it solid lines}) and ordinary mode ({\it dashed lines}) using Equations~(\ref{eqExtV79}) and (\ref{eqOrdV79}). The value of the propagation angle $\theta$ is indicated. Note the strong resonance at the cyclotron energy for the extraordinary mode. These results are consistent with Figure~2 from \citet{Ventura1979}.
\label{fig:elecscat}}
\end{figure}

In the absence of vacuum polarization effects, and assuming that $\epsilon \gg \epsilon_{p}$, \citet{Ventura1979} demonstrated that the pure-plasma electron scattering cross sections for extraordinary and ordinary mode photons, denoted by $\sigma_{_{P1}}$ and $\sigma_{_{P2}}$, respectively, can be written as
\begin{equation}
\frac{\sigma_{_{P1}}}{\sigmaT} = \left[1+\alpha^2(\theta)\right]^{-1} \left\{\alpha^2(\theta)\sin^2\theta
+ \frac{1}{2} \left[\frac{1+\alpha(\theta)\cos\theta}{1+u^{1/2}}\right]^2
+ \frac{1}{2} \left[\frac{1-\alpha(\theta)\cos\theta}{1-u^{1/2}}\right]^2\right\}
\ ,
\label{eqExtV79}
\end{equation}
and
\begin{equation}
\frac{\sigma_{_{P2}}}{\sigmaT} = \left[1+\alpha^2(\theta)\right]^{-1} \left\{\sin^2\theta
+ \frac{1}{2} \left[\frac{\cos\theta - \alpha(\theta)}{1+u^{1/2}}\right]^2
+ \frac{1}{2} \left[\frac{\cos\theta + \alpha(\theta)}{1-u^{1/2}}\right]^2\right\}
\ ,
\label{eqOrdV79}
\end{equation}
where $\theta$ is the angle between the magnetic field vector and the photon propagation direction, and the function $\alpha(\theta)$ is defined by
\begin{equation}
\alpha(\theta) \equiv \frac{-b}{1+(1+b^2)^{1/2}} \ ,
\label{eq2V79}
\end{equation}
with
\begin{equation}
b = 2 \, u^{-1/2} \, \frac{\cos\theta (1-v)}{\sin^2\theta} \ ,
\label{eq3V79}
\end{equation}
and
\begin{equation}
u \equiv \left(\frac{\epsilon}{\epsilon_c}\right)^{-2} \ ,
\qquad v \equiv \left(\frac{\epsilon}{\epsilon_p}\right)^{-2} \ .
\label{eq4V79}
\end{equation}
We note that $v$ is essentially zero in the X-ray pulsar application since $\epsilon \gg \epsilon_p$. In order to validate our implementation of the cross sections, in Figure~\ref{fig:elecscat} we plot $\sigma_{_{P1}}$ and $\sigma_{_{P2}}$, evaluated using Equation~(\ref{eqExtV79}) and (\ref{eqOrdV79}), respectively. The results agree with Figure~2 from \citet{Ventura1979}. The extraordinary mode cross section exhibits a strong resonance at the cyclotron energy, as expected, while the ordinary mode cross section is non-resonant. Near the resonance, the extraordinary mode cross section greatly exceeds the Thomson value, whereas the ordinary mode cross section remains essentially sub-Thomson at all photon energies. We shall see below that the results for the scattering cross section are qualitatively different once the effects of vacuum polarization are considered.

\subsubsection{Effects of Vacuum Polarization}
\label{sec:EVP}

The results obtained by \citet{Ventura1979} and presented in Equations~(\ref{eqExtV79}) and (\ref{eqOrdV79}) are applicable when the effects of vacuum polarization are negligible, which corresponds to the energy range $\epsilon \lesssim \epsilon_{\rm vac}$, where the vacuum polarization energy, $\epsilon_{\rm vac}$, is given by \citep{MeszarosandVentura1979}
\begin{equation}
\epsilon_{\rm vac} \equiv \frac{h \omega_{\rm vac}}{2 \pi }
= 1.32 \left(\frac{n_e}{10^{20}\,{\rm cm}^{-3}}\right)^{1/2} \left(\frac{B}{10^{12}\,{\rm G}}\right)^{-1} \, {\rm keV}
\ ,
\label{eqVacEnergy1}
\end{equation}
and the vacuum polarization frequency, $\omega_{\rm vac}$, is defined by
\begin{equation}
\omega_{\rm vac} \equiv \left(\frac{15 \pi}{\alpha_F}\right)^{1/2} \left(\frac{B}{B_c}\right)^{-1} \omega_p
\ .
\label{eqVacEnergy2}
\end{equation}
Here, $\alpha_F$ and $\omega_{p}$ denote the fine-structure constant and the electron plasma frequency (Equation~(\ref{eqElecPlasma2})), respectively, and $B_c$ represents the critical magnetic field strength, given by
\begin{equation}
B_c \equiv \frac{2 \pi m_e^2 c^3}{e h} = 4.415 \times 10^{13} \, {\rm G}
\ ,
\label{eqCritField}
\end{equation}
which is obtained by setting $\epsilon_c = m_e c^2$ (see Equation~(\ref{eq2.1})).
Vacuum polarization effects will strongly influence the electron scattering cross section when the photon energy $\epsilon \gtrsim \epsilon_{\rm vac}$. In the X-ray pulsar accretion columns of interest here, $B \sim 10^{12}\,$G, and the electron number density, $n_e$, lies in the range $10^{18}\,{\rm cm}^{-3} \lesssim n_e \lesssim 10^{24}\,{\rm cm}^{-3}$, where the lower and upper limits correspond to X Per and Her X-1, respectively. According to Equation~(\ref{eqVacEnergy1}), the corresponding range for the vacuum polarization energy is therefore $0.1\,{\rm keV} \lesssim \epsilon_{\rm vac} \lesssim 100\,{\rm keV}$. Focusing on the case of Her X-1, with $\epsilon_{\rm vac} \sim 100\,{\rm keV}$, it is clear that vacuum polarization effects will be unimportant for the X-ray continuum in this source. On the other hand, in the case of X~Per, with $\epsilon_{\rm vac} \sim 0.1\,{\rm keV}$, we note that vacuum polarization effects will be very important for the formation of the entire X-ray continuum. We summarize the results for the electron scattering cross sections including the effects of vacuum polarization below.

\citet{MeszarosandVentura1979} derived expressions for the electron scattering cross sections including the modifications due to vacuum polarization. The results they obtained for the cross sections for extraordinary and ordinary mode photons, denoted by $\sigma_{_{V1}}$ and $\sigma_{_{V2}}$, respectively, can be written as
\begin{equation}
\begin{aligned}
\frac{\sigma_{_{V1}}}{\sigmaT} = (1+K_1^2)^{-1} &\bigg\{\frac{1}{2} (1-K_1 \cos\theta)^2[(1-u^{1/2})^2
+ \gamma^2]^{-1} \cr
&+ \frac{1}{2} (1+K_1 \cos\theta)^2 [1+u^{1/2}]^{-2} + K_1^2 \sin^2\theta\bigg\}
\ ,
\end{aligned}
\label{eqExtMV79}
\end{equation}
and
\begin{equation}
\begin{aligned}
\frac{\sigma_{_{V2}}}{\sigmaT} = (1+K_2^2)^{-1} &\bigg\{\frac{1}{2} (1-K_2 \cos\theta)^2[(1-u^{1/2})^2
+ \gamma^2]^{-1} \cr
&+ \frac{1}{2} (1+K_2 \cos\theta)^2 [1+u^{1/2}]^{-2} + K_2^2 \sin^2\theta\bigg\}
\ ,
\end{aligned}
\label{eqOrdMV79}
\end{equation}
where the radiation damping constant $\gamma$ for photon frequency $\omega$ is defined by
\begin{equation}
\gamma \equiv \frac{2}{3} \frac{e^2 \omega}{m_e c^3} \ ,
\label{eq2MV79}
\end{equation}
and the quantities $K_1$ and $K_2$ are computed using
\begin{equation}
K_1 = K_2^{-1} = \alpha_V(\theta) \ ,
\label{eq3MV79}
\end{equation}
with
\begin{equation}
\alpha_V(\theta) \equiv \frac{-1}{{b_V+(1+b_V^2)^{1/2}}} \ ,
\label{eq4MV79}
\end{equation}
and
\begin{equation}
b_V = \frac{u^{1/2} \sin^2\theta}{2 \cos\theta (1-v)} \left[1+\frac{3\delta(1-u)}{v u}\right] \ .
\label{eq5MV79}
\end{equation}
The quantities $u$ and $v$ are defined in Equation~(\ref{eq4V79}), and the parameter $\delta$ is computed using
\begin{equation}
\delta \equiv \frac{2 e^2}{45  hc} \left(\frac{B}{B_c}\right)^2
= 5.16 \times 10^{-5} \left(\frac{B}{B_c}\right)^2 \ ,
\label{eq6MV79}
\end{equation}
where the critical field, $B_c$, is defined in Equation~(\ref{eqCritField}).

\begin{figure}[t]
\begin{center}
\epsfig{file=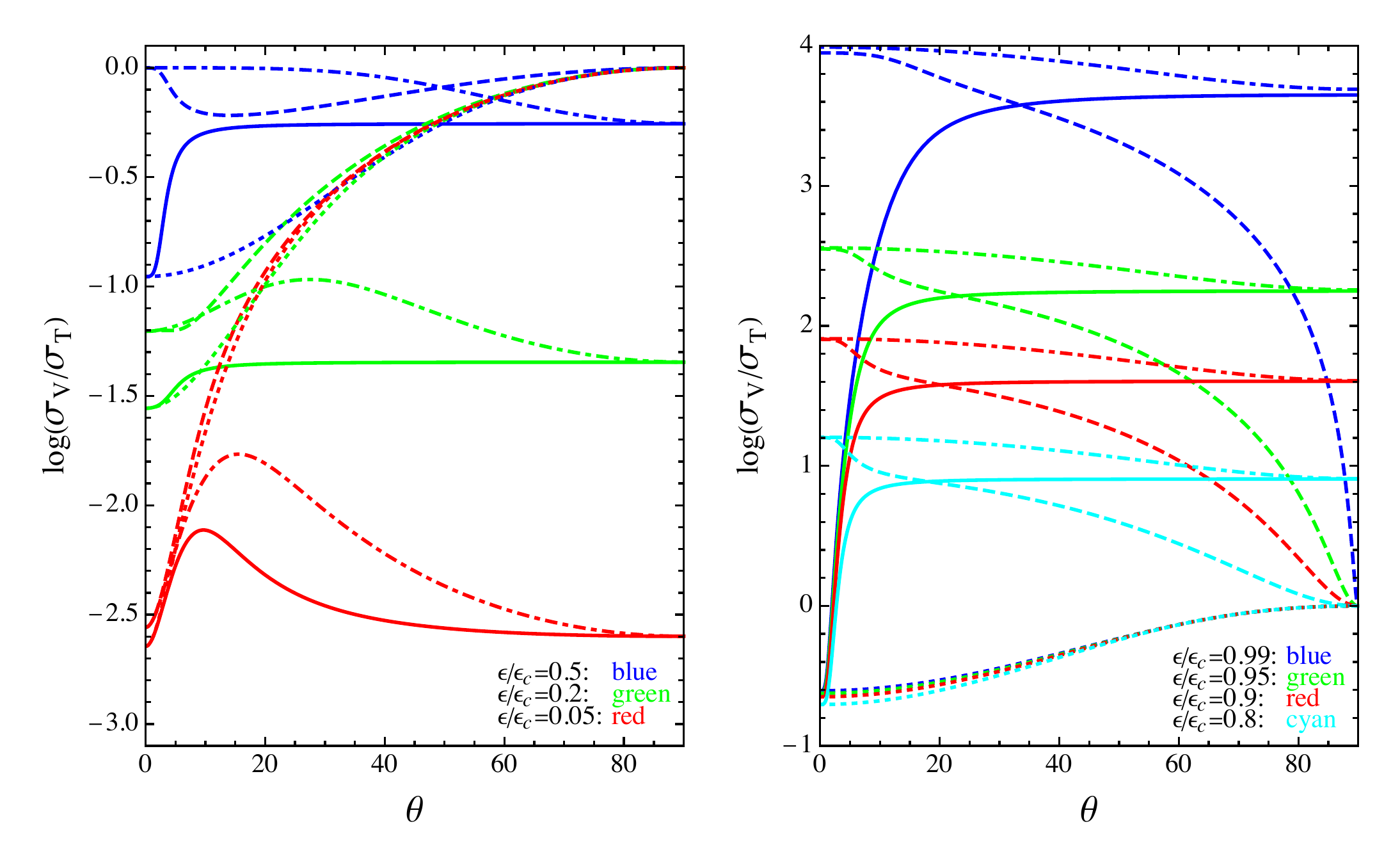,height=9.0cm}
\end{center}
\vskip-0.5truein
\caption{Vacuum-modified electron scattering cross sections, $\sigma_{_{V1}}$ and $\sigma_{_{V2}}$, plotted as functions of the propagation angle $\theta$ relative to the Thomson value $\sigma_{_{\rm T}}$ for extraordinary mode ({\it solid lines}) and ordinary mode ({\it dashed lines}) photons using Equations~(\ref{eqExtMV79}) and (\ref{eqOrdMV79}) for $n_e = 1.89 \times 10^{21}\,{\rm cm}^{-3}$ and $B = 4.41 \times 10^{12}\,$G, which are the values used in Figure~4 from \citet{MeszarosandVentura1979}. The value of the photon energy ratio $\epsilon/\epsilon_c$ is indicated. For comparison, the plasma-only cross sections are also plotted for the extraordinary mode ({\it dot-dashed lines}) and ordinary mode ({\it dotted lines}) using Equations~(\ref{eqExtV79}) and (\ref{eqOrdV79}).
\label{fig:secondelecscatplot}}
\end{figure}

\begin{figure}[t]
\begin{center}
\epsfig{file=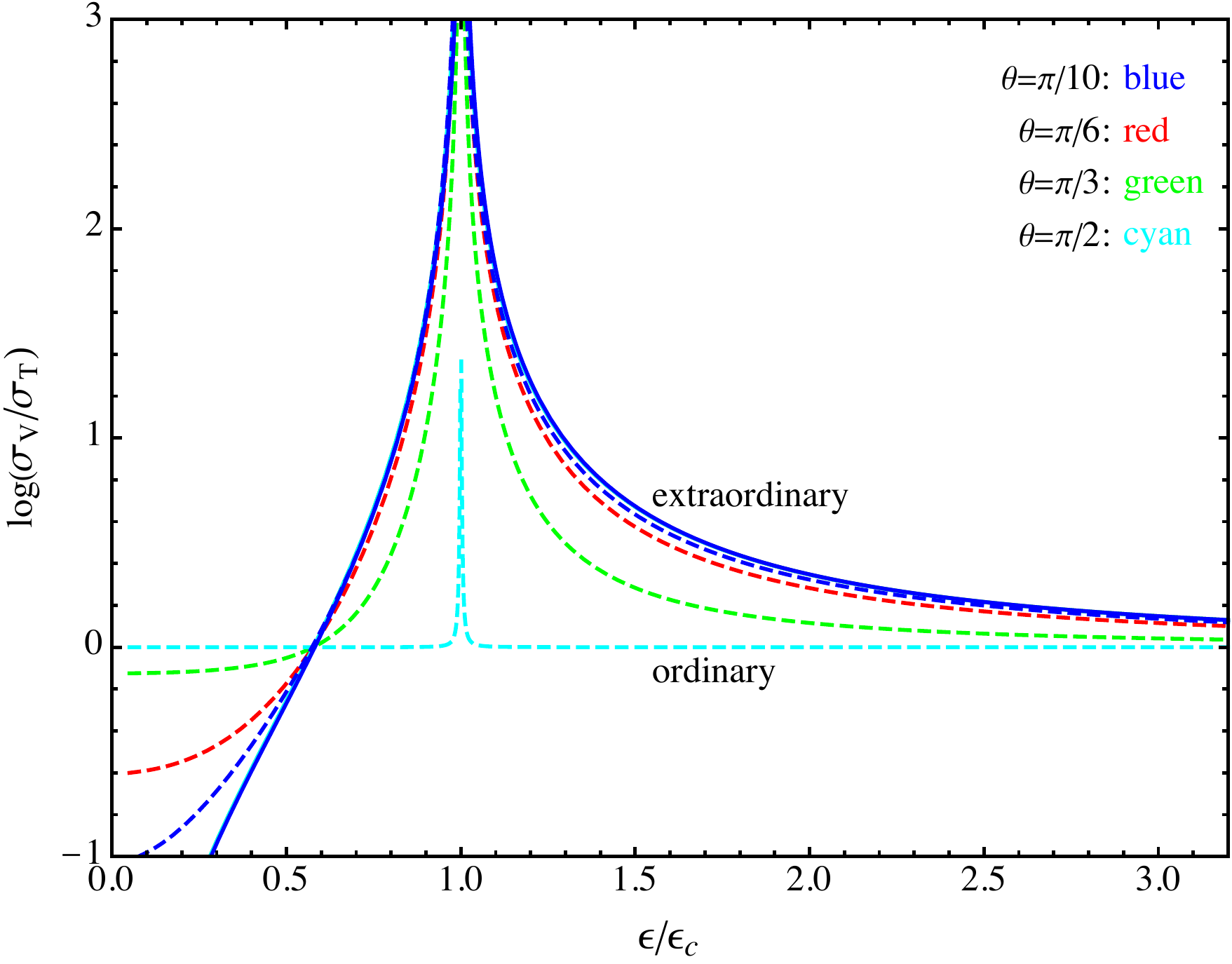,height=9.0cm}
\end{center}
\vskip-0.3truein
\caption{Vacuum-modified electron scattering cross sections, $\sigma_{_{V1}}$ ({\it solid lines}) and $\sigma_{_{V2}}$ ({\it dashed lines}), plotted as functions of the photon energy ratio $\epsilon/\epsilon_c$ relative to the Thomson value $\sigma_{_{\rm T}}$ using Equations~(\ref{eqExtMV79}) and (\ref{eqOrdMV79}) for $n_e = 1.89 \times 10^{21}\,{\rm cm}^{-3}$ and $B = 4.41 \times 10^{12}\,$G. The value of the propagation angle $\theta$ is indicated. Note that in the angular range considered here, $\sigma_{_{V1}}$ is independent of $\theta$.
\label{fig:secondelecscatplot2}}
\end{figure}

In order to validate our implementation of the vacuum-modified cross sections, in Figure~\ref{fig:secondelecscatplot} we utilize Equations~(\ref{eqExtMV79}) and (\ref{eqOrdMV79}) to plot $\sigma_{_{V1}}$ and $\sigma_{_{V2}}$ using the parameter values $n_e = 1.89 \times 10^{21}\,{\rm cm}^{-3}$ and $B = 4.41 \times 10^{12}\,$G, which are the values used by \citet{MeszarosandVentura1979} in their Figure~4. For these parameters, we obtain $\epsilon_c = 51\,$keV and $\epsilon_{\rm vac} = 1.3\,$keV, according to Equations~(\ref{eq2.1}) and (\ref{eqVacEnergy1}), respectively. Hence in this case vacuum polarization will strongly modify the scattering cross sections for photons with energy $\epsilon \gtrsim 1.3\,$keV, which comprises the entire X-ray band. This behavior is clearly seen in Figure~\ref{fig:secondelecscatplot}, especially for energies close to the cyclotron resonance, where we note that the vacuum-modified extraordinary mode cross section (solid lines) displays a rapid increase as a function of the propagation angle $\theta$, from sub-Thomson values for $\theta \sim 0$ to highly super-Thomson values for $\theta \sim 90^\circ$. This behavior differs qualitatively from that of the plasma-only extraordinary-mode cross section (dot-dashed lines), which does not display a strong dependence on $\theta$ for any value of the photon energy.

In Figure~\ref{fig:secondelecscatplot2} we use Equations~(\ref{eqExtMV79}) and (\ref{eqOrdMV79}) to plot the vacuum-modified electron scattering cross sections for the extraordinary and ordinary mode photons as functions of the photon energy ratio, $\epsilon/\epsilon_c$, for the same values of $n_e$ and $B$ used in Figure~\ref{fig:secondelecscatplot}. Due to the effect of vacuum polarization, the extraordinary mode cross section is now a very weak function of the propagation angle $\theta$, for $\theta \gtrsim 20^\circ$, and the ordinary mode now displays a resonance at the cyclotron energy, which was absent in the plasma-only cross section plotted in Figure~\ref{fig:elecscat}. It is interesting to note that the value of the electron number density used in Figures~\ref{fig:secondelecscatplot} and \ref{fig:secondelecscatplot2} lies between the values expected at the base of the accretion columns in Her X-1 and X Per, and therefore the behaviors displayed in these two figures are likely to be intermediate between the two sources.
We will consider the cross section values in more detail in Section~\ref{sec:modselfcon}.

\subsubsection{Simplified Scattering Cross Sections}

The detailed energy and angular dependences of the electron scattering cross sections for the extraordinary and ordinary polarization modes, including the effects of vacuum polarization, are given by Equations~(\ref{eqExtMV79}) and (\ref{eqOrdMV79}), respectively. Unfortunately, it is not possible to include the full complexity of these expressions into the model developed here, and we will therefore follow BW07 and \citet{WangandFrank1981} by introducing a set of approximate scattering cross sections, averaged over the photon energy and the two polarization modes. The cross sections for photons propagating parallel and perpendicular to the radial direction are denoted by $\sigpar$ and $\sigperp$, respectively, and the angle-averaged cross section is denoted by $\sigbar$. The magnitudes of these cross sections are fairly well understood for relatively luminous sources such as Her X-1, in which case one usually finds that $\sigperp \sim \sigmaT$ and $\sigperp \gg \sigbar \gg \sigpar$ \citep[e.g.,][]{Wolff_etal2016}. The hierarchy of these cross section values stems from the fact that $\epsilon < \epsilon_c < \epsilon_{\rm vac}$ in Her X-1, and therefore the super-Thomson cross section values near the cyclotron resonance are avoided (see Figures~\ref{fig:elecscat} and \ref{fig:secondelecscatplot}).
Conversely, in the case of low-luminosity sources such as X Per, the situation is reversed, and we generally find that $\epsilon_{\rm vac} < \epsilon < \epsilon_c$. The change in the energy hierarchy leads to a greatly increased contribution from the cyclotron resonance, which results in cross section values $\sigperp \gg \sigmaT$ and $\sigpar \lesssim \sigbar \ll \sigperp$. This issue is further discussed in Section~\ref{sec:modselfcon}.

\section{RADIATIVE TRANSFER}
\label{sec:radxfer}

The exact solution for the radiation spectrum inside a cylindrical X-ray pulsar accretion column was obtained by BW07 under the assumption that the flow velocity of the accreting gas approaches zero at the stellar surface, as a result of passage through a radiative, radiation-dominated shock wave. This assumption leads to a theoretical model that produces X-ray spectra that agree quite well with the relatively flat power-law spectra observed from luminous X-ray pulsars such as Her X-1 and LMC X-4. However, attempts to fit the same model to low-luminosity sources such as X Per, with steeper spectra, yield parameter values that are difficult to justify physically. We suspect that the problems with the low-luminosity sources stem from a need to generalize the boundary conditions utilized in the model. This observation has motivated a comprehensive reexamination of the boundary conditions, which in turn led to a complete reformulation of the model that not only generalizes the boundary conditions, but also incorporates a more realistic conical geometry for the accretion column. In addition, the new model also includes a more realistic velocity profile, and the capability to separately compute the radiation components emitted through the walls and the top of the accretion column. The transport equation introduced below includes the effects of special relativity up to first order in $\vel/c$, where $\vel$ is the accretion velocity. However, the effects of general relativity are not included, as this would greatly complicate the model, and the resulting corrections are not likely to be important at the level of approximation employed here. We present the core elements of the new model in this section.

\subsection{Transport Equation}

The time-dependent transport equation governing the propagation, scattering, and escape of radiation in a pulsar accretion column with an arbitrary geometry can be written in the vector form
\begin{eqnarray}
\frac{\partial f}{\partial t} + \vec \vel \cdot \vec\nabla f
&=& (\vec\nabla \cdot \vec \vel) \, \frac{\epsilon}{3} \,
\frac{\partial f}{\partial\epsilon}
+ \vec\nabla \cdot \left(\frac{c}{3 n_e \sigpar} \, \vec\nabla\cdot f \right)
\nonumber
\\
&+& \frac{n_e \sigbar c}{m_e c^2} \frac{1}{\epsilon^2}
\frac{\partial}{\partial\epsilon}\left[\epsilon^4\left(f
+ k T_e \, \frac{\partial f}{\partial\epsilon}\right)\right]
+ Q
\ ,
\label{eq3.1a}
\end{eqnarray}
where $\vec \vel$ is the accretion velocity field, $f(\vec r, \epsilon, t)$ is the photon distribution function, $\vec r$ is the spatial vector, $t$ is time, $\epsilon$ is the photon energy, $n_e$ is the electron number density, $T_e$ is the electron temperature, $k$ is Boltzmann's constant, $\sigpar$ represents the electron scattering cross section for photons propagating in the radial direction, and $\sigbar$ denotes the angle-averaged scattering cross section. The distribution function $f$ is related to the occupation number $\bar n$ via $f = 8 \pi \, \bar n / (c^3 h^3)$, and the quantity $\epsilon^2 f(\vec r, \epsilon, t) \, d\epsilon$ gives the number density of photons in the energy range between $\epsilon$ and $\epsilon+d\epsilon$. Hence the total photon number density, $n_r$, and energy density, $U_r$, are computed using the integrals
\begin{equation}
n_r(\vec r, t) = \int_0^\infty \epsilon^2 \,
f(\vec r, \epsilon, t) \, d\epsilon
\ , \qquad
U_r(\vec r, t) = \int_0^\infty \epsilon^3 \,
f(\vec r, \epsilon, t) \, d\epsilon
\ .
\label{eq3.2xx}
\end{equation}
The left-hand side of Equation~(\ref{eq3.1a}) denotes the comoving time derivative of $f$, and the terms on the right-hand side represent first-order Fermi energization (``bulk Comptonization''), spatial diffusion, thermal Comptonization, and the radiation source term, $Q$, respectively. We note  that equation~(\ref{eq3.1a}) is valid in the region of the accretion column above the thermal mound, where the opacity is dominated by electron scattering (see Section~\ref{sec:thermalmound}).

BW07 solved Equation~(\ref{eq3.1a}) under the assumption of a cylindrical accretion column geometry. Here, we adopt a {\it hollow conical geometry}, in which the solid angle of the column, $\Omega$, is given by
\begin{equation}
\Omega = 2 \pi (\cos\Theta_2 - \cos\Theta_1) \ ,
\label{eq3.2c}
\end{equation}
where $\Theta_1$ and $\Theta_2$ denote the opening angles for the outer and inner walls of the column, respectively. The conical shape is a more reasonable approximation of the true dipole geometry of the column, especially in the lower region close to the stellar surface. In the conical accretion column geometry, the accretion rate, $\dot M$, is related to the electron number density, $n_e$, the radius measured from the center of the neutron star, $R$, and the radial accretion velocity, $\vel<0$, via
\begin{equation}
\dot M = \Omega R^2 n_e m_p |\vel|
\ ,
\label{eq3.2b}
\end{equation}
where we have assumed that the accreting gas is composed of pure, fully-ionized hydrogen.

In order to treat a variety of photon injection scenarios, such as bremsstrahlung, cyclotron, and blackbody emission, it is convenient to compute the Green's function, $\green$, which represents the photon distribution inside the accretion column, resulting from the continual injection of $\dot N_0$ monochromatic seed photons per unit time, each with energy $\epsilon_0$, from a source located at radius $R_0$. In a conical geometry, the equation satisfied by $\green$ can be written as
\begin{eqnarray}
\frac{\partial \green}{\partial t} + \vel \, \frac{\partial \green}{\partial R}
&=& \frac{1}{R^2}\frac{\partial}{\partial R}\left(R^2 \, \vel \right)\,\frac{\epsilon}{3} \,
\frac{\partial \green}{\partial\epsilon}
+ \frac{1}{R^2} \frac{\partial}{\partial R}
\left(\frac{R^2 \, c}{3 n_e \sigpar}\,\frac{\partial \green}{\partial R}\right)
- \frac{\green}{t_{\rm esc}}
\nonumber
\\
&+& \frac{n_e \sigbar c}{m_e c^2} \frac{1}{\epsilon^2}
\frac{\partial}{\partial\epsilon}\left[\epsilon^4\left(\green
+ k T_e \, \frac{\partial \green}{\partial\epsilon}\right)\right]
+ \frac{\dot N_0 \, \delta(\epsilon-\epsilon_0) \, \delta(R-R_0)}{\Omega R_0^2 \, \epsilon_0^2}
\ ,
\label{eq3.1}
\end{eqnarray}
where $t_{\rm esc}$ represents the mean time for photons to escape through the walls of the accretion column. The escape of radiation through the top of the accretion column is implemented using a free-streaming boundary condition as discussed in Section~\ref{sec:SBC}. Equation~(\ref{eq3.1}) is obtained by adopting a conical geometry in Equation~(\ref{eq3.1a}) and then averaging over the azimuthal direction in the column. Hence the only spatial variable appearing in Equation~(\ref{eq3.1}) is the central radius $R$. We are interested in solving the steady-state version of Equation~(\ref{eq3.1}) and therefore we will set $\partial\green/\partial t = 0$.

The escape of radiation through the walls of the conical accretion column is modeled using an escape-probability formalism, based on the mean escape timescale, $t_{\rm esc}$, appearing in Equation~(\ref{eq3.1}). The value of $t_{\rm esc}$ is equal to the average time required for photons to diffuse through the walls of the column, which is computed using the expressions
\begin{equation}
t_{\rm esc}(R) = \frac{r_{\perp}}{\vel_{\rm diff}} \ , \qquad
\vel_{\rm diff}(R) = \frac{c}{\tauperp} \ ,
\label{eq3.3}
\end{equation}
where $\vel_{\rm diff}$ represents the radiation diffusion velocity. The perpendicular scattering optical thickness, $\tau_\perp$, and the perpendicular radius of the accretion column, $r_\perp$, are given by
\begin{equation}
\tau_\perp(R) = n_e \, \sigperp \, \colrad \ , \qquad
\colrad(R) = b \, R \ ,
\label{eq3.4}
\end{equation}
where the constant $b$ is given by
\begin{equation}
b = \tan\Theta_1 - \tan\Theta_2 \ .
\label{eq3.4b}
\end{equation}
The diffusion approximation employed in Equation~(\ref{eq3.3}) is valid provided the perpendicular scattering optical thickness $\tauperp > 1$, and it is important to confirm that this is the case in our applications. By combining relations, Equation~(\ref{eq3.3}) for the mean escape timescale can be rewritten as
\begin{equation}
t_{\rm esc}(R) = \frac{b^2 \sigperp \dot M}{\Omega \, m_p \, c \, |\vel|}
\ .
\label{eq3.5}
\end{equation}

Following BW07 and \citet{LyubarskiiandSyunyaev1982}, we shall assume that the electron distribution is isothermal, with constant temperature $T_e$, throughout the emission region. This is an acceptable approximation, since the electron temperature is expected to be regulated in a fairly small interval via thermal Comptonization \citep[e.g.,][]{SunyaevandTitarchuk1980,West_etal2017a,West_etal2017b}. Under the assumption of a constant electron temperature, it is convenient to work in terms of the dimensionless energy variable, $\chi$, defined by
\begin{equation}
\chi(\epsilon) \equiv \frac{\epsilon}{kT_e}
\ .
\label{eq3.9}
\end{equation}
It is also useful to transform the spatial variable from the radius, $R$, to the scattering optical depth, $\tau$, measured in the radial direction, starting from the stellar surface. The differential $d\tau$ is related to $dR$ via
\begin{equation}
d\tau = n_e(R) \, \sigpar \, dR
\ ,
\label{eq3.10a}
\end{equation}
which can be rewritten in terms of the dimensionless radius, $y$, defined by
\begin{equation}
y(R) \equiv \frac{R}{R_*}
\ ,
\label{eq3.10y}
\end{equation}
to obtain
\begin{equation}
d\tau = n_e(y) \, \sigpar R_* \, dy
\ .
\label{eq3.10}
\end{equation}
Integration of Equation~(\ref{eq3.10}) yields
\begin{equation}
\tau(y) = \int_{1}^y n_e(y') \, \sigpar R_* \, dy'
\ ,
\label{eq3.10tau}
\end{equation}
so that $\tau = 0$ at the stellar surface, where $y = 1$.

Making the change of variable from $(R,\epsilon)$ to $(\tau,\chi)$ in Equation~(\ref{eq3.1}) and using Equation~(\ref{eq3.5}) to substitute for the escape timescale $t_{\rm esc}$, we find after some algebra that the transport equation satisfied by the steady-state Green's function can be written in the form
\begin{eqnarray}
u \, \frac{\partial \green}{\partial \tau}
&=& \frac{1}{y^2}\,\frac{d}{d\tau}(y^2 u)\,\frac{\chi}{3} \,
\frac{\partial \green}{\partial\chi}
+ \frac{1}{3 y^2}\,\frac{\partial}{\partial\tau} \left(y^2 \, \frac{\partial\green}{\partial\tau}\right)
- \xi^2 u^2 y^2 \green
\nonumber
\\
&+& \frac{\sigbar}{\sigpar} \, \frac{kT_e}{m_e c^2}
\frac{1}{\chi^2}\frac{\partial}{\partial\chi}\left[\chi^4\left(\green
+ \frac{\partial\green}{\partial\chi}\right)\right]
+ \frac{\dot N_0 \, \delta(\chi-\chi_0) \, \delta(\tau-\tau_0)}{\Omega \, c \, R_0^2 \chi_0^2 (k T_e)^3}
\ ,
\label{eq3.11}
\end{eqnarray}
where $\tau_0$ and $\chi_0$ denote the injection optical depth and the dimensionless injection energy, respectively, and we have introduced the dimensionless accretion velocity, $u < 0$, defined by
\begin{equation}
u(\tau) \equiv \frac{\vel(\tau)}{c}
\ .
\label{eq3.11d}
\end{equation}
The dimensionless escape constant, $\xi$, introduced in Equation~(\ref{eq3.11}), parameterizes the rate of diffusion of photons through the walls of the conical accretion column, and is defined by
\begin{equation}
\xi \equiv \frac{\Omega \, R_* m_p c}{b \, \dot M (\sigpar \sigperp)^{1/2}}
\ ,
\label{eq3.12a}
\end{equation}
which can be rewritten as
\begin{equation}
\xi = \frac{\Omega}{4 \pi b} \left(\frac{\dot M}{\dot M_{\rm E}}\right)^{-1}
\left(\frac{\sigperp}{\sigmaT}\right)^{-1/2}
\left(\frac{\sigpar}{\sigmaT}\right)^{-1/2} \left(\frac{R_*}{R_g}\right)
\ ,
\label{eq3.12b}
\end{equation}
where $R_g=GM_*/c^2$ denotes the gravitational radius of the neutron star, and the Eddington accretion rate is defined by
\begin{equation}
\dot M_{\rm E} \equiv \frac{4 \pi G M_* m_p}{\sigmaT c}
= 1.40 \times 10^{17} \, \left(\frac{M_*}{\msun}\right) \ {\rm g \, s}^{-1}
\ .
\label{eqEddingtonMdot}
\end{equation}

\subsection{Separability}

It is important to identify the physical situations that result in the separability of the transport equation (Equation~(\ref{eq3.11})), since separability allows one to obtain the exact analytical solution for the steady-state Green's function $\green(\tau,\chi)$. Equation~(\ref{eq3.11}) is separable if each of the terms containing energy operators has the same dependence on the scattering optical depth, $\tau$. In this case, separability can be accomplished if the accretion velocity, $u$, satisfies the equation
\begin{equation}
\frac{1}{y^2} \frac{d}{d\tau}(y^2 u) = - \alpha \ ,
\label{eq3.13}
\end{equation}
where $\alpha$ is a positive constant. By utilizing Equations~(\ref{eq3.2b}), (\ref{eq3.10}), and (\ref{eq3.10tau}) to transform from $d\tau$ to $dy$ in Equation~(\ref{eq3.13}) and rearranging the resulting expression, we can show that
\begin{equation}
\frac{d}{dy}(y^2 u)^2
= \frac{2 \alpha \dot M \sigpar}{\Omega R_* m_p c} \, y^2 \ ,
\label{eq3.13i}
\end{equation}
which can be integrated to obtain
\begin{equation}
(y^2 u)^2
= \left(\frac{2 R_g}{R_*}\right) [k_\infty^2 (y^3-1) + k_0^2] \ ,
\label{eq3.13y}
\end{equation}
where $k_0$ is a constant of integration, and the parameter $\alpha$ is related to $k_\infty$ via
\begin{equation}
\alpha = \frac{3 k_\infty^2 \Omega \, R_g \, m_p c}{\sigpar \dot M} \ .
\label{eq3.13m}
\end{equation}
The expression for the dimensionless constant $\alpha$ can also be written as
\begin{equation}
\alpha = \frac{3 k_\infty^2 \Omega}{4 \pi}
\left(\frac{\dot M}{\dot M_{\rm E}}\right)^{-1} \left(\frac{\sigpar}{\sigmaT}\right)^{-1}
\ ,
\label{eqAlphaNew2}
\end{equation}
where the Eddington accretion rate $\dot M_{\rm E}$ is defined in Equation~(\ref{eqEddingtonMdot}). Based on Equation~(\ref{eq3.13y}), we find that the general expression for the accretion velocity profile, $u(y)$, resulting from the separability condition can be written in the form
\begin{equation}
u(y) = \frac{\vel(y)}{c}
= - \left(\frac{2 R_g}{R_*}\right)^{1/2} \, \left[\frac{k_\infty^2 (y^3-1) + k_0^2}{y^4}\right]^{1/2}
\ .
\label{eq3.13l}
\end{equation}

It is important to discuss the physical interpretation of the dimensionless constants $k_0$ and $k_\infty$ introduced in Equation~(\ref{eq3.13y}), which are connected with the asymptotic behaviors of the velocity profile function $u(y)$ as $y \to \infty$ or $y \to 1$. We note that at the stellar surface ($y=1$), Equation~(\ref{eq3.13l}) yields
\begin{equation}
u(y) \to - k_0 \left(\frac{2 G M_*}{R_* c^2}\right)^{1/2} \equiv \frac{\vel_*}{c} \ , \qquad y \to 1 \ ,
\label{eq3.13q}
\end{equation}
where $\vel_*$ is the impact velocity. The constant $k_0$ is equal to the ratio of the impact velocity divided by the Newtonian free-fall velocity, and we therefore refer to $k_0$ as the ``impact velocity constant.'' Next we note that as $y \to \infty$, the asymptotic behavior of Equation~(\ref{eq3.13l}) is given by
\begin{equation}
u(y) \to - k_\infty \left(\frac{2 G M_*}{R \, c^2}\right)^{1/2} \ , \qquad y \to \infty \ ,
\label{eq3.13w}
\end{equation}
which establishes that the constant $k_\infty$ is equal to the ratio of the accretion velocity divided by the Newtonian free-fall velocity in the upstream region, far from the neutron star. Hence we refer to $k_\infty$ as the ``velocity at infinity constant.'' When treating high-luminosity (supercritical) pulsars, such as Her X-1, we set $k_\infty = 1$ so that the velocity of the accreting gas merges smoothly with the Newtonian free-fall velocity profile as $y \to \infty$. On the other hand, setting $k_\infty \sim 0.25$, for example, results in a sub-free-fall velocity profile above the star. This may be appropriate when treating low-luminosity (subcritical) sources such as X Per, in which a collisionless shock is thought to be located at the top of the accretion column, so that the downstream (post-shock) velocity is reduced from the free-fall value by a factor of 4 \citep{LangerandRappaport1982}.

The $P$d$V$ work done on the radiation field via bulk Comptonization in the accretion column depends sensitively on the amount of compression occurring near the stellar surface, where the gas decelerates and impacts against the star. Lower impact velocities lead to higher levels of compression, and enhanced bulk Comptonization, resulting in flatter power-law continuum spectra (BW07). Since the parameter $k_0$ expresses the impact velocity relative to the free-fall velocity at the stellar surface, we can expect that the value of $k_0$ will be connected with the power-law slope of the X-ray continuum spectrum escaping through the walls and the top of the accretion column. High-luminosity sources such as Her X-1, with relatively flat continuum spectra, are expected to have values of $k_0$ close to zero, which maximizes the compression occurring at the base of the accretion column. Conversely, low-luminosity sources sources such as X Per, which display relatively steep power-law continuum spectra, are expected to have non-zero values of $k_0$.

\begin{figure}[t]
\begin{center}
\epsfig{file=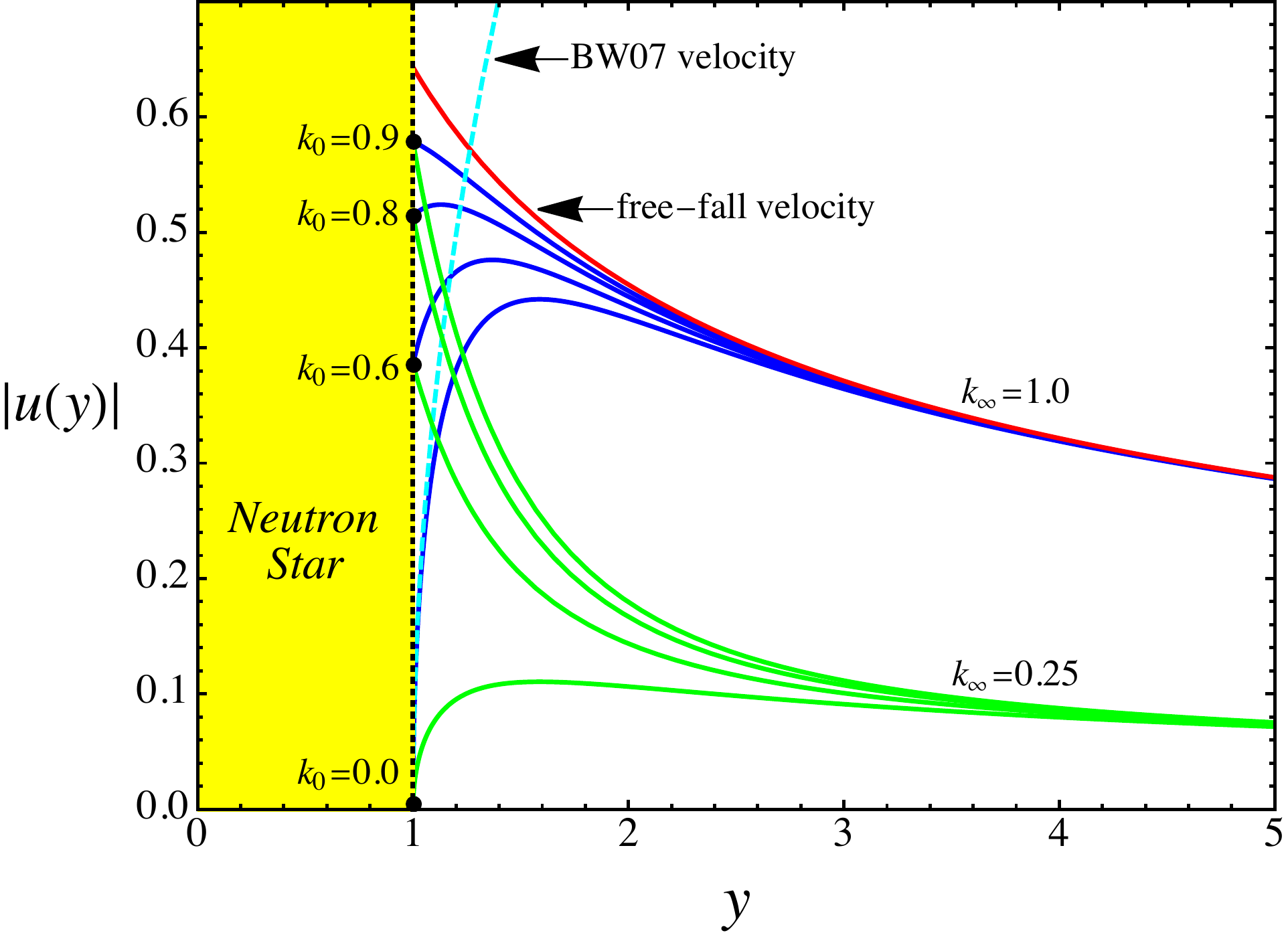,height=8.0cm}
\end{center}
\vskip-0.2truein
\caption{Various examples of the dimensionless flow velocity profile, $u$, for gas accreting in a conical accretion column, plotted as a function of $y = R/R_*$ using Equation~(\ref{eq3.13l}), for the indicated values of the impact velocity parameter, $k_0$. The cases with $k_\infty = 1$ ({\it blue lines}) merge with the free-fall profile as $y \to \infty$, and the cases with $k_\infty = 0.25$ ({\it green lines}) approximate the velocity profile below a shock at the top of the column. We have also included the BW07 velocity profile ({\it cyan line}), and the Newtonian free-fall profile ({\it red line}). The yellow region on the left-hand side represents the interior of the neutron star.
\label{fig:dimlessflvel}}
\end{figure}

In Figure~\ref{fig:dimlessflvel} we plot several examples of the input velocity function, $u(y)$, computed using Equation~(\ref{eq3.13l}). Note that a wide variety of accretion scenarios can be modeled, depending on the values selected for the impact velocity constant, $k_0$, and the velocity at infinity constant, $k_\infty$. For example, setting $k_0 = 0$ results in a gradual settling of the gas onto the neutron star surface, appropriate for high-luminosity sources such as Her X-1, in which radiation pressures dominates the accretion dynamics. On the other hand, setting $k_0 \sim 0.4$ may provide a reasonable description of the flow velocity near the stellar surface in low-luminosity sources such as X Per, in which radiation pressure is probably insufficient to decelerate the gas, and instead the gas impacts on the stellar surface with a high residual velocity, equal to $0.64\,k_0 \,c$. In this situation, the final merger with the stellar crust occurs within the final few cm above the neutron star surface via Coulombic deceleration \citep[e.g.,][]{Sokolova-Lapa_etal2021}. It is also interesting to examine the dependence of the input velocity profile (Equation~(\ref{eq3.13l})) on the parameter $k_\infty$.
The cases with $k_\infty = 1$ plotted in Figure~\ref{fig:dimlessflvel} merge smoothly with the Newtonian free-fall profile as $y \to \infty$, and the cases with $k_\infty = 0.25$ approximate the velocity profile in the post-shock region below a standing discontinuous shock located at the top of the accretion column.

\subsection{Optical Depth Variation}

\begin{figure}[t]
\begin{center}
\hskip-0.4truein
\epsfig{file=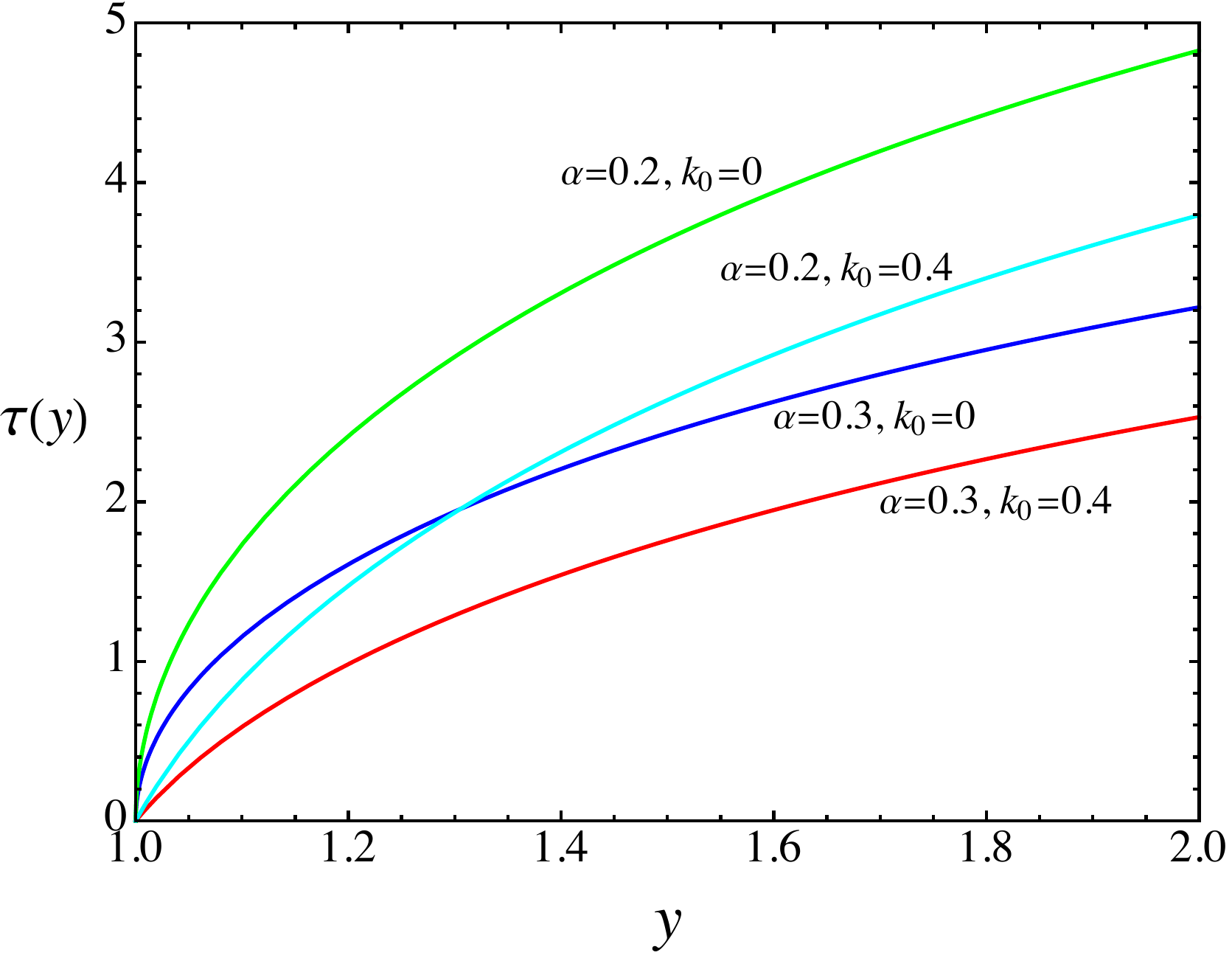,height=9.0cm,angle=0}
\end{center}
\vskip-0.2truein
\caption{Variation of the electron scattering optical depth, $\tau$, measured upwards from the bottom of the accretion column, for photons propagating in the radial direction, computed using Equation~(\ref{eq3.3.6}). The four examples plotted correspond to $\alpha = 0.3$, $k_0 = 0$, ({\it blue line}); $\alpha = 0.3$, $k_0 = 0.4$, ({\it red line}); $\alpha = 0.2$, $k_0 = 0$, ({\it green line}); and $\alpha = 0.2$, $k_0 = 0.4$ ({\it cyan line}). In each case, we set $k_\infty = 1$ so that the accretion velocity approaches Newtonian free-fall as $R \to \infty$.
\label{fig:varelecscatopdpth}}
\end{figure}

It is useful to develop an expression for the variation of the electron scattering optical depth, $\tau$, measured from the stellar surface, for photons propagating in the radial direction, as a function of the radius $R$. We will work in terms of the dimensionless radius $y \equiv R/R_*$ introduced in Equation~(\ref{eq3.10y}). Starting with Equation~(\ref{eq3.10}) for the differential optical depth, $d\tau$, we can use Equation~(\ref{eq3.2b}) to substitute for the electron number density, $n_e$, to obtain
\begin{equation}
d\tau = \frac{\dot M \sigpar}{\Omega R_* y^2 m_p |\vel(y)|} \, dy
\ .
\label{eq3.3.1a}
\end{equation}
Next, we can utilize Equation~(\ref{eq3.13m}) to eliminate $\dot M$ in Equation~(\ref{eq3.3.1a}), which yields
\begin{equation}
d\tau = \frac{3 k_\infty^2 R_g}{\alpha R_* y^2 |u(y)|} \, dy
\ ,
\label{eq3.3.1b}
\end{equation}
where $u=\vel/c$ according to Equation~(\ref{eq3.11d}), and $\alpha$ is given by Equation~(\ref{eqAlphaNew2}). Equation~(\ref{eq3.3.1b}) can be integrated with respect to $y$ to obtain
\begin{equation}
\tau(y) = \int_{1}^y
\frac{3 k_\infty^2 R_g}{\alpha R_* y'^2 |u(y')|} \, dy'
\ .
\label{eq3.3.2}
\end{equation}
We can use Equation~(\ref{eq3.13l}) to substitute for the velocity profile, $u(y)$, in Equation~(\ref{eq3.3.2}), which yields
\begin{equation}
\tau(y) = \frac{3}{\alpha \sqrt{2}} \sqrt{\frac{R_g}{R_*}}
\int_{1}^y
\frac{k_\infty^2}{\sqrt{k_\infty^2 (y'^3-1) + k_0^2}} \, dy'
\ .
\label{eq3.3.5}
\end{equation}
The elliptic integral in Equation~(\ref{eq3.3.5}) can be evaluated analytically with the assistance of Equations~(15.2.5) and (15.3.5) from \citet{AbramowitzandStegun1970}, giving the closed-form result
\begin{equation}
\tau(y) = \frac{3 \sqrt{2} \, k_\infty}{\alpha} \left(\frac{R_g}{R_*}\right)^{1/2}
\left[G(k_0,k_\infty,1) - G(k_0,k_\infty,y)\right]
\ ,
\label{eq3.3.6}
\end{equation}
where the function $G$ is defined by
\begin{equation}
G(k_0,k_\infty,y) \equiv \, _2F_1\left(\frac{1}{6},\frac{1}{2};\frac{7}{6};
\frac{1-k_0^2/k_\infty^2}{y^3}\right) \frac{1}{\sqrt{y}}
\ ,
\label{eq3.3.7}
\end{equation}
and \!$_2F_1$ denotes the hypergeometric function \citep{AbramowitzandStegun1970}. Note that according to Equation~(\ref{eq3.3.6}), $\tau=0$ at the stellar surface ($y=1$) as required. In Figure~\ref{fig:varelecscatopdpth} we plot the variation of the optical depth $\tau(y)$ for a few representative values of the parameters $\alpha$ and $k_0$. In each of the cases plotted in Figure~\ref{fig:varelecscatopdpth}, we have set $k_\infty = 1$, so that the flow velocity merges smoothly with the Newtonian free-fall profile as $y \to\infty$.

\section{SOLUTION FOR THE GREEN'S FUNCTION}
\label{sec:solgrnfun}

Adopting the velocity profile given by Equation~(\ref{eq3.13l}), we can utilize Equation~(\ref{eq3.13}) to rewrite the transport equation (Equation~(\ref{eq3.11})) in the form
\begin{eqnarray}
u \, \frac{\partial \green}{\partial \tau}
&=& - \alpha\,\frac{\chi}{3} \,
\frac{\partial \green}{\partial\chi}
+ \frac{1}{3 y^2}\,\frac{\partial}{\partial\tau} \left(y^2 \, \frac{\partial\green}{\partial\tau}\right)
- \xi^2 u^2 y^2 \green
\nonumber
\\
&+& \frac{\sigbar}{\sigpar} \, \frac{kT_e}{m_e c^2}
\frac{1}{\chi^2}\frac{\partial}{\partial\chi}\left[\chi^4\left(\green
+ \frac{\partial\green}{\partial\chi}\right)\right]
+ \frac{\dot N_0 \, \delta(\chi-\chi_0) \, \delta(\tau-\tau_0)}{\Omega \, c \, R_0^2 \chi_0^2 (k T_e)^3}
\ .
\label{eq3.11qqq}
\end{eqnarray}
This equation is separable if we restrict attention to the case with $\chi \ne \chi_0$, so that the source term vanishes. In this situation, we can reorganize the transport equation to obtain
\begin{eqnarray}
\alpha \, \frac{\chi}{3} \, \frac{\partial \green}{\partial\chi}
&-& \frac{\sigbar}{\sigpar} \, \frac{kT_e}{m_e c^2}
\frac{1}{\chi^2}\frac{\partial}{\partial\chi}\left[\chi^4\left(\green
+ \frac{\partial\green}{\partial\chi}\right)\right]
\nonumber
\\
&=& \frac{1}{3 y^2}\,\frac{\partial}{\partial\tau} \left(y^2 \, \frac{\partial\green}{\partial\tau}\right)
- u \, \frac{\partial \green}{\partial \tau} - \xi^2 u^2 y^2 \green
\ ,
\label{eq3.11yyy}
\end{eqnarray}
which has been rearranged so that all of the energy operators are all on the left-hand side and all of the spatial operators are on the right-hand side. Following BW07, Equation~(\ref{eq3.11yyy}) can be separated in energy and space using the functions
\begin{equation}
f_\lambda(\tau,\chi) \equiv g(\lambda,\tau) \ h(\lambda,\chi) \ ,
\label{eq4.2}
\end{equation}
where $\lambda$ is the separation constant and $g$ and $h$ denote the spatial and energy separation functions, respectively. Setting $\green=f_\lambda=g h$ in Equation~(\ref{eq3.11yyy}) yields
\begin{eqnarray}
\frac{1}{h}\,\bigg\{\alpha \, \frac{\chi}{3} \, \frac{d h}{d\chi}
&-& \frac{\sigbar}{\sigpar} \, \frac{kT_e}{m_e c^2}
\frac{1}{\chi^2}\frac{d}{d\chi}\left[\chi^4\left(h
+ \frac{d h}{d\chi}\right)\right]\bigg\}
\nonumber
\\
&=& \frac{1}{g}\,\bigg[\frac{1}{3 y^2}\,\frac{d}{d\tau} \left(y^2 \, \frac{d g}{d\tau}\right)
- u \, \frac{d g}{d\tau} - \xi^2 u^2 y^2 g\bigg] = - \frac{\alpha}{3}\lambda
\ ,
\label{eq4.2b}
\end{eqnarray}
from which we conclude that the functions $g$ and $h$ satisfy the differential equations
\begin{equation}
\frac{1}{3 y^2}\,\frac{d}{d \tau} \left(y^2 \, \frac{d g}{d \tau}\right)
- u \, \frac{d g}{d \tau} + \left(\frac{\alpha \lambda}{3} - \xi^2 u^2 y^2\right) g = 0
\ ,
\label{eq4.3}
\end{equation}
and
\begin{equation}
\frac{1}{\chi^2}\frac{d}{d\chi}\left[\chi^4
\left(h + \frac{d h}{d\chi}\right)\right]
- \deltapar \, \chi \,
\frac{d h}{d\chi}
- \deltapar \, \lambda \, h = 0
\ ,
\label{eq4.4}
\end{equation}
respectively, where the constant $\deltapar$ is defined by
\begin{equation}
\deltapar \equiv \frac{\alpha}{3} \,
\frac{\sigpar}{\sigbar} \, \frac{m_e c^2}{kT_e}
\ ,
\label{eq4.5}
\end{equation}
which is equivalent to Equation~(38) from BW07. By combining Equations~(\ref{eqAlphaNew2}) and (\ref{eq4.5}), we can rewrite the dimensionless constant $\deltapar$ as
\begin{equation}
\deltapar = \frac{k_\infty^2 \Omega}{4 \pi}
\left(\frac{\dot M}{\dot M_{\rm E}}\right)^{-1}
\left(\frac{\sigbar}{\sigmaT}\right)^{-1} \,
\left(\frac{kT_e}{m_e c^2}\right)^{-1}
\ ,
\label{eqDeltaPar2}
\end{equation}
where the Eddington accretion rate $\dot M_{\rm E}$ is defined in Equation~(\ref{eqEddingtonMdot}).

\subsection{Spatial Boundary Conditions}
\label{sec:SBC}

The spatial separation functions $g(\lambda,\tau)$, must satisfy Equation~(\ref{eq4.3}) in combination with suitable boundary conditions. There are a variety of boundary conditions that can be employed in the application to X-ray pulsar accretion columns, which are based on placing various constraints on the radiation flux at the stellar surface and at the top of the accretion column. We discuss the relevant formulations here, and motivate the specific choice we make in carrying out the derivation of the Green's function $\green$ obtained in this paper.

The spatial boundary conditions are based fundamentally on the behavior of the specific radiation flux, $F$, also called the ``streaming function,'' which is related to the radiation distribution function, $f$, via the expression \citep[e.g.,][]{Becker1992}
\begin{equation}
F(R,\epsilon) \equiv - \frac{\epsilon \vel}{3} \, \frac{\partial f}{\partial \epsilon}
- \kappa \, \frac{\partial f}{\partial R} \quad \propto \quad
{\rm s^{-1}\,cm^{-2}\,erg^{-3}} \ ,
\label{eq4.5a}
\end{equation}
where the spatial diffusion coefficient, $\kappa$, is given by
\begin{equation}
\kappa(R) = \frac{c}{3 \, \sigpar n_e(R)} \ .
\label{eq4.5b}
\end{equation}
The first and second terms on the right-hand side of Equation~(\ref{eq4.5a}) correspond to advection and spatial diffusion, respectively. The corresponding total number flux of the radiation, $F_\#$, is computed from $F$ using the energy integration
\begin{equation}
F_\#(R) = \int_0^\infty \epsilon^2 F(R,\epsilon) \, d\epsilon \quad \propto \quad
{\rm s^{-1}\,cm^{-2}} \ ,
\label{eq4.5c}
\end{equation}
which yields
\begin{equation}
F_\#(R) = \vel \, n_r - \kappa \, \frac{\partial n_r}{\partial R} \ ,
\label{eq4.5d}
\end{equation}
where $n_r$ is the radiation number density defined in Equation~(\ref{eq3.2xx}). Likewise, the total radiation energy flux, $F_{\rm en}$, is computed using the integral
\begin{equation}
F_{\rm en}(R) = \int_0^\infty \epsilon^3 F(R,\epsilon) \, d\epsilon \quad \propto \quad
{\rm ergs\,s^{-1}\,cm^{-2}} \ ,
\label{eq4.5e}
\end{equation}
from which we obtain
\begin{equation}
F_{\rm en}(R) = \frac{4}{3} \, \vel \, U_r - \kappa \, \frac{\partial U_r}{\partial R} \ ,
\label{eq4.5f}
\end{equation}
where $U_r$ is the radiation energy density, evaluated using Equation~(\ref{eq3.2xx}). At the top of the accretion column, we will utilize a free-streaming boundary condition to ensure that spatial diffusion inside the column makes a proper transition when the accreting gas becomes optically thin to electron scattering. At the lower surface of the accretion column, where the gas merges with the neutron star crust, there are three primary scenarios for imposing boundary conditions on the radiation flux, as discussed below.

\subsubsection{Free-Streaming Upper Boundary Condition}

At the upper surface of the accretion column, the gas becomes optically thin to electron scattering, and therefore the radiation transport makes a transition from spatial diffusion (a three-dimensional random walk) to radial free streaming of the escaping photons at the speed of light. The free-streaming boundary condition applicable at the top of the accretion column can be written as
\begin{equation}
- \kappa \, \frac{\partial f}{\partial R} = c f \ , \qquad R \to R_{\rm top} \ ,
\label{eq4.9a}
\end{equation}
where $R_{\rm top}$ denotes the radius at the top of the column, which is a free parameter in our model. By employing Equations~(\ref{eq3.10a}) and (\ref{eq4.5b}), we can transform the spatial variable using
\begin{equation}
- \kappa \, \frac{\partial}{\partial R} = - \frac{c}{3} \, \frac{\partial}{\partial\tau} \ ,
\label{eq4.9c}
\end{equation}
which allows us to rewrite the free-streaming upper boundary condition as
\begin{equation}
- \frac{1}{3} \, \frac{\partial f}{\partial \tau} = f \ , \qquad \tau \to \tau_{\rm top} \ .
\label{eq4.9b}
\end{equation}
Here, $\tau_{\rm top}$ denotes the scattering optical depth at the top of the accretion column, which is related to $R_{\rm top}$ via (see Equation~(\ref{eq3.3.6}))
\begin{equation}
\tau_{\rm top} = \tau(y_{\rm top})
\ ,
\label{eqTauTop}
\end{equation}
where $y_{\rm top} \equiv R_{\rm top}/R_*$ is the dimensionless radius at the top of the accretion column (see Equation~(\ref{eq3.10y})).

\subsubsection{Zero Diffusion Flux Lower Boundary Condition}

There are three possible prescriptions for setting the radiation boundary condition imposed at the lower surface of the accretion column, where the gas merges with the crust of the neutron star. The simplest boundary condition is to set the {\it diffusive flux} equal to zero at the stellar surface, i.e.,
\begin{equation}
- \kappa \, \frac{\partial f}{\partial R} = 0 \ , \qquad R \to R_* \ ,
\label{eq4.6a}
\end{equation}
which can be expressed in terms of the optical depth as
\begin{equation}
- \frac{1}{3} \, \frac{\partial f}{\partial \tau} = 0 \ , \qquad \tau \to 0 \ .
\label{eq4.6b}
\end{equation}
This boundary condition is applicable in situations in which the flow smoothly decelerates to rest at the base of the accretion column, as expected in luminous sources such as Her X-1, in which the gas passes through a radiative, radiation-dominated shock before setting onto the stellar surface (BW07). However, in low-luminosity sources such as X Per, this boundary condition may not be applicable, since the gas is expected to strike the stellar surface with a substantial residual velocity due to the lack of sufficient radiation pressure to accomplish smooth deceleration. In this case, the final merger occurs via Coulombic deceleration \citep{Sokolova-Lapa_etal2021}.

\subsubsection{Zero Number Flux Lower Boundary Condition}

The second possibility is that the net flux of the {\it photon number} vanishes at the lower boundary of the accretion column, which is the photon-conservation scenario. In this case, we can use see Equation~(\ref{eq4.5d}) to write the radiation boundary condition at the surface of the neutron star as
\begin{equation}
F_\#(R) = \vel \, n_r - \kappa \, \frac{\partial n_r}{\partial R} = 0 \ , \qquad R \to R_* \ ,
\label{eq4.7a}
\end{equation}
or, in terms of the optical depth,
\begin{equation}
F_\#(\tau) = \vel \, n_r - \frac{c}{3} \, \frac{\partial n_r}{\partial \tau} = 0 \ , \qquad \tau \to 0 \ .
\label{eq4.7b}
\end{equation}
The advantage of this boundary condition is that photon conservation is guaranteed at the stellar surface. However, unfortunately, this condition does not guarantee the conservation of the photon {\it energy flux} at the surface of the neutron star. That problem leads to the third possibility for the spatial boundary condition at the stellar surface.

\subsubsection{Zero Energy Flux Lower Boundary Condition}

The third alternative for the spatial boundary condition at the lower boundary of the accretion column is based on the requirement that the net flux of the {\it photon energy} vanishes at the stellar surface, which is an energy-conservation principle. In this scenario, we can use Equation~(\ref{eq4.5f}) to write the radiation boundary condition at the surface of the neutron star in the form
\begin{equation}
F_{\rm en}(R) = \frac{4}{3} \, \vel \, U_r - \kappa \, \frac{\partial U_r}{\partial R} = 0 \ , \qquad R \to R_* \ ,
\label{eq4.8a}
\end{equation}
or, in terms of the optical depth,
\begin{equation}
F_{\rm en}(\tau) = \frac{4}{3} \, \vel \, U_r - \frac{c}{3} \, \frac{\partial U_r}{\partial \tau} = 0 \ , \qquad \tau \to 0 \ .
\label{eq4.8b}
\end{equation}
This condition ensures that there is no net photon energy flux either into or out of the star at the stellar surface, which establishes energy conservation at the base of the accretion column. In practical terms, this means that the luminosity of the radiated X-rays is strictly connected with processes occurring inside the accretion column, and there is no anomalous source of energy emitted from the interior of the neutron star, so that the emergent X-ray luminosity, $L_X$, is related to the accretion rate, $\dot M$, via
\begin{equation}
L_X = \frac{G M_* \dot M}{R_*}
\ .
\label{eq4.8c}
\end{equation}
In our view, this is the overarching requirement of any physical model attempting to explain the formation of the X-ray continuum spectrum in an X-ray pulsar accretion column. Hence we will utilize Equation~(\ref{eq4.8b}) as the fundamental spatial boundary condition for the radiation field at the stellar surface as we develop our solution for the emitted X-ray spectrum.

\subsection{Eigenvalues and Spatial Eigenfunctions}

The solution for the Green's function, $\green$, can be expressed as the infinite series
\begin{equation}
\green(\tau_0,\chi_0,\tau,\chi)
= \sum_{n=0}^\infty \ C_n \, g_n(\tau) \, h_n(\chi)
\ ,
\label{eq4.17a}
\end{equation}
where the expansion coefficients are denoted by $C_n$, and the spatial eigenfunctions, $g_n(\tau)$, are defined by
\begin{equation}
g_n(\tau) \equiv g(\lambda_n,\tau) \ ,
\label{eq4.17b}
\end{equation}
with $\lambda_n$ denoting an eigenvalue of the separation constant $\lambda$. Equation~(\ref{eq4.17a}) is valid provided the spatial eigenfunctions $g_n(\tau)$ form an orthogonal set, which we will establish in Section~\ref{sec:EO}. The eigenvalues $\lambda_n$ are determined by requiring that the spatial separation functions $g_n$ satisfy appropriate physical boundary conditions at the surface of the neutron star ($\tau = 0$), and at the top of the accretion column ($\tau = \tau_{\rm top}$). We discuss the development of the upper and lower boundary conditions for the function $g_n$ below.

\subsubsection{Upper Boundary Condition}

At the top of the accretion column, located at radius $R = R_{\rm top}$ and scattering optical depth $\tau = \tau_{\rm top}$, the radiation distribution must satisfy the free-streaming boundary condition given by Equation~(\ref{eq4.9b}), which can be rewritten in terms of the Green's function $\green$ as
\begin{equation}
\frac{1}{3} \, \frac{\partial \green}{\partial \tau} + \green = 0 \ , \qquad \tau \to \tau_{\rm top} \ .
\label{eq4.18b}
\end{equation}
We can use this expression to derive a related boundary condition for the spatial separation function $g_n$. By substituting Equation~(\ref{eq4.17a}) into Equation~(\ref{eq4.18b}) and interchanging the order of summation and differentiation, we obtain
\begin{equation}
\sum_{n=0}^\infty \ C_n \, h_n(\chi)
\left(\frac{1}{3} \, \frac{d g_n}{d \tau} + g_n\right)
= 0  \ , \qquad \tau \to \tau_{\rm top} \ .
\label{eq4.18dd}
\end{equation}
This result implies that at the top of the accretion column, the spatial separation functions $g_n$ must satisfy the boundary condition
\begin{equation}
\frac{1}{3} \, \frac{d g_n}{\partial \tau} + g_n = 0 \ ,
\qquad \tau \to \tau_{\rm top} \ .
\label{eq4.18e}
\end{equation}
The satisfaction of Equation~(\ref{eq4.18e}) by the spatial separation functions $g_n$ is a necessary and sufficient condition to ensure that the Green's function, $\green$, satisfies the free-streaming boundary condition at the top of the accretion column expressed by Equation~(\ref{eq4.18b}).

\subsubsection{Lower Boundary Condition}

We can use the vanishing of the net photon energy flux at the stellar surface (expressed by Equation~(\ref{eq4.8b})) as the starting point for developing the appropriate physical boundary condition applicable to the spatial separation function $g_n$ at $\tau=0$. By operating on Equation~(\ref{eq4.17a}) with $\int_0^\infty \epsilon^3\,d\epsilon$ and integrating term-by-term, we can show that the solution for the radiation energy density, $U_r$ (Equation~(\ref{eq3.2xx})), is given by the series
\begin{equation}
U_r(\tau)
= (k T_e)^4 \sum_{n=0}^\infty \ C_n \, g_n(\tau) \, \int_0^\infty \chi^3 h_n(\chi)\,d\chi
\ ,
\label{eq4.17c}
\end{equation}
where we have transformed from $\epsilon$ to $\chi$ using Equation~(\ref{eq3.9}). Next, we can substitute Equation~(\ref{eq4.17c}) into Equation~(\ref{eq4.8b}) to show that the radiation energy flux is given by
\begin{equation}
F_{\rm en}(\tau)
= (k T_e)^4 \sum_{n=0}^\infty \ C_n \, \left(\frac{4}{3} \, \vel \, g_n - \frac{c}{3} \, \frac{d g_n}{d \tau}\right)
\, \int_0^\infty \chi^3 h_n(\chi)\,d\chi
\ .
\label{eq4.17d}
\end{equation}
According to Equation~(\ref{eq4.8b}), the radiation energy flux vanishes at the stellar surface, and therefore $F_{\rm en}(\tau) \to 0$ as $\tau \to 0$. In combination with Equation~(\ref{eq4.17d}), this condition implies that the spatial separation functions $g_n$ must satisfy the boundary condition
\begin{equation}
\frac{4}{3} \, \vel \, g_n - \frac{c}{3} \, \frac{d g_n}{d \tau} = 0 \ ,
\qquad \tau \to 0 \ .
\label{eq4.17e}
\end{equation}
The satisfaction of Equation~(\ref{eq4.17e}) by the spatial separation functions $g$ is a necessary and sufficient condition to guarantee that the Green's function $\green$ complies with the zero-energy-flux boundary condition at the bottom of the accretion column expressed by Equation~(\ref{eq4.8b}).

\subsubsection{Spatial Eigenfunctions}

The global eigenfunctions $g_n(\tau)$ are those functions that satisfy the fundamental differential equation given by Equation~(\ref{eq4.3}), in combination with the upper and lower boundary conditions, expressed by Equations~(\ref{eq4.18e}) and Equation~(\ref{eq4.17e}), respectively. Equation~(\ref{eq4.17e}) ensures that the radiation energy flux vanishes at the base of the accretion column, and Equation~(\ref{eq4.18e}) ensures that the radiation diffusion flux correctly transitions to the proper free-streaming form at the top of the accretion column. This combination of relations is only satisfied when the separation constant $\lambda$ equals one of the discrete eigenvalues $\lambda_n$.

The general procedure required to solve for the eigenvalues $\lambda_n$ and the associated eigenfunctions $g_n(\tau)$ involves the bi-directional integration of Equation~(\ref{eq4.3}). The first step is to choose a candidate value for $\lambda$, which we wish to investigate in order to determine whether or not it is an eigenvalue. The second step is to integrate numerically Equation~(\ref{eq4.3}) starting at the stellar surface, $\tau = 0$, and utilizing the lower boundary condition given by Equation~(\ref{eq4.17e}). The solution obtained in this step is referred to as $g_{\rm lower}(\lambda,\tau)$. The third step is to integrate numerically Equation~(\ref{eq4.3}) starting at the top of the accretion column, $\tau = \tau_{\rm top}$, and utilizing the upper boundary condition expressed by Equation~(\ref{eq4.18e}). The solution obtained in this step is referred to as $g_{\rm upper}(\lambda,\tau)$. 
Because $g_{\rm lower}(\lambda,\tau)$ and $g_{\rm upper}(\lambda,\tau)$ are 
continuous functions of $\tau$ we can confirm that the 
candidate value of $\lambda$ is an eigenvalue by establishing that these two 
functions
are linearly dependent, which 
requires evaluation of the Wronskian of the two functions, ${\cal W}$, defined by
\begin{equation}
{\cal W}(\lambda,\tau) \equiv g_{\rm lower}(\lambda,\tau) \, g_{\rm upper}'(\lambda,\tau)
- g_{\rm upper}(\lambda,\tau) \, g_{\rm lower}'(\lambda,\tau)
\ ,
\label{eq4.17u}
\end{equation}
where primes denote differentiation with respect to $\tau$. The two functions $g_{\rm lower}$ and $g_{\rm upper}$ are linearly dependent if the Wronskian ${\cal W}$ vanishes. Hence, the eigenvalues $\lambda_n$ are the roots of the equation
\begin{equation}
{\cal W}(\lambda,\tau_*) = 0 \ ,
\label{eq4.17x}
\end{equation}
where $\tau_*$ is an arbitrary value of $\tau$ within the computational domain, so that $\tau_{\rm top} > \tau_* > 0$. When the separation constant $\lambda$ is equal to one of the eigenvalues $\lambda_n$, then the resulting functions $g_{\rm upper}(\tau)$ and $g_{\rm lower}(\tau)$ describe the same global eigenfunction, $g_n(\tau)$, which is a smooth continuous function throughout the computational domain.

\subsection{Eigenfunction Orthogonality}
\label{sec:EO}

In order to evaluate the Green's function, $\green$, using the series expansion in Equation~(\ref{eq4.17a}), it is essential that we establish the orthogonality of the spatial eigenfunctions $g_n(\tau)$. We begin by noting that the general Sturm-Liouville operator can be written as
\begin{equation}
\frac{d}{d\tau} \left[S(\tau) \frac{dg_n}{d\tau}\right] + \alpha \lambda_n \omega(\tau) g_n(\tau)
+ V(\tau) g_n(\tau) = 0 \ ,
\label{eq4.10b}
\end{equation}
where $\omega(\tau)$ is the weight function and $S(\tau)$ and $V(\tau)$ are auxiliary functions. Expansion and reorganization of Equation~(\ref{eq4.10b}) yields the equivalent differential equation
\begin{equation}
g_n''(\tau) + \frac{S'(\tau)}{S(\tau)} \, g_n'(\tau) + \alpha \lambda_n \frac{\omega(\tau)}{S(\tau)} \, g_n(\tau)
+ \frac{V(\tau)}{S(\tau)} \, g_n(\tau) = 0 \ ,
\label{eq4.10c}
\end{equation}
where primes denote differentiation with respect to $\tau$. Next we note that the fundamental differential equation (Equation~(\ref{eq4.3})) satisfied by the spatial separation functions $g(\tau)$ can be rewritten in the form
\begin{equation}
g_n''(\tau) + \left[\frac{2 y'(\tau)}{y(\tau)} - 3 u(\tau)\right] g_n'(\tau) + \alpha \lambda_n \, g_n(\tau)
- 3 \xi^2 u^2(\tau) y^2(\tau) \, g_n(\tau)  = 0 \ .
\label{eq4.10d}
\end{equation}
Comparison of Equations~(\ref{eq4.10c}) and (\ref{eq4.10d}) yields the identifications
\begin{equation}
\omega(\tau) = S(\tau) = y^2(\tau) \exp\left[-3\int_0^\tau u(\tau')\,d\tau'\right] \ ,
\label{eq4.10e}
\end{equation}
and
\begin{equation}
V(\tau) = - 3 \xi^2 u^2(\tau) y^2(\tau) S(\tau) \ ,
\label{eq4.10f}
\end{equation}
where $\omega(\tau)$ is the weight function.

We can make further progress by invoking the relationship between the scattering optical depth $\tau$ and the dimensionless radius $y$. Based on Equation~(\ref{eq3.3.1b}), we can write
\begin{equation}
\frac{d\tau}{dy} = \frac{3 k_\infty^2 R_g}{\alpha R_* y^2 |u(y)|}
\ ,
\label{eq4.10i}
\end{equation}
which can be utilized to transform the variable of integration in Equation~(\ref{eq4.10e}) from $\tau$ to $y$. After simplification, the result obtained for the weight function is
\begin{equation}
\omega(\tau) = S(\tau) = y^2(\tau) \exp\left\{\frac{9 R_g k_\infty^2}{\alpha R_*} \left[1-\frac{1}{y(\tau)}\right]\right\} \ .
\label{eq4.10j}
\end{equation}
Likewise, Equation~(\ref{eq4.10f}) now becomes
\begin{equation}
V(\tau) = - 3 \, \xi^2 u^2(\tau) \, y^4(\tau) \exp\left\{\frac{9 R_g k_\infty^2}{\alpha R_*} \left[1-\frac{1}{y(\tau)}\right]\right\} \ .
\label{eq4.10k}
\end{equation}

We are now in position to establish the orthogonality of the family of spatial eigenfunctions, $g_n(\tau)$, using a standard analysis procedure based on the Sturm-Liouville form of the transport equation (Equation~(\ref{eq4.10b})). The first step is to replace $g_n$ with $g_m$ in Equation~(\ref{eq4.10b}) to obtain
\begin{equation}
\frac{d}{d\tau} \left[S(\tau) \frac{dg_m}{d\tau}\right] + \alpha \lambda_m \omega(\tau) g_m(\tau)
+ V(\tau) g_m(\tau) = 0 \ .
\label{eq4.11a}
\end{equation}
Next we multiply Equation~(\ref{eq4.10b}) by $g_m$ and Equation~(\ref{eq4.11a}) by $g_n$ and subtract the first from the second, which yields
\begin{equation}
g_n(\tau) \frac{d}{d\tau} \left[S(\tau) \frac{dg_m}{d\tau}\right]
- g_m(\tau) \frac{d}{d\tau} \left[S(\tau) \frac{dg_n}{d\tau}\right]
= (\lambda_n - \lambda_m) \, \alpha\,\omega(\tau) \, g_n(\tau) \, g_m(\tau) \ .
\label{eq4.11b}
\end{equation}
This relation can be integrated by parts from $\tau=0$ to $\tau=\tau_{\rm top}$ to obtain
\begin{equation}
S(\tau) \left[g_n(\tau) \frac{dg_m}{d\tau} - g_m(\tau)\frac{dg_n}{d\tau}\right]
\bigg|_0^{\tau_{\rm top}} = (\lambda_n - \lambda_m) \int_0^{\tau_{\rm top}} \alpha\,\omega(\tau) \, g_n(\tau) \, g_m(\tau) \, d\tau \ .
\label{eq4.11c}
\end{equation}
Based on the boundary conditions for $g(\tau)$ given by Equations~(\ref{eq4.18e}) and (\ref{eq4.17e}), we conclude that in the two limits $\tau \to 0$ and $\tau \to \tau_{\rm top}$, the left-hand side of Equation~(\ref{eq4.11c}) vanishes, which therefore establishes the orthogonality relation
\begin{equation}
\int_0^{\tau_{\rm top}} \omega(\tau) \,
g_n(\tau) \, g_m(\tau) \, d\tau = 0 , \qquad n \ne m \ .
\label{eq4.11d}
\end{equation}
This important result confirms that the family of spatial eigenfunctions $g_n(\tau)$ form an orthogonal set relative to the weight function $\omega(\tau)$ defined in Equation~(\ref{eq4.10j}). This is an essential property for carrying out the series expansion for the Green's function $\green$ in Equation~(\ref{eq4.17a}).

\subsection{Energy Eigenfunctions}

The energy separation functions, $h(\lambda,\chi)$, are the solutions to Equation~(\ref{eq4.4}) that satisfy appropriate physical boundary conditions in the energy space. As $\chi \to 0$, we require that $h$ not increase faster than $\chi^{-3}$ in order to ensure that the Green's function possesses a finite total photon number density (see Equation~(\ref{eq3.2xx})). Conversely, as $\chi \to \infty$, we require that $h$ decrease more rapidly than $\chi^{-4}$ since the Green's function must also contain a finite total photon energy density. We also require that $h$ be continuous at the injection energy $\chi = \chi_0$ in order to avoid an infinite diffusive flux in the energy space. The fundamental solution to Equation~(\ref{eq4.4}) that satisfies the various boundary and continuity conditions can be written as (BW07)
\begin{equation}
h(\lambda,\chi) =
\begin{cases}
\chi^{\kappa-4} \, e^{-\chi/2} \, W_{\kappa,\mu}(\chi_0)
\, M_{\kappa,\mu}(\chi) \ ,
& \chi \le \chi_0 \ , \\
\chi^{\kappa-4} \, e^{-\chi/2} \, M_{\kappa,\mu}(\chi_0)
\, W_{\kappa,\mu}(\chi) \ ,
& \chi \ge \chi_0 \ , \\
\end{cases}
\label{eq4.13}
\end{equation}
where $M_{\kappa,\mu}$ and $W_{\kappa,\mu}$ denote Whittaker functions \citep{AbramowitzandStegun1970},
and we have made the definitions
\begin{equation}
\kappa \equiv \frac{1}{2} \, (\deltapar+4) \ ,
\ \ \ \ \
\mu \equiv \frac{1}{2} \left[(3-\deltapar)^2 + 4 \, \deltapar \lambda \right]
^{1/2}
\ ,
\label{eq4.14}
\end{equation}
with the parameter $\deltapar$ defined in Equation~(\ref{eq4.5}).

When the separation constant $\lambda$ is equal to one of the eigenvalues $\lambda_n$, then $h(\lambda,\chi)$ is an energy eigenfunction, denoted by $h_n(\chi)$, which can be written in the compact form
\begin{equation}
h_n(\chi) \equiv h(\lambda_n,\chi) = \chi^{\kappa-4} \, e^{-\chi/2} \, M_{\kappa,\mu}(\chimin)
\, W_{\kappa,\mu}(\chimax)
\ ,
\label{eq4.15}
\end{equation}
where
\begin{equation}
\chimin \equiv \min(\chi,\chi_0) \ ,
\ \ \ \ \
\chimax \equiv \max(\chi,\chi_0)
\ .
\label{eq4.16}
\end{equation}
We note that each of the eigenvalues $\lambda_n$ results in a different value for $\mu$ by setting $\lambda = \lambda_n$ in Equation~(\ref{eq4.14}).

\subsection{Green's Function Expansion}
\label{subsec:grnfunexp}

Equation~(\ref{eq4.17a}) gives the series solution for the Green's function, $\green$, as a function of the injection optical depth, $\tau_0$, the injection dimensionless energy, $\chi_0$, the evaluation optical depth, $\tau$, and the evaluation dimensionless energy, $\chi$. The computation of the expansion coefficients, $C_n$, in Equation~(\ref{eq4.17a}) is accomplished by exploiting the orthogonality of the spatial eigenfunctions, $g_n(\tau)$, along with the derivative jump condition
\begin{equation}
\lim_{\varepsilon \to 0} \
\bigg\{
\frac{\partial\green}{\partial\chi}\Bigg|_{\chi=\chi_0+\varepsilon}
- \frac{\partial\green}{\partial\chi}\Bigg|_{\chi=\chi_0-\varepsilon} \ 
\bigg\}
= - \, \frac{3 \, \deltapar \dot N_0 \, \delta(\tau-\tau_0)}
{\alpha \, \Omega \, c \, R_*^2 \, y_0^2 \, \chi_0^4 \, (k T_e)^3}
\ ,
\label{eq4.18}
\end{equation}
which is obtained by integrating Equation~(\ref{eq3.11qqq}) with respect to $\chi$ in a very small range surrounding the injection energy $\chi_0$, and we have introduced the dimensionless injection radius, $y_0 \equiv R_0/R_*$. The parameter $\deltapar$ appearing in Equation~(\ref{eq4.18}) is defined in Equation~(\ref{eq4.5}).
Combining Equations~(\ref{eq4.17a}) and (\ref{eq4.18}) yields
\begin{equation}
\lim_{\varepsilon \to 0} \
\sum_{n=0}^\infty \ C_n \, g_n(\tau) \, \left[h_n'(\chi_0+\varepsilon)
- h_n'(\chi_0-\varepsilon)\right]
= - \, \frac{3 \, \deltapar \dot N_0 \, \delta(\tau-\tau_0)}
{\alpha \, \Omega \, c \, R_*^2 \, y_0^2 \, \chi_0^4 \, (k T_e)^3}
\ ,
\label{eq4.19}
\end{equation}
where primes denote differentiation with respect to $\chi$. By using Equation~(\ref{eq4.15}) to substitute for $h_n(\chi)$, we find that
\begin{equation}
\sum_{n=0}^\infty \ C_n \, g_n(\tau) \, \chi_0^{\kappa-4}
\, e^{-\chi_0/2} \mathfrak W(\chi_0)
= - \, \frac{3 \, \deltapar \dot N_0 \, \delta(\tau-\tau_0)}
{\alpha \, \Omega \, c \, R_*^2 \, y_0^2 \, \chi_0^4 \, (k T_e)^3}
\ ,
\label{eq4.20}
\end{equation}
where $\mathfrak W$ denotes the Wronskian of the Whittaker functions, defined by
\begin{equation}
\mathfrak W(\chi_0) \equiv M_{\kappa,\mu}(\chi_0) \,
W'_{\kappa,\mu}(\chi_0) - W_{\kappa,\mu}(\chi_0) \,
M'_{\kappa,\mu}(\chi_0)
\ .
\label{eq4.21}
\end{equation}
The Wronskian can be evaluated analytically using Equation~(55) from BW07, which yields
\begin{equation}
\mathfrak W(\chi_0) = - \, \frac{\Gamma(1+2\mu)}{\Gamma(\mu-\kappa+1/2)}
\ .
\label{eq4.22}
\end{equation}

Using Equation~(\ref{eq4.22}) to substitute for $\mathfrak W(\chi_0)$ in Equation~(\ref{eq4.20}), and rearranging factors, we find that
\begin{equation}
\sum_{n=0}^\infty \ \frac{\Gamma(1 + 2 \mu) \, C_n \, g_n(\tau)}{\Gamma(\mu-\kappa+1/2)}
= \, \frac{3 \, \deltapar \dot N_0 e^{\chi_0/2} \, \delta(\tau-\tau_0)}
{\alpha \, \Omega \, c \, R_*^2 \, y_0^2 \, \chi_0^\kappa \, (k T_e)^3}
\ ,
\label{eq4.23}
\end{equation}
where $\mu$ is a function of $\lambda = \lambda_n$ according to Equation~(\ref{eq4.14}).
We can derive an expression for the expansion coefficients $C_n$ by exploiting the orthogonality of the spatial eigenfunctions $g_n(\tau)$. Multiplying both sides of
Equation~(\ref{eq4.23}) by the weight function $\omega(\tau)$ and the eigenfunction $g_m(\tau)$ and integrating with respect to $\tau$ from $\tau = 0$ to $\tau = \tau_{\rm top}$, we find that, according to the orthogonality relation in Equation~(\ref{eq4.11d}), only the $n=m$ term in the sum survives, and we therefore obtain
\begin{equation}
C_n = \frac{3 \, \deltapar \dot N_0 e^{\chi_0/2} \, \omega(\tau_0)}
{\alpha \, \Omega \, c \, R_*^2 \, y_0^2 \, \chi_0^\kappa \, (k T_e)^3}
\, \frac{\Gamma(\mu-\kappa+1/2) \,  g_n(\tau_0)}{{\cal I}_n \, \Gamma(1+2\mu)}
\ ,
\label{eq4.24}
\end{equation}
where we have introduced the quadratic normalization integrals, ${\cal I}_n$, defined by
\begin{equation}
{\cal I}_n \equiv \int_0^{\tau_{\rm top}} g_n^2(\tau) \omega(\tau) \, d\tau \ .
\label{eq4.24b}
\end{equation}

The final closed-form solution for the Green's function is obtained by combining Equations~(\ref{eq4.17a}), (\ref{eq4.15}), and (\ref{eq4.24}), which yields
\begin{eqnarray}
\green(\tau_0,\tau,\chi_0,\chi)
&=& \frac{3 \, \deltapar \dot N_0 \, \omega(\tau_0)
\, \chi^{\kappa-4} \, e^{(\chi_0-\chi)/2}}{\alpha \, \Omega \, c \, R_*^2 \, y_0^2 \, \chi_0^\kappa \, (k T_e)^3}
\sum_{n=0}^\infty \
\frac{\Gamma(\mu-\kappa+1/2)}{{\cal I}_n \, \Gamma(1+2\mu)}
\nonumber
\\
&\times& g_n(\tau_0) \, g_n(\tau)
\, M_{\kappa,\mu}(\chimin)
\, W_{\kappa,\mu}(\chimax)
\ ,
\label{eq4.25}
\end{eqnarray}
where the parameters $\kappa$ and $\mu$ are computed using Equation~(\ref{eq4.14}), and $\chimin$ and $\chimax$ are defined in Equation~(\ref{eq4.16}). The Green's function can also be expressed directly in terms of the photon energy $\epsilon$ by writing
\begin{eqnarray}
\green(\tau_0,\tau,\epsilon_0,\epsilon)
&=& \frac{3 \, \deltapar \dot N_0 \, \omega(\tau_0) \, k T_e
\, \epsilon^{\kappa-4} \, e^{(\epsilon_0-\epsilon)/(2kT_e)}}
{\alpha \, \Omega \, c \, R_*^2 \, y_0^2 \, \epsilon_0^\kappa}
\sum_{n=0}^\infty \
\frac{\Gamma(\mu-\kappa+1/2)}{{\cal I}_n \, \Gamma(1+2\mu)}
\nonumber
\\
&\times& g_n(\tau_0) \, g_n(\tau)
\, M_{\kappa,\mu}\left(\frac{\epsmin}{k T_e}\right)
W_{\kappa,\mu}\left(\frac{\epsmax}{k T_e}\right)
\ ,
\label{eq4.26}
\end{eqnarray}
where
\begin{equation}
\epsmin \equiv \min(\epsilon,\epsilon_0) \ ,
\ \ \ \ \
\epsmax \equiv \max(\epsilon,\epsilon_0)
\ .
\label{eq4.27}
\end{equation}
Equations~(\ref{eq4.25}) and (\ref{eq4.26}) provide the exact solution for the steady-state Green's function, $\green$, which represents the radiation spectrum inside the accretion column, for selected values of the photon energy $\epsilon$ and the scattering optical depth $\tau$, resulting from the continual injection of $\dot N_0$ monochromatic seed photons per unit time, each with energy $\epsilon_0$, from a source located at optical depth $\tau_0$. The series expansions in Equations~(\ref{eq4.25}) and (\ref{eq4.26}) converge rapidly, and we generally obtain three significant figures of accuracy in our calculations of $\green$ using the first 5-10 terms in the series.

\subsection{Radiation Distribution for Arbitrary Source Term}

The fundamental transport equation (Equation~(\ref{eq3.1})) is linear, and therefore the exact solution obtained for the Green's function $\green$ in Section~\ref{subsec:grnfunexp} can be used to compute the particular solution, $f$, representing the radiation distribution inside the accretion column resulting from an arbitrary source term $Q$ in the fundamental transport equation (Equation~(\ref{eq3.1a})). The particular solution for the radiation distribution, $f$, associated with an arbitrary source function $Q$ is given by the integral convolution \citep{Becker2003}
\begin{equation}
f(R,\epsilon) = \int_{R_*}^{R_{\rm top}}\int_0^\infty
\frac{1}{\dot N_0} \, \green(R_0,R,\epsilon_0,\epsilon)
\, \Omega R_0^2 \, Q(R_0,\epsilon_0) \, d\epsilon_0 \, dR_0
\ ,
\label{eq3.2}
\end{equation}
where the Green's function, $\green$, is computed using Equation~(\ref{eq4.26}), and the source function $Q$ is normalized so that the quantity $\epsilon_0^2 \, \Omega R_0^2 \, Q(R_0,\epsilon_0) \, d\epsilon_0 \, dR_0$ gives the number of seed photons injected per unit time with energy between $\epsilon_0$ and $\epsilon_0 + d\epsilon_0$ from the radial zone between $R_0$ and $R_0 + dR_0$. Examples of source functions relevant for accretion-powered X-ray pulsars include bremsstrahlung, cyclotron, and blackbody emission.

Equation~(\ref{eq4.26}) expresses the Green's function, $\green$, in terms of the scattering optical depth, rather than the radius, and therefore it is convenient to transform the spatial variable of integration in Equation~(\ref{eq3.2}) from $R_0$ to $\tau_0$ using the expression (cf. Equation~(\ref{eq3.3.1b}))
\begin{equation}
dR_0 = R_* \, dy_0 = \frac{\alpha R_*^2 y_0^2 |u|}{3 k_\infty^2 R_g} \, d\tau_0
\ ,
\label{eq3.3.2bbb}
\end{equation}
which yields for the particular solution
\begin{equation}
f(\tau,\epsilon) = \int_{0}^{\tau_{\rm top}}\int_0^\infty
\frac{1}{\dot N_0} \, \green(\tau_0,\tau,\epsilon_0,\epsilon)
\, Q(\tau_0,\epsilon_0) \,
\frac{\alpha \Omega R_*^4 \, y_0^4 |u|}{3 k_\infty^2 R_g} \, d\epsilon_0 \, d\tau_0
\ ,
\label{eq3.2ggg}
\end{equation}
where $y_0$ is related to $\tau_0$ via Equation~(\ref{eq3.3.6}). Equation~(\ref{eq3.2ggg}) gives the particular solution for the radiation distribution inside the accretion column, $f$, resulting from the continual emission of photons from a source with distribution $Q$. In Section~\ref{sec:phsources} we will use Equation~(\ref{eq3.2ggg}) to develop particular solutions for the radiation distribution $f$ inside the accretion column resulting from a variety of physical emission mechanisms.

\section{EMITTED RADIATION SPECTRUM}
\label{sec:emradspec}

Our transport equation formalism includes both an escape-probability formalism to model the diffusion of radiation through the column walls (Equation~(\ref{eq3.5})), as well as a free-streaming boundary condition to model the escape of radiation through the top of the accretion column (Equation~(\ref{eq4.9a})). Since these two processes are included in our formalism, we can therefore self-consistently compute the radiation emission components due to the escape of photons through either the walls or the top of the column.
The capability to compute these two emission components separately facilitates more detailed simulations of X-ray pulsar phase-averaged spectra than are possible using the formalism of BW07, who only considered the escape of radiation through the column walls. Phase-averaged spectra evaluated using the new two-component model are presented in Section~\ref{sec:astroapps}. Here we discuss the specific procedures used to compute the two separate emission components.

\subsection{Green's Function for Column Wall Spectrum}

In the escape-probability approach employed here, the Green's function describing the number spectrum of the photons {\it escaping through the walls} of the conical accretion column is given by
\begin{equation}
\greenphoton(R_0,R,\epsilon_0,\epsilon) \equiv \frac{\Omega \, R^2
\epsilon^2}{t_{\rm esc}(\tau)} \, \green(R_0,R,\epsilon_0,\epsilon)
\ ,
\label{eq5.1}
\end{equation}
where the escape timescale, $t_{\rm esc}$, is evaluated using Equation~(\ref{eq3.5}). The quantity $\greenphoton \, dR \, d\epsilon$ represents the number of photons escaping per unit time, with energy between $\epsilon$ and $\epsilon + d\epsilon$, from the disk-shaped volume between radii $R$ and $R + dR$.
We can combine Equations~(\ref{eq3.5}), (\ref{eq3.12a}), and (\ref{eq3.13m}) to show that the escape timescale, $t_{\rm esc}$, can also be written in the form
\begin{equation}
t_{\rm esc}(R) = \frac{\alpha R_*^2}{3 k_\infty^2 \xi^2 R_g |\vel|}
\ ,
\label{eq3.5yyy}
\end{equation}
which can be used to substitute for $t_{\rm esc}$ in Equation~(\ref{eq5.1}), yielding
\begin{equation}
\greenphoton(R_0,R,\epsilon_0,\epsilon) = \Omega \, R^2
\epsilon^2 \, \frac{3 k_\infty^2 \xi^2 R_g |\vel|}{\alpha R_*^2}
\, \green(R_0,R,\epsilon_0,\epsilon)
\ .
\label{eq5.1b}
\end{equation}
Transforming the spatial variable from the radius $R$ to the scattering optical depth $\tau$, we can combine Equations~(\ref{eq4.26}) and (\ref{eq5.1b}) to express the exact solution for Green's function describing the photon number spectrum escaping through the column walls as
\begin{eqnarray}
\greenphoton(\tau_0,\tau,\epsilon_0,\epsilon)
&=&
\frac{9 k_\infty^2 y^2 \xi^2 R_g |\vel| \deltapar \dot N_0 \, \omega(\tau_0) \, k T_e
\, \epsilon^{\kappa-2}  \, e^{(\epsilon_0-\epsilon)/(2kT_e)}}
{\alpha^2 c R_*^2 \, y_0^2 \, \epsilon_0^\kappa}
\nonumber
\\
&\times& \sum_{n=0}^\infty \
\frac{\Gamma(\mu-\kappa+1/2)}{{\cal I}_n \, \Gamma(1+2\mu)}
g_n(\tau_0) \, g_n(\tau)
\, M_{\kappa,\mu}\left(\frac{\epsmin}{k T_e}\right)
W_{\kappa,\mu}\left(\frac{\epsmax}{k T_e}\right)
\ ,
\label{eq5.1c}
\end{eqnarray}
where $\epsmin$ and $\epsmax$ are defined in Equations~(\ref{eq4.27}), and the ${\cal I}_n$ integrals are computed using Equation~(\ref{eq4.24b}).

In many situations, we do not need to consider the detailed distribution of photons escaping through the column walls as a function of $R$, and instead it is sufficient to consider the {\it column-integrated} Green's function, $\greencolumn$, which describes the number distribution of the radiation escaping through the walls of the {\it entire} conical accretion column, from the stellar surface at $R = R_*$ up to the top of the column at $R = R_{\rm top}$. The function $\greencolumn$ is related to the distribution $\greenphoton$ via
\begin{equation}
\greencolumn(R_0,\epsilon_0,\epsilon) \equiv \int_{R_*}^{R_{\rm top}}
\greenphoton(R_0,R,\epsilon_0,\epsilon) \, dR
\ ,
\label{eq5.1d}
\end{equation}
and the quantity $\greencolumn \, d\epsilon$ gives the number of photons escaping through the column walls per unit time with energy between $\epsilon$ and $\epsilon + d\epsilon$. Since Equation~(\ref{eq5.1c}) gives $\greenphoton$ as a function of the scattering optical depth $\tau$, it is convenient to transform the variable of integration in Equation~(\ref{eq5.1d}) from $R$ to $\tau$ using the expression (cf. Equation~(\ref{eq3.3.2bbb}))
\begin{equation}
dR = R_* \, dy = \frac{\alpha R_*^2 y^2 |u|}{3 k_\infty^2 R_g} \, d\tau
\ ,
\label{eq3.3.2xxx}
\end{equation}
which yields
\begin{equation}
\greencolumn(\tau_0,\epsilon_0,\epsilon) \equiv \int_{0}^{\tau_{\rm top}}
\greenphoton(\tau_0,\tau,\epsilon_0,\epsilon) \, \frac{\alpha R_*^2 y^2 |u|}{3 k_\infty^2 R_g} \, d\tau
\ ,
\label{eq5.1e}
\end{equation}
where $y$ is related to $\tau$ via Equation~(\ref{eq3.3.6}). Next, we employ Equation~(\ref{eq5.1c}) to substitute for $\greenphoton$ in Equation~(\ref{eq5.1e}) and interchange the order of summation and integration to obtain
\begin{eqnarray}
\greencolumn(\tau_0,\epsilon_0,\epsilon)
&=&
\frac{6 \deltapar \xi^2 \dot N_0 R_g \omega(\tau_0) \, k T_e
\, \epsilon^{\kappa-2}  \, e^{(\epsilon_0-\epsilon)/(2kT_e)}}
{\alpha y_0^2 R_* \epsilon_0^\kappa}
\nonumber
\\
&\times&
\sum_{n=0}^\infty \
\frac{\Gamma(\mu-\kappa+1/2)}{{\cal I}_n \, \Gamma(1+2\mu)}
g_n(\tau_0) \, I_n
\, M_{\kappa,\mu}\left(\frac{\epsmin}{k T_e}\right)
W_{\kappa,\mu}\left(\frac{\epsmax}{k T_e}\right)
\ ,
\label{eq5.1f}
\end{eqnarray}
where the quantity $I_n$ denotes the spatial integral
\begin{equation}
I_n \equiv \left(\frac{R_*}{2 R_g}\right) \int_0^{\tau_{\rm top}} g_n(\tau) \, y^4(\tau) \, u^2(\tau) \, d\tau \ ,
\label{eq5.1g}
\end{equation}
and the ${\cal I}_n$ integrals are defined in Equation~(\ref{eq4.24b}). Equation~(\ref{eq5.1g}) can be simplified by using Equation~(\ref{eq3.13l}) to substitute for the dimensionless velocity, $u(\tau)$, which yields
\begin{equation}
I_n = \int_0^{\tau_{\rm top}} g_n(\tau) \, [k_0^2 + k_\infty^2(y^3-1)] \, d\tau
\ .
\label{eq5.1h}
\end{equation}
Once the set of spatial eigenfunctions $g_n(\tau)$ has been obtained, the integrals in Equation~(\ref{eq5.1h}) can be evaluated numerically.

\subsection{Wall Spectrum for a General Source}

Once the particular solution $f$ is determined using Equation~(\ref{eq3.2}), we can show by analogy with Equation~(\ref{eq5.1}) that the corresponding height-dependent number distribution describing the radiation spectrum escaping through the column walls for an arbitrary source term $Q$ can be computed using
\begin{equation}
\dot N_\epsilon(R,\epsilon) \equiv \frac{\Omega \, R^2 \,
\epsilon^2}{t_{\rm esc}(R)} \, f(R,\epsilon)
\ .
\label{eq5.6}
\end{equation}
By combining Equations~(\ref{eq3.2}), (\ref{eq5.1}), and (\ref{eq5.6}), we find that the particular solution for the emitted photon spectrum can be written as
\begin{equation}
\dot N_\epsilon(R,\epsilon) = \int_{R_*}^{R_{\rm top}}\int_0^\infty
\frac{\greenphoton(R_0,R,\epsilon_0,\epsilon)}{\dot N_0} \ \epsilon_0^2
\, \Omega R_0^2 \, Q(R_0,\epsilon_0) \, d\epsilon_0 \, dR_0
\ ,
\label{eq5.7}
\end{equation}
where the quantity $\dot N_\epsilon \, dR \, d\epsilon$ represents the number of photons emitted per unit time with energy between $\epsilon$ and $\epsilon + d\epsilon$ from the disk-shaped volume between radii $R$ and $R + dR$. It is convenient to transform the spatial variable of integration in Equation~(\ref{eq5.7}) from $R_0$ to $\tau_0$ using Equation~(\ref{eq3.3.2bbb}), because $\greenphoton$ is evaluated as a function of $(\tau_0,\tau,\epsilon_0,\epsilon)$ according to Equation~(\ref{eq5.1c}). The change of variables yields for the height-dependent escaping photon number distribution the general expression
\begin{equation}
\dot N_\epsilon(R,\epsilon) = \int_{0}^{\tau_{\rm top}}\int_0^\infty
\frac{\greenphoton(\tau_0,\tau,\epsilon_0,\epsilon)}{\dot N_0} \ \epsilon_0^2
\, Q(\tau_0,\epsilon_0) \, \frac{\alpha \Omega R_*^4 \, y_0^4 |u|}{3 k_\infty^2 R_g} \, d\epsilon_0 \, d\tau_0
\ ,
\label{eq5.7tau}
\end{equation}
where the relationship between $y_0$ and $\tau_0$ is given by Equation~(\ref{eq3.3.6}). Equation~(\ref{eq5.7tau}) allows us to calculate the number distribution of the photons escaping through the walls of the accretion column at any radius $R$ as a function of the photon energy $\epsilon$ for an arbitrary source distribution $Q$.

By analogy with Equation~(\ref{eq5.1d}), we can also obtain the {\it column-integrated} photon spectrum escaping through the walls of the conical accretion column due to a general photon source $Q$ using the integral
\begin{equation}
\Phi_\epsilon(\epsilon) \equiv \int_{R_*}^{R_{\rm top}}
\dot N_\epsilon(R,\epsilon) \, dR
\ ,
\label{eq5.13}
\end{equation}
where the quantity $\Phi_\epsilon \, d\epsilon$ gives the number of photons escaping through the column walls per unit time with energy between $\epsilon$ and $\epsilon + d\epsilon$ due to the photon source function $Q$.
By combining Equations~(\ref{eq5.1d}), (\ref{eq5.7}), and (\ref{eq5.13}), we can show that the particular solution for the column-integrated photon number spectrum escaping through the column walls for an arbitrary source function $Q$ is given by
\begin{equation}
\Phi_\epsilon(\epsilon) = \int_{R_*}^{R_{\rm top}}\int_0^\infty
\frac{\greencolumn(R_0,\epsilon_0,\epsilon)}{\dot N_0} \ \epsilon_0^2
\, \Omega R_0^2 \, Q(R_0,\epsilon_0) \, d\epsilon_0 \, dR_0
\ .
\label{eq5.14}
\end{equation}
Equation~(\ref{eq5.1f}) gives $\greencolumn$ as a function of $(\tau_0,\epsilon_0,\epsilon)$, and therefore it is convenient to transform the spatial variable of integration in Equation~(\ref{eq5.14}) from $R_0$ to $\tau_0$ using Equation~(\ref{eq3.3.2bbb}), which yields
\begin{equation}
\Phi_\epsilon(\epsilon) = \int_{0}^{\tau_{\rm top}}\int_0^\infty
\frac{\greencolumn(\tau_0,\epsilon_0,\epsilon)}{\dot N_0} \ \epsilon_0^2
\, Q(\tau_0,\epsilon_0)
\, \frac{\alpha \Omega R_*^4 \, y_0^4 |u|}{3 k_\infty^2 R_g} \, d\epsilon_0 \, d\tau_0
\ ,
\label{eq5.14c}
\end{equation}
where $y_0$ is related to $\tau_0$ via Equation~(\ref{eq3.3.6}). Equation~(\ref{eq5.14c}) allows us to calculate the column-integrated photon number spectrum escaping through the walls of the accretion column, $\Phi_\epsilon$, as a function of the photon energy, $\epsilon$, for an arbitrary photon source function $Q$. In Section~\ref{sec:phsources} we will use Equation~(\ref{eq5.14c}) to derive the particular solution for $\Phi_\epsilon$ associated with each of the primary photon emission mechanisms thought to be important in the formation of the observed spectra from accretion-powered X-ray pulsars.

\subsection{Green's Function for Column Top Spectrum}

The escape of photons out of the top of the conical accretion column is mediated by the process of spatial diffusion, in which photons execute a random walk through the plasma by scattering off electrons. At the top of the column, the gas becomes optically thin, and the escape of radiation through the upper surface of the column is therefore described by the free-streaming boundary condition expressed by Equation~(\ref{eq4.9a}). It follows that the Green's function describing the number distribution of the photons diffusing through the upper surface of the accretion column, denoted b $\dot {\cal N}^{\,\rm G}_\epsilon$, is given by
\begin{equation}
\dot {\cal N}^{\,\rm G}_\epsilon(\tau_0,\epsilon_0,\epsilon) = c \,  \Omega \, R^2_{\rm top}
\, \epsilon^2 \green(\tau_0,\tau_{\rm top},\epsilon_0,\epsilon)
\ ,
\label{eq5.7xxx}
\end{equation}
where $R_{\rm top}$ is the radius at the column top, $\tau_{\rm top}$ denotes the scattering optical depth from the stellar surface  to the column top, and the function $\green$ is evaluated using Equation~(\ref{eq4.26}). The quantity $\dot {\cal N}^{\,\rm G}_\epsilon \, d\epsilon$ gives the number of photons with energy between $\epsilon$ and $\epsilon+d\epsilon$ escaping per unit time through the top of the accretion column. By combining Equation~(\ref{eq4.26}) and (\ref{eq5.7xxx}), we can show that the closed-form expression for $\dot {\cal N}^{\,\rm G}_\epsilon$ is given by
\begin{eqnarray}
\dot {\cal N}^{\,\rm G}_\epsilon(\tau_0,\epsilon_0,\epsilon) &=&
\frac{3 \, \deltapar \dot N_0 \, \omega(\tau_0) \, k T_e
\, R^2_{\rm top} \,
\epsilon^{\kappa-2} \, e^{(\epsilon_0-\epsilon)/(2kT_e)}}
{\alpha R_*^2 \, y_0^2 \, \epsilon_0^\kappa}
\sum_{n=0}^\infty \
\frac{\Gamma(\mu-\kappa+1/2)}{{\cal I}_n \, \Gamma(1+2\mu)}
\nonumber
\\
&\times& g_n(\tau_0) \, g_n(\tau_{\rm top})
\, M_{\kappa,\mu}\left(\frac{\epsmin}{k T_e}\right)
W_{\kappa,\mu}\left(\frac{\epsmax}{k T_e}\right)
\ ,
\label{eq4.26xxx}
\end{eqnarray}
where $\epsmin$ and $\epsmax$ are evaluated using Equations~(\ref{eq4.27}). Equation~(\ref{eq4.26xxx}) facilitates the computation of the Green's function for the number distribution of the radiation escaping through the column top, resulting from the continual injection of seed photons with energy $\epsilon_0$ at optical depth $\tau_0$. In the next section we will consider the convolution of the Green's function with an arbitrary source term.

\subsection{Column Top Spectrum for a General Source}

In our applications to accretion-powered X-ray pulsars, we must consider the injection of seed photons via the processes of cyclotron, bremsstrahlung, and blackbody emission. By analogy with Equation~(\ref{eq5.7}) for the wall emission, we can show that the particular solution for the number distribution of the photons escaping through the top of the accretion column, denoted by $\dot {\cal N}_\epsilon$, is given by the convolution
\begin{equation}
\dot {\cal N}_\epsilon(\epsilon) = \int_{R_*}^{R_{\rm top}}\int_0^\infty
\frac{\dot {\cal N}^{\,\rm G}_\epsilon(R_0,\epsilon_0,\epsilon)}{\dot N_0} \ \epsilon_0^2
\, \Omega R_0^2 \, Q(R_0,\epsilon_0) \, d\epsilon_0 \, dR_0
\ ,
\label{eq5.7ggg}
\end{equation}
where $\dot {\cal N}^{\,\rm G}_\epsilon$ is evaluated using Equation~(\ref{eq4.26xxx}). Since Equation~(\ref{eq4.26xxx}) gives $\dot {\cal N}^{\,\rm G}_\epsilon$ as a function of optical depth rather than radius, it is convenient to transform the spatial variable of integration in Equation~(\ref{eq5.7ggg}) from $R_0$ to $\tau_0$ using Equation~(\ref{eq3.3.2bbb}), which yields (cf. Equation~(\ref{eq3.2ggg}))
\begin{equation}
\dot {\cal N}_\epsilon(\epsilon) = \int_{0}^{\tau_{\rm top}}\int_0^\infty
\frac{\dot {\cal N}^{\,\rm G}_\epsilon(\tau_0,\epsilon_0,\epsilon)}{\dot N_0} \ \epsilon_0^2
\, Q(\tau_0,\epsilon_0) \,
\frac{\alpha \Omega R_*^4 \, y_0^4 |u|}{3 k_\infty^2 R_g} \, d\epsilon_0 \, d\tau_0
\ ,
\label{eq5.7hhh}
\end{equation}
where $y_0$ and $\tau_0$ are related via Equation~(\ref{eq3.3.6}), and $\dot {\cal N}^{\,\rm G}_\epsilon$ is computed using Equation~(\ref{eq4.26xxx}). The quantity $\dot {\cal N}_\epsilon \, d\epsilon$ gives the number of photons with energy between $\epsilon$ and $\epsilon+d\epsilon$ escaping through the top of the accretion column per unit time due to the photon source function $Q$. Hence we can use Equation~(\ref{eq5.7hhh}) to compute the number distribution of the photons escaping through the top of the conical accretion column for any desired source function $Q$.

\section{MODEL PARAMETERS AND CONSTRAINTS}
\label{sec:modparcon}

If we adopt canonical values for the neutron star mass $M_*$ and radius $R_*$, then the remaining fundamental physical free parameters in our model comprise the accretion rate, $\dot M$, the electron temperature, $T_e$, the magnetic field strength, $B$, the dimensionless radius at the column top, $y_{\rm top} \equiv R_{\rm top}/R_*$, the electron scattering cross sections, $\sigperp$, $\sigpar$, and $\sigbar$, the dynamical constants $k_0$ and $k_\infty$, and the outer and inner column wall angles, $\Theta_1$ and $\Theta_2$, respectively. The theory also includes three dimensionless auxiliary parameters that are related to the physical parameters mentioned above, namely $\xi$ (Equation~(\ref{eq3.12b})), $\alpha$ (Equation~(\ref{eqAlphaNew2})), and $\deltapar$ (Equation~(\ref{eqDeltaPar2})). In this section, we explore the relationships between these three dimensionless quantities and the fundamental physical free parameters. We also consider the thermodynamic structure of the accretion column and establish a framework for determining the location of the thermal mound.

\subsection{Scattering Cross Sections}

The detailed energy and angular dependences of the vacuum-modified electron scattering cross sections for the extraordinary and ordinary polarization modes are given by Equations~(\ref{eqExtMV79}) and (\ref{eqOrdMV79}), respectively. Since it is not possible to include the full complexity of these expressions into the model developed here, we have followed BW07 and \citet{WangandFrank1981} by introducing a set of approximate scattering cross sections, averaged over the photon energy and the two polarization modes. The cross sections for photons propagating parallel and perpendicular to the radial direction are denoted by $\sigpar$ and $\sigperp$, respectively, and the angle-averaged cross section is denoted by $\sigbar$. In our applications to specific sources, we shall generally treat the dimensionless constants $\alpha$, $\xi$, and $\deltapar$ as free parameters, and derive the set of associated scattering cross sections $\sigperp$, $\sigpar$, and $\sigbar$ using Equations~(\ref{eq3.12b}), (\ref{eqAlphaNew2}), and (\ref{eqDeltaPar2}). The procedure is described below.

First, we can derive an explicit expression for $\sigpar$ by rearranging Equation~(\ref{eqAlphaNew2}) to obtain
\begin{equation}
\frac{\sigpar}{\sigmaT} = \frac{3 k_\infty^2}{\alpha}
\left(\frac{\Omega}{4 \pi}\right)
\left(\frac{\dot M}{\dot M_{\rm E}}\right)^{-1}
\ ,
\label{eqAlphaNew4}
\end{equation}
where the solid angle $\Omega$ is a function of the outer and inner column wall angles $\Theta_1$ and $\Theta_2$ via Equation~(\ref{eq3.2c}). Next, we can combine Equations~(\ref{eq3.12b}) and (\ref{eqAlphaNew4}) to obtain an expression for $\sigperp$, which yields
\begin{equation}
\frac{\sigperp}{\sigmaT} = \frac{\alpha}{3 k_\infty^2 b^2 \xi^2}
\left(\frac{\Omega}{4 \pi}\right)
\left(\frac{\dot M}{\dot M_{\rm E}}\right)^{-1}
\left(\frac{R_*}{R_g}\right)^2
\ ,
\label{eqXiParNew2}
\end{equation}
where the parameter $b$ is a function of $\Theta_1$ and $\Theta_2$ via Equation~(\ref{eq3.4b}). Finally, we can obtain an explicit expression for $\sigbar$ by rearranging Equation~(\ref{eqDeltaPar2}) to obtain
\begin{equation}
\frac{\sigbar}{\sigmaT} = \frac{k_\infty^2}{\deltapar}
\left(\frac{\Omega}{4 \pi}\right)
\left(\frac{\dot M}{\dot M_{\rm E}}\right)^{-1}
\left(\frac{kT_e}{m_e c^2}\right)^{-1}
\ .
\label{eqDeltaPar4}
\end{equation}
For given values of the free parameters $\dot M$, $T_e$, $\Theta_1$, $\Theta_2$, $k_\infty$, $\alpha$, $\xi$, and $\deltapar$, we can
compute the set of scattering cross sections $\sigpar$, $\sigperp$, and $\sigbar$ using Equations~(\ref{eqAlphaNew4}), (\ref{eqXiParNew2}), and (\ref{eqDeltaPar4}), respectively, and this is the procedure used to determine the cross sections in our model.
In Section~\ref{sec:modselfcon}, we will confirm that the values obtained for the cross sections in our applications to specific pulsars are reasonably consistent with the results obtained using the detailed scattering cross sections given by Equations~(\ref{eqExtMV79}) and (\ref{eqOrdMV79}).

\subsection{Thermal Mound Properties}
\label{sec:thermalmound}

The ``thermal mound'' in an X-ray pulsar accretion column is the effective surface for photon creation and destruction, which occurs mainly via free-free emission and absorption. Inside the thermal mound, we expect thermodynamic equilibrium to prevail, and above the mound, the opacity is dominated by electron scattering. We can therefore define the top of the thermal mound as the radius at which the free-free absorption optical thickness across the column is equal to unity, so that
\begin{equation}
\tau^{\rm ff} \equiv \colrad \, \alpha_{\rm R}^{\rm ff} = 1
\ ,
\label{eq6.3.1}
\end{equation}
where $r_\perp$ is the radius of the accretion column, measured perpendicular to the column axis, and the Rosseland mean absorption coefficient for the free-free process in pure, fully-ionized hydrogen gas, is given in cgs units by \citep{RybickiandLightman1979}
\begin{equation}
\alpha_{\rm R}^{\rm ff} = A_0 \ \Tmound^{-7/2}
\ \rhomound^2
\ ,
\label{eq6.3.2}
\end{equation}
where
\begin{equation}
A_0 \equiv 6.10 \times 10^{22}
\ ,
\label{eq6.3.2b}
\end{equation}
and $T_{\rm th}$ and $\rho_{\rm th}$ denote the temperature and density at the top of the thermal mound, respectively. Using Equation~(\ref{eq3.4}) to substitute for $r_\perp$ in Equation~(\ref{eq6.3.1}) yields
\begin{equation}
b \, R_{\rm th} \, \alpha_{\rm R}^{\rm ff} = 1
\ ,
\label{eq6.3.1b}
\end{equation}
where the geometrical constant $b$ is computed using Equation~(\ref{eq3.4b}), and $R_{\rm th}$ denotes the radius at the top of the thermal mound.

The temperature of the gas in the thermal mound, $T_{\rm th}$, is computed using the energy conservation relation (see Equation~(91) from BW07)
\begin{equation}
\sigma_{_{\rm SB}} \, \Tmound^4 = \frac{1}{2} \, J_{\rm th} \, \vmound^2
\ ,
\label{eq6.3.1c}
\end{equation}
where $\vel_{\rm th} \equiv \vel(R_{\rm th})$ denotes the accretion velocity at the top of the thermal mound, $\sigma_{\rm SB}$ is the Stephan-Boltzmann constant, and the mass flux at the top of the thermal mound, $J_{\rm th}$, is given by
\begin{equation}
J_{\rm th} = \rhomound \, |\vmound|
\ .
\label{eq6.3.1dd}
\end{equation}
Using Equation~(\ref{eq6.3.1dd}) to eliminate $\rho_{\rm th}$ in Equation~(\ref{eq6.3.2}) and combining the result with Equation~(\ref{eq6.3.1b}) yields the condition
\begin{equation}
\vel^2_{\rm th} (R_{\rm th} b A_0)^{-8/15} J_{\rm th}^{-3/5} (2 \sigma_{\rm SB})^{-7/15} = 1
\ ,
\label{eq6.3.1e}
\end{equation}
which is satisfied at the top of the thermal mound. In the conical accretion flow treated here, with solid angle $\Omega$, the mass flux at the top of the thermal mound, $J_{\rm th}$, is related to the accretion rate, $\dot M$, via
\begin{equation}
J_{\rm th} = \frac{\dot M}{\Omega R_{\rm th}^2}
\ .
\label{eq6.3.1d}
\end{equation}

The accretion velocity at the top of the thermal mound, $\vel_{\rm th}$, is related to the thermal mound radius, $R_{\rm th}$, via Equation~(\ref{eq3.13l}), which yields
\begin{equation}
\vel_{\rm th}^2 \equiv \vel^2(y_{\rm th})
= c^2 \left(\frac{2 R_g}{R_*}\right) \, \frac{k_\infty^2 (y_{\rm th}^3-1) + k_0^2}{y_{\rm th}^4}
\ ,
\label{eq3.13lxxx}
\end{equation}
where the dimensionless thermal mound radius, $y_{\rm th}$, is related to $R_{\rm th}$ via
\begin{equation}
y_{\rm th} \equiv \frac{R_{\rm th}}{R_*}
\ .
\label{eq:yth}
\end{equation}
Combining Equations~(\ref{eq6.3.1e}) and (\ref{eq6.3.1d}) and utilizing Equation~(\ref{eq3.13lxxx}) to substitute for the accretion velocity, $\vel_{\rm th}$, yields an equation satisfied by $y_{\rm th}$. The result obtained is
\begin{equation}
y_{\rm th}^{-10/3}[(y_{\rm th}^3-1)k_\infty^2+k_0^2] \left(\frac{2 R_g}{R_*}\right) c^2 (R_* b A_0)^{-8/15}
\left(\frac{\dot M}{R_*^2 \Omega}\right)^{-3/5} (2 \sigma_{\rm SB})^{-7/15} =1
\ .
\label{eq6.3.1f}
\end{equation}
The dimensionless thermal mound radius, $y_{\rm th}$, is the root of this equation, if one exists for $y_{\rm th}>1$. If no such root exists, then the thermal mound is a thin ``hot spot'' on the stellar surface, and therefore $y_{\rm th}=1$. The associated thermal mound altitude, $z_{\rm th}$, is defined by
\begin{equation}
z_{\rm th} \equiv (y_{\rm th} - 1) R_*
\ .
\label{eqMoundZ}
\end{equation}

Once the value of $y_{\rm th}$ has been determined, we can compute the associated accretion velocity, $\vel_{\rm th}$, using Equation~(\ref{eq3.13lxxx}). Thereafter, the thermal mound density, $\rho_{\rm th}$, can be evaluated by combining Equations~(\ref{eq6.3.1dd}), (\ref{eq6.3.1d}), and (\ref{eq:yth}) to obtain
\begin{equation}
\rhomound = \frac{\dot M}{\Omega \, R_*^2 \, y_{\rm th}^2 \, \vel_{\rm th}}
\ .
\label{eq6.3.3}
\end{equation}
Next, we can compute the thermal mound temperature, $T_{\rm th}$, by combining Equations~(\ref{eq6.3.1c}) and (\ref{eq6.3.1dd}), which yields
\begin{equation}
T_{\rm th} = \left(\frac{\rhomound \, \vmound^3}{2 \, \sigma_{\rm SB}}\right)^{1/4}
\ .
\label{eq6.3.3dd}
\end{equation}
The electron scattering optical depth measured from the surface of the neutron star to the top of the thermal mound is defined by
\begin{equation}
\taumound \equiv \tau(y_{\rm th})
\ ,
\label{eq6.3.5}
\end{equation}
which is evaluated using Equation~(\ref{eq3.3.6}). We will use $\tau_{\rm th}$ as the lower limit for the spatial integrations performed in Section~\ref{sec:phsources} when we calculate the emergent spectra resulting from the Comptonization of bremsstrahlung and cyclotron seed photons.

\section{PHOTON SOURCES AND ASSOCIATED SPECTRA}
\label{sec:phsources}

The X-ray spectrum observed from an accretion-powered X-ray pulsar is generated via the thermal and bulk Comptonization of seed photons injected into the accreting gas via the cyclotron, blackbody, and bremsstrahlung emission processes (BW07), which are each represented by different source functions $Q$ in the fundamental transport equation (Equation~(\ref{eq3.1a})). Because the transport equation is linear, the emergent spectrum can be treated as a superposition of the individual spectra resulting from the reprocessing of the various seed photon populations. For a given source function $Q$, the general expression for the height-dependent photon number distribution escaping through the column walls, $\dot N_\epsilon$, is given by the integral convolution in Equation~(\ref{eq5.7tau}). Likewise, the corresponding particular solution for the {\it column-integrated} photon number distribution, $\Phi_\epsilon$, escaping through the walls of the accretion column, is computed using the integral convolution given by Equation~(\ref{eq5.14c}). Finally, the particular solution for the photon number distribution, $\Psi_\epsilon$, escaping through the {\it top} of the accretion column is given by the integral convolution in Equation~(\ref{eq5.7hhh}).

\subsection{Seed Photon Source Functions}

The primary sources of seed photons in the context of accretion-powered X-ray pulsars are blackbody, cyclotron, and bremsstrahlung emission \citep{AronsKleinandLea1987}. Blackbody emission occurs at the top of the ``thermal mound,'' which is the ``photosphere'' for the accretion column. The top of the thermal mound is expected to be located near the stellar surface, at the base of the accretion column, where the gas achieves sufficient optical thickness to thermal absorption. Blackbody emission is a broad-band source of seed photons, but it is localized in space, since it is concentrated at the top of the thermal mound. On the other hand, cyclotron emission occurs throughout the accretion column as the result of electron-ion collisions, which can excite electrons to the $n = 1$ Landau level. Radiative deexcitation back to the ground state results in the emission of cyclotron photons. Finally, bremsstrahlung emission also occurs throughout the accretion column, but in contrast to cyclotron emission, bremsstrahlung is a broad-band process that generates a roughly uniform energy distribution up to a high-energy exponential turnover at a photon energy comparable to the electron thermal energy. Photons produced via any of these three mechanisms are subsequently Comptonized before escaping from the accretion column.

The radiation source function $Q$ appearing in the transport Equation~(\ref{eq3.1a}) is related to the photon emissivity, $\dot n_\epsilon$, via
\begin{equation}
\epsilon^2 Q(R,\epsilon) = \dot n_\epsilon(R,\epsilon)
\ ,
\label{eq7.1}
\end{equation}
where $\dot n_\epsilon(R,\epsilon) \, d\epsilon$ expresses the number of photons injected into the accretion column per unit time per unit volume at radius $R$ with energy between $\epsilon$ and $\epsilon+d\epsilon$. In this section we will use Equation~(\ref{eq7.1}) to derive the source functions $Q$ for bremsstrahlung, cyclotron, and blackbody emission based on the associated emissivity functions $\dot n_\epsilon$.

\subsection{Cyclotron Radiation}

Cyclotron photons are produced when electrons are excited to the $n=1$ Landau level as a result of collisions with protons. The excitations are rapidly followed by a radiative decay back to the ground state ($n=0$). Estimates indicate that in X-ray pulsars, radiative deexcitation dominates over collisional deexcitation, and therefore the production of cyclotron photons tends to cool the gas \citep[e.g.,][]{Nagel1980}. The cyclotron source function is essentially a monochromatic function centered at the cyclotron energy $\epsilon_c$ (Equation~(\ref{eq2.1})), although thermal effects can lead to significant broadening of the monochromatic source, and additional broadening can also occur as a result of magnetic field gradients. However, the dominant broadening mechanism is thermal Comptonization, which tends to drive any source function towards a Wien spectrum \citep{RybickiandLightman1979}.

Assuming that the accreting gas is composed of pure, fully-ionized hydrogen, we can use Equations (7) and (11) from \citet{AronsKleinandLea1987} or Equation~(114) from BW07 to obtain for the cyclotron photon emissivity in cgs units
\begin{equation}
\dot n_\epsilon^{\rm cyc} = 2.10 \times 10^{36} \, \rho^2 \,
B_{12}^{-3/2} \, H\left(\frac{\epsilon_c}{kT_e}\right) \, e^{-\epsilon_c/kT_e}
\, \delta(\epsilon-\epsilon_c)
\ ,
\label{eq7.2}
\end{equation}
where $\rho = m_p n_e$ denotes the mass density, the cyclotron energy $\epsilon_c$ is given by Equation~(\ref{eq2.1}),
and the function $H$ is defined by
\begin{equation}
H\left(\frac{\epsilon_c}{kT_e}\right) \equiv
\begin{cases}
0.41 \ , & \epsilon_c/kT_e > 7.5 \ , \\
0.15 \, \sqrt{\epsilon_c/kT_e} \ , & \epsilon_c/kT_e < 7.5 \ . \\
\end{cases}
\label{eq7.3}
\end{equation}
The associated source function, $Q^{\rm cyc}$, is given by the expression
\begin{equation}
Q^{\rm cyc}(R,\epsilon)
= 2.10 \times 10^{36} \, \rho^2
\, B_{12}^{-7/2} \, H\left(\frac{\epsilon_c}{kT_e}\right)
\, \epsilon^{-2} \, e^{-\epsilon_c/kT_e} \, \delta(\epsilon-\epsilon_c)
\ ,
\label{eq7.4}
\end{equation}
obtained by combining Equations~(\ref{eq7.1}) and (\ref{eq7.2}). The cyclotron source function $Q^{\rm cyc}(R,\epsilon)$ is operative between the thermal mound radius, $R_{\rm th}$, and the top of the accretion column, $R_{\rm top}$, and it is concentrated near the stellar surface due to the appearance of the factor $\rho^2$, which reflects the two-body nature of the excitation process. Because of the concentration of the emission process near the stellar surface, we can assume a roughly constant value for the magnetic field strength in the cyclotron emission region.

The particular solution for the column-integrated photon number spectrum resulting from the escape of reprocessed cyclotron seed photons through the {\it walls}  of the accretion column, denoted by $\Phi^{\rm cyc}_\epsilon(\epsilon)$, is obtained by using Equation~(\ref{eq7.4}) to substitute for $Q$ in Equation~(\ref{eq5.14c}). Due to the delta-function dependence on the photon energy $\epsilon$, the energy integration is trivial. By integrating over $\tau_0$ term-by-term, after some algebra we find that the particular solution for the column-integrated escaping photon number spectrum due to Comptonized cyclotron emission is given in cgs units by
\begin{eqnarray}
\Phi^{\rm cyc}_\epsilon(\epsilon)
&=& \frac{1.15 \times 10^{-12} \dot M^2 \, T_e \, \deltapar \, \xi^2 \, H(\chi_c) \, \epsilon^{\kappa-2}}{k_\infty^2
\sqrt{R_g R_*} \, \Omega \, e^{(\epsilon+\epsilon_c)/(2k T_e)} \, \epsilon_c^{\kappa+3/2}}
\sum_{n=0}^\infty \frac{\Gamma(\mu - \kappa +1/2)}{\Gamma(1 + 2 \mu)}
\frac{I_n \, J_n}{{\cal I}_n} 
\nonumber
\\
\quad
&\times&
M_{\kappa,\mu}[\min(\chi,\chi_c)] \,
W_{\kappa,\mu}[\max(\chi,\chi_c)]
\ ,
\phantom{\left(\frac{\epsmin}{k T_e}\right)}
\label{eq7.5}
\end{eqnarray}
where 
\begin{equation}
\chi \equiv \frac{\epsilon}{k T_e} \ , \qquad
\chi_c \equiv \frac{\epsilon_c}{k T_e}
\ ,
\label{eq7.5b}
\end{equation}
and the quantity $J_n$ represents the spatial integral
\begin{equation}
J_n \equiv \int_{\taumound}^{\tau_{\rm top}}
\frac{\omega(\tau_0) \, g_n(\tau_0)}{\sqrt{(y_0^3-1)k_\infty^2 + k_0^2}} \, d\tau_0
\ .
\label{eq7.6}
\end{equation}
The weight function $\omega(\tau)$ appearing in Equation~(\ref{eq7.6}) is defined in Equation~(\ref{eq4.10j}), and the quantities $\tau_{\rm top}$ (Equation~(\ref{eqTauTop})) and $\tau_{\rm th}$ (Equation~(\ref{eq6.3.5})) represent the scattering optical depths at the top of the accretion column and at the surface of the thermal mound, respectively. The integrals ${\cal I}_n$ and $I_n$ in Equation~(\ref{eq7.5}) are defined in Equations~(\ref{eq4.24b}) and (\ref{eq5.1g}), respectively. The quantity $\Phi^{\rm cyc}_\epsilon(\epsilon)$ gives the number of photons escaping through the walls of the accretion column per unit time in the energy range between $\epsilon$ and $\epsilon+d\epsilon$.

We can also obtain the particular solution for the photon number spectrum resulting from the escape of Comptonized cyclotron emission through the {\it top} of the accretion column, denoted by $\dot {\cal N}_\epsilon^{\rm cyc}(\epsilon)$, by using Equation~(\ref{eq7.4}) to substitute for $Q$ in Equation~(\ref{eq5.7hhh}). After simplification, the result obtained in cgs units is
\begin{eqnarray}
\dot {\cal N}_\epsilon^{\rm cyc}(\epsilon)
&=& \frac{5.76 \times 10^{-13} \dot M^2 \, T_e \, y_{\rm top}^2 \, \deltapar \sqrt{R_g R_*} \, H(\chi_c) \, \epsilon^{\kappa-2}}
{\Omega R_g^2 \, k_\infty^2
\, e^{(\epsilon+\epsilon_c)/(2 k T_e)} \, \epsilon_c^{\kappa+3/2}}
\sum_{n=0}^\infty \frac{\Gamma(\mu - \kappa +1/2)}{\Gamma(1 + 2 \mu)}
\frac{J_n \, g_n(\tau_{\rm top})}{{\cal I}_n} 
\nonumber
\\
\quad
&\times&
M_{\kappa,\mu}[\min(\chi,\chi_c)] \,
W_{\kappa,\mu}[\max(\chi,\chi_c)]
\ ,
\phantom{\left(\frac{\epsmin}{k T_e}\right)}
\label{eq7.7b}
\end{eqnarray}
where $\chi$ and $\chi_c$ are defined in Equation~(\ref{eq7.5b}). The quantity $\dot {\cal N}_\epsilon^{\rm cyc}(\epsilon)$ gives the number of photons escaping through the top of the accretion column per unit time in the energy range between $\epsilon$ and $\epsilon+d\epsilon$.

\subsection{Blackbody Radiation}

At the base of the accretion column, the gas becomes sufficiently dense that thermodynamic equilibrium is established, in the region referred to as the thermal mound. The upper surface of the thermal mound radiates a blackbody distribution, with surface energy flux equal to $\pi$ times the Planck intensity function, $B_\epsilon(\epsilon)$, given by \citep{RybickiandLightman1979}
\begin{equation}
B_\epsilon(\epsilon) = \frac{2 \, \epsilon^3}{c^2 h^3}
\, \frac{1}{e^{\epsilon/k\Tmound} - 1}
\ ,
\label{eq7.9}
\end{equation}
where $\Tmound$ is the temperature of the gas in the thermal mound. Note that the function $B_\epsilon$ denotes the blackbody ``energy intensity,'' with units ${\rm ergs \ s^{-1} \, ster^{-1} \, cm^{-2} \, erg^{-1}}$. Following \citet{BeckerandWolff2005b}, and BW07, we define the thermal mound surface emission function, $S(\epsilon)$, using
\begin{equation}
\epsilon^3 S(\epsilon) \, d\epsilon
\equiv \pi \, B_\epsilon(\epsilon) \, d\epsilon
\ .
\label{eq7.8}
\end{equation}
According to Equation~(\ref{eq7.8}), the quantity $\epsilon^3 S(\epsilon) \, d\epsilon$ gives the energy emitted per unit time per unit area from the upper surface of the thermal mound in the energy range between $\epsilon$ and $\epsilon + d\epsilon$.

The function $S$ is related to the source term $Q$ appearing in the transport Equation~(\ref{eq3.1a}) via
\begin{equation}
Q^{\rm bb}(R,\epsilon) \equiv S(\epsilon) \, \delta(R-R_{\rm th})
\ ,
\label{eq7.10}
\end{equation}
which can be combined with Equations~(\ref{eq7.9}) and (\ref{eq7.8}) to obtain
\begin{equation}
Q^{\rm bb}(R,\epsilon) = \frac{2 \pi}{c^2 h^3}
\, \frac{\delta(R - R_{\rm th})}{e^{\epsilon/k\Tmound} - 1}
\ .
\label{eq7.11}
\end{equation}
This result for the blackbody source term is localized in physical space, but it possesses a distributed (broad-band) energy dependence, which is the exact opposite of the cyclotron source given by Equation~(\ref{eq7.4}).

We can obtain the particular solution for the column-integrated photon number spectrum resulting from the escape of Comptonized blackbody radiation through the {\it walls} of the accretion column by using Equation~(\ref{eq7.11}) to substitute for $Q$ in Equation~(\ref{eq5.14}). The resulting spatial integration is trivial, and only the energy integral remains, which reduces to
\begin{equation}
\Phi^{\rm bb}_\epsilon(\epsilon) = \frac{2 \pi \, \Omega R_{\rm th}^2}{c^2 h^3}
\int_0^\infty
\frac{\greencolumn(R_{\rm th},\epsilon_0,\epsilon)}{\dot N_0} \
\frac{\epsilon_0^2}{e^{\epsilon_0/k\Tmound}-1} \ d\epsilon_0
\ ,
\label{eq7.12}
\end{equation}
where $\greencolumn$ is computed using Equation~(\ref{eq5.1f}). Likewise, we can obtain the particular solution for the photon number spectrum resulting from the escape of Comptonized blackbody photons through the {\it top}  of the accretion column by combining Equations~(\ref{eq5.7ggg}) and (\ref{eq7.11}), which yields
\begin{equation}
\dot {\cal N}_\epsilon^{\rm bb}(\epsilon) = \frac{2 \pi \, \Omega R_{\rm th}^2}{c^2 h^3} \int_0^\infty
\frac{\dot {\cal N}^{\,\rm G}_\epsilon(R_{\rm th},\epsilon_0,\epsilon)}{\dot N_0} \ 
\, \frac{\epsilon_0^2}{e^{\epsilon_0/k\Tmound} - 1} \ d\epsilon_0
\ ,
\label{eq7.11b}
\end{equation}
where $\dot {\cal N}^{\,\rm G}_\epsilon$ is evaluated using Equation~(\ref{eq4.26xxx}).

\subsection{Bremsstrahlung Radiation}

Bremsstrahlung is the dominant contributor to the seed photon distribution in luminous accretion-powered X-ray pulsars such as Her X-1, LMC X-4, and Cen X-3 \citep[e.g.,][]{BeckerandWolff2007,Wolff_etal2016,West_etal2017a,West_etal2017b}. Following BW07, we can use Equation~(5.14b) from \citet{RybickiandLightman1979} to write the bremsstrahlung (free-free) photon production rate per unit volume per unit energy in cgs units as
\begin{equation}
\dot n_\epsilon^{\rm ff} = 3.68 \times 10^{36} \, \rho^2 \,
T_e^{-1/2} \, \epsilon^{-1} \, e^{-\epsilon/kT_e}
\ ,
\label{eq7.13}
\end{equation}
for a plasma composed of pure, fully-ionized hydrogen. Equation~(\ref{eq7.13}) can be combined with Equation~(\ref{eq7.1}) to obtain the free-free source function
\begin{equation}
Q^{\rm ff}(R,\epsilon)
= 3.68 \times 10^{36} \, \rho^2
\, T_e^{-1/2} \, \epsilon^{-3} \, e^{-\epsilon/kT_e}
\ , \ \ \ \ \ 
R_{\rm th} < R < R_{\rm top}
\ ,
\label{eq7.14}
\end{equation}
where $R_{\rm th}$ and $R_{\rm top}$ denote the radii at the surface of the thermal mound and at the top of the accretion column, respectively. Equation~(\ref{eq7.14}) gives the spectrum of the injected, optically-thin bremsstrahlung radiation for photons with energy $\epsilon > \epsilonabs$, where $\epsilonabs$ denotes the self-absorption cutoff, below which the plasma is optically thick and therefore the radiation is thermalized. The value of $\epsilonabs$ is related to the gas density $\rho$ and the electron temperature $T_e$ via Equation~(127) from BW07, which gives
\begin{equation}
\frac{\epsilonabs}{k T_e} = 6.08 \times 10^{12} \, T_e^{-7/4} \, \rho^{1/2}
\ .
\label{eq7.14b}
\end{equation}
In our applications, we set $\epsilonabs/kT_e = 0.05$, which is a reasonable approximation for the calculations performed here, since the emergent spectrum has a weak dependence on this parameter.

The bremsstrahlung source term given by Equation~(\ref{eq7.14}) has non-trivial dependences on both the radius $R$ and the photon energy $\epsilon$ and hence it is more complex to implement computationally than either the cyclotron or blackbody sources given by Equations~(\ref{eq7.4}) and (\ref{eq7.11}), respectively. Using Equation~(\ref{eq7.14}) to substitute for $Q$ in Equation~(\ref{eq5.14}), we find that, after performing the spatial and energy integrals and simplifying, the result obtained for the column-integrated photon number spectrum resulting from the escape of Comptonized bremsstrahlung through the {\it walls} of the accretion column can be written in cgs units as
\begin{equation}
\Phi^{\rm ff}_\epsilon(\epsilon)
= \frac{6.80 \times 10^{7} \, \dot M^2 \, \deltapar \, \xi^2 \, \epsilon^{\kappa-2} \,
e^{-\epsilon/(2 k T_e)}}{k_\infty^2 \sqrt{R_g R_*} \, \Omega 
\, (k T_e)^{\kappa-1/2}} \ \sum_{n=0}^\infty \ \frac{\Gamma(\mu-\kappa+1/2)}
{\Gamma(1+2\mu)} \ \frac{I_n \, J_n}{{\cal I}_n} \, B_n
\ ,
\label{eq7.15}
\end{equation}
where ${\cal I}_n$, $I_n$, and $J_n$ are computed using Equations~(\ref{eq4.24b}), (\ref{eq5.1h}), and (\ref{eq7.6}), respectively. The quantity $B_n$ introduced in Equation~(\ref{eq7.15}) denotes the
energy integral
\begin{equation}
B_n \equiv \int_{\chiabs}^\infty \chi_0^{-1-\kappa}
\, e^{-\chi_0/2} \, M_{\kappa,\mu}[\min(\chi,\chi_0)] \,
W_{\kappa,\mu}[\max(\chi,\chi_0)]
\, d\chi_0
\ ,
\label{eq7.16}
\end{equation}
where $\chi\equiv \epsilon/kT_e$, $\chi_0\equiv \epsilon_0 / kT_e$,
and $\chiabs \equiv \epsilonabs/kT_e$.

We can also derive the particular solution for the photon number spectrum escaping through the {\it top} of the accretion column as a result of Comptonized bremsstrahlung emission by combining Equations~(\ref{eq5.7hhh}) and (\ref{eq7.14}). After some algebra, the result obtained in cgs units is
\begin{eqnarray}
\dot {\cal N}_\epsilon^{\rm ff}(\epsilon)
&=& \frac{3.40 \times 10^{7} \, \dot M^2 \, y_{\rm top}^2 \, \deltapar \, \epsilon^{\kappa-2} \,
e^{-\epsilon/(2 k T_e)}\sqrt{R_g R_*}}{k_\infty^2 R_g^2 \, \Omega 
\, (k T_e)^{\kappa-1/2}}
\nonumber
\\
&\times& \ \sum_{n=0}^\infty \ \frac{\Gamma(\mu-\kappa+1/2) }
{\Gamma(1+2\mu)} \ \frac{g_n(\tau_{\rm top}) \, J_n}{{\cal I}_n} \, B_n
\ .
\label{eq7.15b}
\end{eqnarray}

\section{ASTROPHYSICAL APPLICATIONS}
\label{sec:astroapps}

The results from the previous sections in the paper can be combined to compute the spectrum emitted by an X-ray pulsar accretion column due to the thermal and bulk Comptonization of cyclotron, bremsstrahlung, and blackbody seed photons. Because the theoretical model developed here computes separate emission components for photons escaping through either the walls or the top of the accretion column, we are able to accomplish more detailed calculations of the approximate phase-averaged spectrum than were presented by BW07. In this section we develop the formalism for calculating the contributions of the wall and top emission components to the observed X-ray spectrum, and we apply the new model to Her X-1 and X Per.

The theoretical column-integrated photon count rate spectrum observed at Earth due to the escape of Comptonized photons through the column {\it walls}, denoted by $F^{\rm wall}_\epsilon(\epsilon)$, is computed using the expression
\begin{equation}
F^{\rm wall}_\epsilon(\epsilon) \equiv \frac{\columntotal(\epsilon)}{4 \pi D^2}
\ ,
\label{eq8.1}
\end{equation}
where
\begin{equation}
\columntotal(\epsilon) \equiv \left[
\Phi^{\rm cyc}_\epsilon(\epsilon) +
\Phi^{\rm bb}_\epsilon(\epsilon) +
\Phi^{\rm ff}_\epsilon(\epsilon) \right] A_c(\epsilon)
\label{eq8.2}
\end{equation}
denotes the total photon number spectrum escaping through the column walls, $D$ is the distance to the source, and the cyclotron absorption feature commonly seen in X-ray pulsar spectra is represented by the function $A_c(\epsilon)$, defined using the standard Gaussian form
\citep[e.g.,][]{HeindlandChakrabarty1999,Orlandini_etal1998,Soong_etal1990}
\begin{equation}
A_c(\epsilon) \equiv 1 - \frac{d_c}{\sigma_c
\sqrt{2 \pi}}
\, e^{-(\epsilon-\epsilon_c)^2
/(2\sigma^2_c)}
\ .
\label{eq8.3}
\end{equation}
The quantities on the right-hand side of Equation~(\ref{eq8.2}) express the contributions to the escaping column-wall spectrum due to Comptonized cyclotron emission ($\Phi^{\rm cyc}_\epsilon$; Equation~(\ref{eq7.5})), blackbody emission ($\Phi^{\rm bb}_\epsilon$; Equation~(\ref{eq7.12})), and bremsstrahlung emission ($\Phi^{\rm ff}_\epsilon$; Equation~(\ref{eq7.15})), respectively.

We can also use the theoretical formalism developed here to compute the observed spectrum due to the escape of Comptonized photons through the {\it top} of the accretion column. The total column-top spectrum observed at Earth, denoted by $F^{\rm top}_\epsilon(\epsilon)$, is calculated using the expression
\begin{equation}
F^{\rm top}_\epsilon(\epsilon) \equiv \frac{\dot {\cal N}_\epsilon(\epsilon)}{4 \pi D^2}
\ ,
\label{eq8.1b}
\end{equation}
where
\begin{equation}
\dot {\cal N}^{\rm top}_\epsilon(\epsilon) \equiv \left[
\dot {\cal N}^{\rm cyc}_\epsilon(\epsilon) +
\dot {\cal N}^{\rm bb}_\epsilon(\epsilon) +
\dot {\cal N}^{\rm ff}_\epsilon(\epsilon) \right] A_c(\epsilon)
\ ,
\label{eq8.2b}
\end{equation}
where $A_c(\epsilon)$ is defined in Equation~(\ref{eq8.3}). The quantities on the right-hand side of Equation~(\ref{eq8.2b}) denote the contributions to the column-top spectrum due to Comptonized cyclotron emission ($\dot {\cal N}^{\rm cyc}_\epsilon$; Equation~(\ref{eq7.7b})), blackbody emission ($\dot {\cal N}^{\rm bb}_\epsilon$; Equation~(\ref{eq7.11b})), and bremsstrahlung emission ($\dot {\cal N}^{\rm ff}_\epsilon$; Equation~(\ref{eq7.15b})), respectively.

Following \citet{West_etal2017b}, we can use our model to obtain an approximation of the observed phase-averaged X-ray spectrum by adding the contributions emitted through the walls of the accretion column (Equation~(\ref{eq8.1})) and through the column top (Equation~(\ref{eq8.1b})). The approximate phase-averaged spectrum, denoted by $F^{\rm total}_\epsilon(\epsilon)$, is therefore given by the sum
\begin{equation}
F^{\rm total}_\epsilon(\epsilon) \equiv F^{\rm wall}_\epsilon(\epsilon)
+ F^{\rm top}_\epsilon(\epsilon)
\ .
\label{eq8.1c}
\end{equation}
We will use this approach to compute theoretical predictions for the observed phase-averaged X-ray spectra of Her X-1 and X Per. These two sources were selected because they span a very wide range in X-ray luminosity, and because good quality X-ray spectra are available for both sources in the published literature. The spectral results presented here
are considered example calculations, rather than detailed quantitative fits, which will require further software development and are beyond the scope of this paper.
In each case we adopt canonical values for the stellar mass and radius, with $M_* = 1.4 \, \Msun$ and $R_* = 10\,$km, respectively. The accretion rate, $\dot M$, is set using observational estimates of the X-ray luminosity, combined with Equation~(\ref{eq4.8c}), and the distance to the source, $D$, is adopted from published results. The remaining fundamental theoretical free parameters that must be set in order to evaluate the X-ray spectrum using our model are $\alpha$, $\xi$, $\deltapar$, $B$, $\Theta_1$, $\Theta_2$, $k_0$, $k_\infty$, $y_{\rm top}$, and $T_e$, which are listed in
Table~1 for each of the models considered here. These fundamental free parameters are varied in order to achieve a satisfactory qualitative fit to the observed phase-averaged X-ray spectrum for a given source. In addition, the free parameters are also constrained via analysis of the thermodynamic and hydrodynamic structure of the accretion column, as discussed in Section~\ref{sec:modselfcon}.

Once the fundamental free parameters $\alpha$, $\xi$, $\deltapar$, $B$, $\Theta_1$, $\Theta_2$, $k_0$, $k_\infty$, $y_{\rm top}$, and $T_e$ are determined via the qualitative fitting process, a number of additional auxiliary parameters are computed, with values reported in Table~2. These include the perpendicular electron scattering cross section, $\sigperp$, the parallel electron scattering cross section, $\sigpar$, the angle-averaged scattering cross section, $\sigbar$, the thermal mound temperature, $T_{\rm th}$, the thermal mound altitude, $z_{\rm th}$, the thermal mound scattering optical depth, $\tau_{\rm th}$, and the optical depth at the top of the column, $\tau_{\rm top}$. We also include the radius of the magnetic cap at the base of the accretion column, defined by
\begin{equation}
r_0 \equiv R_* \, \Theta_1 \ .
\label{eq:r0}
\end{equation}
For the two sources treated here, we find that the best results are obtained by setting the inner column wall angle, $\Theta_2$, equal to zero, so that the accretion columns are both completely filled cones, with no hollow interior region. In addition to the requirement of an acceptable qualitative fit to the observed X-ray spectrum, we also impose additional constraints related to the dynamical and thermal structure of the accretion column (see Section~\ref{sec:modselfcon}).

The accretion dynamics in Her X-1 is expected to be dominated by radiation pressure (BW07), in which case the accretion column contains a smooth, radiative, radiation-dominated shock. In this source, most of the kinetic energy of the accreting gas is radiated away through the walls of the accretion column, and the gas settles onto the stellar surface with approximately zero velocity. Conversely, in the case of X Per, it is thought that a strong, discontinuous, gas-mediated standing shock is located at the top of the accretion column \citep{LangerandRappaport1982}. As the gas passes through the shock, its velocity drops by a factor of 4 below the incident free-fall value, and in the downstream region below the shock, the gas may reaccelerate until it collides with the surface of the star. The gas may approach the stellar surface at a large fraction of the speed of light, with the final deceleration occurring via Coulomb collisions as the gas merges with the stellar crust.

Once all of the theoretical parameters are specified for a given source, the resulting X-ray count-rate spectrum is computed using Equations~(\ref{eq8.1}) and (\ref{eq8.1b}), which give the column-wall and the column-top spectra, respectively. The spectral results
presented here were computed using the first 5-10 terms in the analytical expansions, which we find yields at least three decimal digits of accuracy in the resulting X-ray spectra. Photon conservation is a necessary condition for the steady-state model considered here, and therefore the integrity of the calculation is checked by confirming that the total number of photons escaping through the column walls and column top per unit time is equal to the number of seed photons injected per unit time. This check was carried out independently for each photon source mechanism considered here (bremsstrahlung, blackbody, and cyclotron).

\subsection{Her X-1}

\begin{figure}[t]
\begin{center}
\epsfig{file=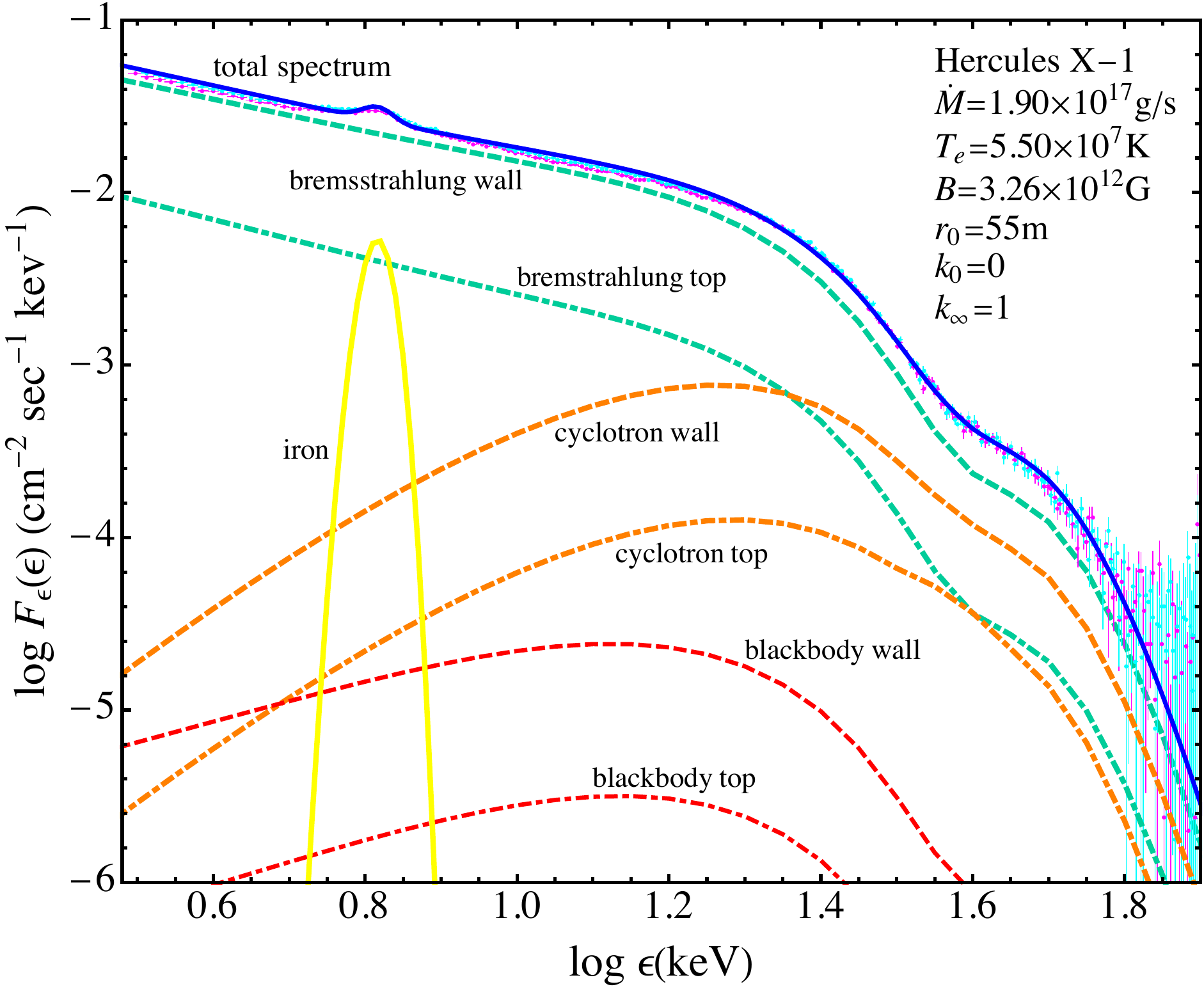,height=9.0cm}
\end{center}
\vskip-0.2truein
\caption{Theoretical column-integrated count rate spectrum $F_\epsilon(\epsilon)$ computed using Equation~(\ref{eq8.1c}) based on the model~1 parameters listed in Table~1, compared with the deconvolved (incident) X-ray spectrum for Her X-1 reported by Wolff et al. (2016; {\it magenta and cyan points and lines}). The plot includes the total theoretical spectrum as well as the individual spectral components due to the column wall and column top emission of Comptonized bremsstrahlung, cyclotron, and blackbody seed radiation, as indicated. An additional iron emission line feature is also included. Note that the total spectrum is dominated by the bremsstrahlung component emitted through the column walls.
\label{fig:herx1details}}
\end{figure}

In Figure~\ref{fig:herx1details} we plot the theoretical count-rate spectrum for Her X-1, computed using Equations~(\ref{eq8.1}), (\ref{eq8.1b}), and (\ref{eq8.1c}), and compare it with the phase-averaged {\it NuSTAR} spectrum reported by \citet{Wolff_etal2016}, which is based on an XSPEC analysis of an observation by \citet{Fuerst_etal2013}. The deconvolved data from the {\it NuSTAR} FPMA and FPMB modules are indicated by the magenta and cyan lines, respectively. Results are presented for the total theoretical phase-averaged spectrum, as well as for the individual contributions to the observed flux due to the Comptonization of cyclotron, blackbody, and bremsstrahlung seed photons. The spectrum for each mechanism is further separated into column-wall and column-top emission components.

\begin{figure}[t]
\begin{center}
\epsfig{file=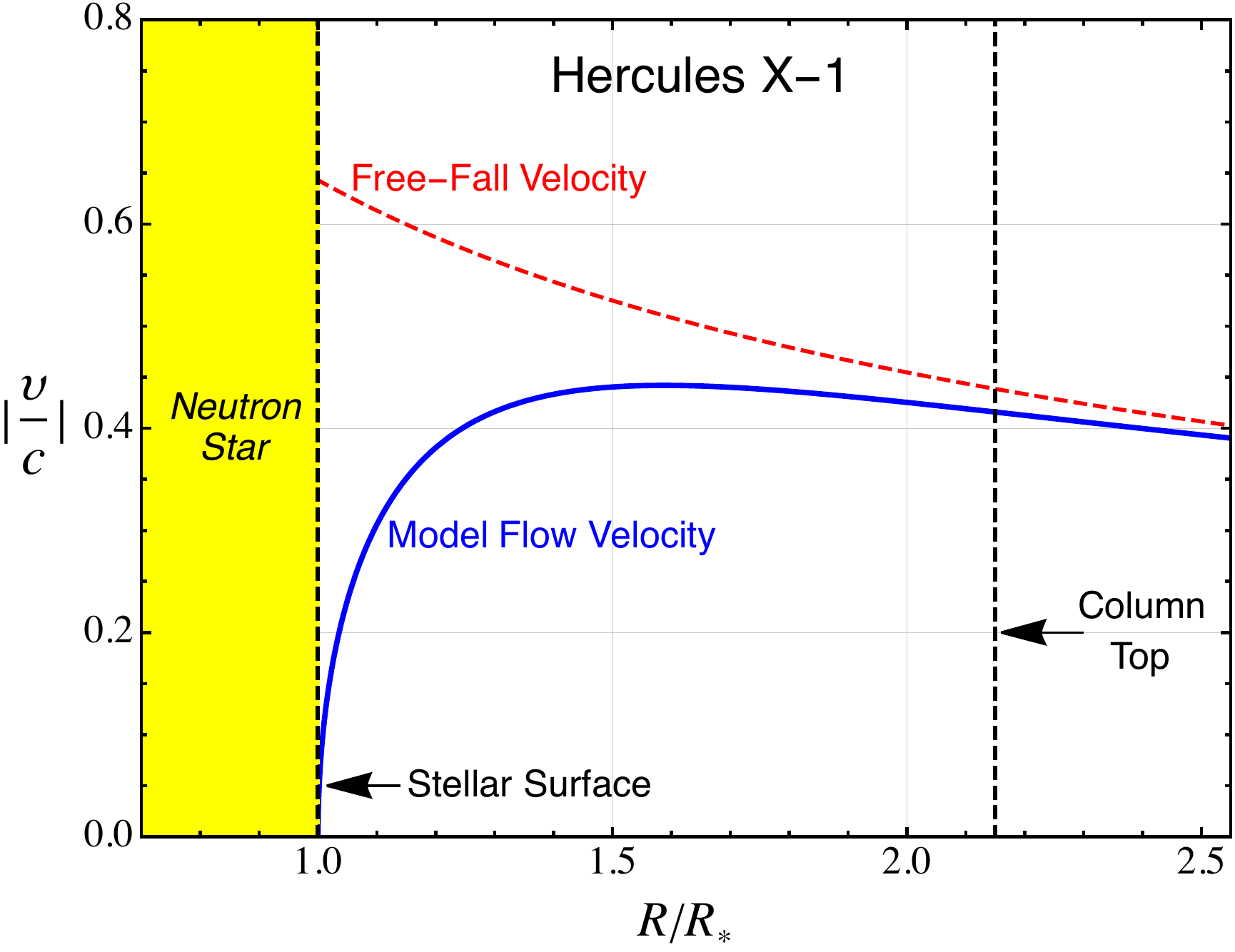,height=9.0cm}
\end{center}
\vskip-0.2truein
\caption{Model velocity profile $\vel$ for Her X-1 ({\it blue line}), computed using Equation~(\ref{eq3.13l}) with $k_0 = 0$ and $k_\infty = 1$. The flow settles onto the star with zero velocity at the surface, and approaches Newtonian free-fall far from the star.
\label{fig:herx1vel}}
\end{figure}

The input values for the fundamental theory parameters used to compute the spectral results for Her X-1 correspond to model~1, with $\dot M = 1.90 \times 10^{17}\,{\rm g\,s^{-1}}$, $T_e = 5.50 \times 10^7\,$K, $B = 3.26 \times 10^{12}\,$G, $D = 6.6\,$kpc, $\alpha = 0.35$, $\xi = 1.14$, $\deltapar = 1.42$, $k_0 = 0$, $k_\infty = 1$, $\Theta_1 = 0.315^\circ$, $\Theta_2 = 0$, and $y_{\rm top} = 2.15$, as indicated in Table~1. The associated values for the various auxiliary parameters are listed in Table~2. In particular, the values obtained for the scattering cross sections are $\sigperp/\sigmaT = 0.537$, $\sigpar/\sigmaT = 6.68 \times 10^{-5}$, and $\sigbar/\sigmaT = 5.90 \times 10^{-4}$. Table~2 also includes values for the scattering optical depth from the stellar surface to the top of the thermal mound, $\tau_{\rm th} = 0.08$, the scattering optical depth from the surface to the top of the column, $\tau_{\rm top} = 2.91$, the distance from the stellar surface to the top of the thermal mound, $z_{\rm th} = 692\,$cm, the radius of the magnetic polar cap at the base of the flow, $r_0 = 55\,$m, the impact velocity at the stellar surface, $\vel_* = 0$, the accretion velocity at the top of the thermal mound, $\vel_{\rm th} = -0.03\,$c, and the derived temperature of the thermal mound, $T_{\rm th} = 6.07 \times 10^7\,$K, which is the temperature used to compute the blackbody seed photon distribution. The velocity profile for the Her X-1 accretion column is computed by substituting the parameter values $k_0 = 0$ and $k_\infty = 1$ into Equation~(\ref{eq3.13l}), and the resulting velocity profile is plotted in Figure~\ref{fig:herx1vel}.

It is clear from the spectral components for the Her X-1 theoretical model plotted in Figure~\ref{fig:herx1details} that the phase-averaged X-ray spectrum for this source is dominated by Comptonized bremsstrahlung emission, radiated primarily through the column walls rather than through the top of the column. However, the cyclotron emission components make a significant contribution to the observed spectrum in the region around the cyclotron absorption feature, at photon energy $\epsilon_c \sim 38\,$keV. The power-law shape of the continuum, combined with the lack of a strong Wien peak in the spectrum, indicates that thermal Comptonization is unsaturated in this source. We note that the blackbody emission components make a negligible contribution to the observed X-ray spectrum for Her X-1.

The values obtained for the electron scattering cross sections in the application to Her X-1, listed in Table~2, satisfy the relation $\sigperp \sim \sigmaT \gg \sigbar \gg \sigpar$, which indicates that the cross sections are not significantly influenced by the effects of vacuum polarization, as discussed in Section~\ref{sec:radproc}. This is reasonable, since the electron number density near the base of the accretion column in Her X-1 is $n_e \sim 10^{24}\,{\rm cm}^{-3}$, which yields for the vacuum energy $\epsilon_{\rm vac} \sim 100\,$keV (see Equation~(\ref{eqVacEnergy1})). Recalling that vacuum polarization effects are unimportant for photon energies $\epsilon \lesssim \epsilon _{\rm vac}$, it is clear that the X-ray continuum in Her X-1 is unaffected by this process. In addition to the effect of vacuum polarization, one must also consider the distinction between the scattering cross sections experienced by ordinary and extraordinary mode photons propagating in the strong magnetic field. According to Figure~\ref{fig:elecscat}, the extraordinary mode photons will experience a resonance in the scattering cross section at photon energy $\epsilon \sim \epsilon_c$. Hence, if extraordinary mode photons are dominant in Her X-1, then we may expect to see super-Thomson cross section values for photons energies close to $\epsilon_c$. However, the
dominance of bremsstrahlung emission (which is non-resonant) in the spectrum of Her X-1 implies that ordinary mode photons dominate the seed photon population in this source. Furthermore, detailed consideration of the cross sections for mode-change scattering indicates that the ordinary mode photons produced via bremsstrahlung will primarily experience ``mode-coherent'' scattering, and will therefore remain as ordinary mode photons. This argument supports the cross section relation $\sigperp \sim \sigmaT \gg \sigbar \gg \sigpar$ which we obtain in our application of the model to Her X-1. This issue will be discussed in more detail in Section~\ref{sec:modselfcon}.

The relatively low value obtained for the thermal mound velocity, $\vel_{\rm th} = -0.03\,$c, in Her X-1 indicates that radiation pressure has nearly removed all of the kinetic energy from the gas by the time it reaches the top of the thermal mound. All of the remaining kinetic energy is used to power the blackbody ``hotspot'' at the base of the column, which generates the blackbody seed photons, as the gas continues to decelerate due to radiation pressure. By the time the gas reaches the stellar surface, the impact velocity is $\vel_* = 0$, as implied by the value $k_0 = 0$ listed in Table~1 (see Equations~(\ref{eq3.13l}) and (\ref{eq3.13q})). The bremsstrahlung and cyclotron seed photons are generated throughout the entire accretion column, but their production is concentrated near the base of the flow since the density reaches its maximum value there.
While bremsstrahlung and cyclotron emission both make important contributions to the observed X-ray spectrum in Her X-1, the blackbody component makes a negligible contribution because very little kinetic energy is left by the time the gas enters the thermal mound.

In the case of Her X-1, the spectrum plotted in Figure~\ref{fig:herx1details} also includes a Gaussian Fe K$\alpha$ emission line profile computed using the function
\begin{equation}
F^{\rm K}_\epsilon(\epsilon) \equiv \frac{d_{\rm K}}{\sigma_{\rm K}
\sqrt{2 \pi}} \, e^{-(\epsilon-
\epsilon_{\rm K})^2/(2\sigma^2_{\rm K})}
\ ,
\label{eq8.5}
\end{equation}
where $d_{\rm K}$ is the total iron line photon flux measured at Earth. It is important to note that in our approach, the iron line is simply added to the final spectrum rather than being subject to Comptonization inside the accretion column. A more sophisticated approach is beyond the scope of this paper due to the complexity of the spectral formation process for the iron line photons. The associated auxiliary parameter value we used to model the iron line for Her X-1 are $\epsilon_{\rm K} = 6.55\,$keV, $d_{\rm K} = 4.00 \times 10^{-3}\,{\rm s^{-1}\,cm^{-2}}$, and $\sigma_{\rm K} = 0.30\,$keV, and the parameter values used to treat cyclotron absorption via Equation~(\ref{eq8.3}) are are $\epsilon_c = 37.7\,$keV, $d_c = 18.0\,$keV, and $\sigma_c = 10.0\,$keV. We note that most of the parameter values obtained here are very similar to those reported by BW07 and by \citet{Wolff_etal2016} in their simulations of the phase-averaged X-ray spectrum of Her X-1.

\subsection{X Per}

\begin{figure}[t]
\begin{center}
\epsfig{file=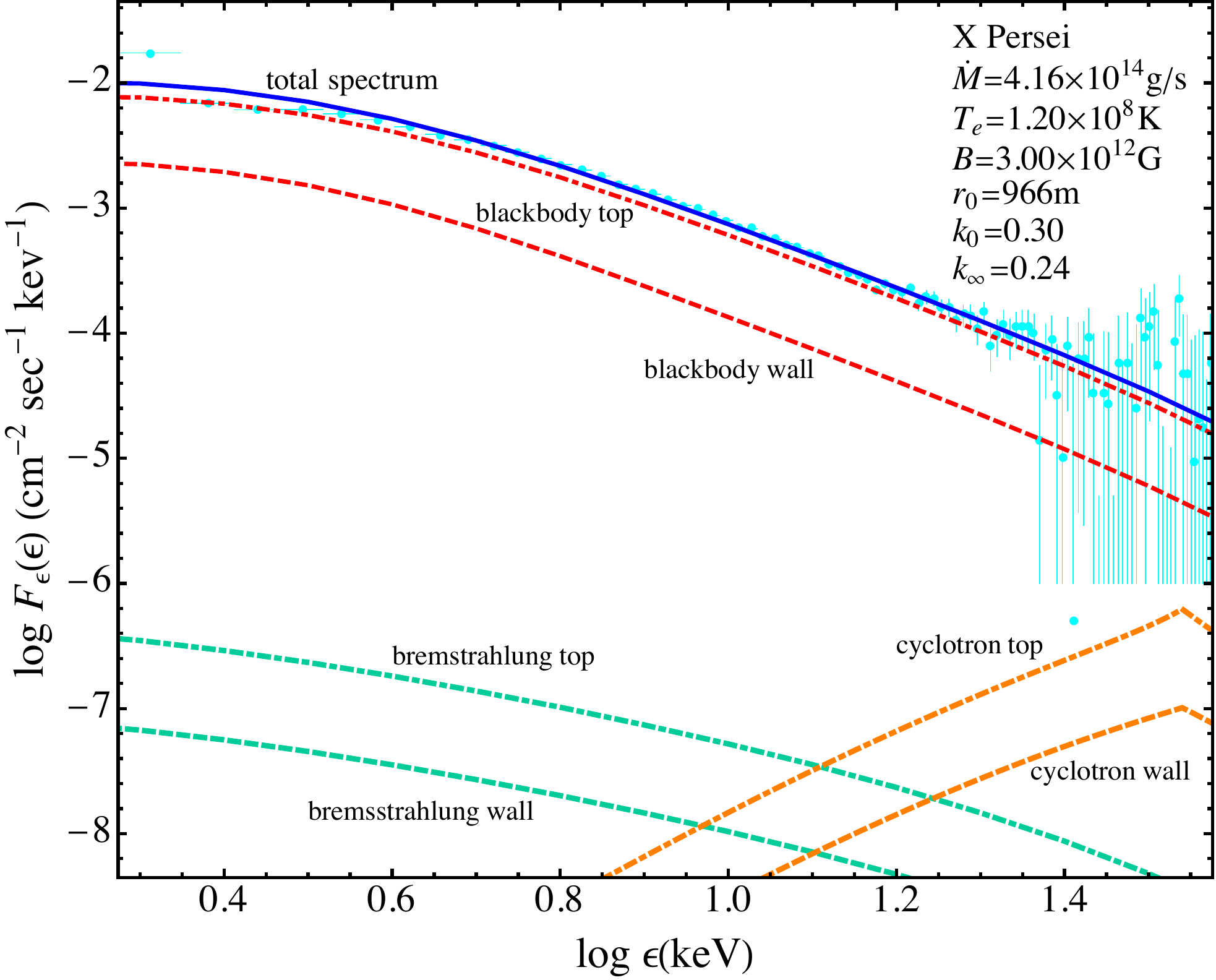,height=9.0cm}
\end{center}
\caption{Same as Figure~\ref{fig:herx1details}, except the data correspond to X Per and the theoretical spectra were computed based on the model~2 parameters listed in Table~1. The data were reported by Burderi et al. (2000; {\it circles and crosses}). This case also includes interstellar absorption with hydrogen column density $N_{\rm H} = 3.0 \times 10^{21}\,{\rm cm}^{-2}$.
\label{fig:xperspec}}
\end{figure}

In Figure~\ref{fig:xperspec} we compare the theoretical count-rate spectrum computed using Equations~(\ref{eq8.1}), (\ref{eq8.1b}), and (\ref{eq8.1c}) with the deconvolved, phase-averaged {\it RXTE} spectrum for X Per, which is based on an XSPEC analysis of an archival observation taken in 1998 July and reported by \citet{Delgado-Marti_etal2001}. The {\it RXTE} spectrum considered here is the same one analyzed by \citet{BeckerandWolff2005a,BeckerandWolff2005b} using their bulk Comptonization model. The theoretical results plotted in Figure~\ref{fig:xperspec} include the total phase-averaged spectrum, and the individual column-wall and column-top contributions to the observed spectrum due to the Comptonization of cyclotron, blackbody, and bremsstrahlung seed photons.

The magnetic field strength in X Per has been the subject of intense debate for several decades, and it has been estimated using a variety of techniques. For example, \citet{GhoshandLamb1979} used their model for torque-driven pulsar spin-down to estimate the surface magnetic field strength in X Per as $B = 4.8 \times 10^{12}\,$G, assuming a canonical neutron star with radius $R_* = 10\,$km. \citet{DiSalvo_etal1998} used an analysis of {\it BeppoSAX} observations of X Per to obtain the similar estimate $B = 5.6 \times 10^{12}\,$G, and \citet{Coburn_etal2001} used the detection of a putative cyclotron absorption feature in the {\it RXTE} spectrum of X Per to conclude that the surface magnetic field strength is $B = 3.3 \times 10^{12}\,$G. \citet{Maitra_etal2017} used an analysis of {\it Suzaku} data to obtain the similar value $B = 3.4 \times 10^{12}\,$G. On the other hand, \citet{Yatabe_etal2018} have argued that application of the \citet{GhoshandLamb1979} theory to the interpretation of {\it MAXI} data yields a best fit that indicates a much higher estimate for the magnetic field strength in the range $B \sim 10^{13-14}\,$G. Given the uncertainty in the data and the lack of consensus regarding the strength of the magnetic field in this source, we have chosen to adopt the representative value $B = 3.0 \times 10^{12}\,$G.

The theoretical spectrum for X Per plotted in Figure~\ref{fig:xperspec} is based on model~2, with fundamental theory parameters given by $\dot M = 4.16 \times 10^{14}\,{\rm g\,s^{-1}}$, $T_e = 1.20 \times 10^8\,$K, $B = 3.00 \times 10^{12}\,$G, $D = 0.725\,$kpc, $\alpha = 0.08$, $\xi = 1.20$, $\deltapar = 0.75$, $k_0 = 0.3$, $k_\infty = 0.244$, $\Theta_1 = 5.536^\circ$, $\Theta_2 = 0$, and $y_{\rm top} = 2.20$, as listed in Table~1. The associated auxiliary parameters are listed in Table~2. In particular, the values obtained for the scattering cross sections are $\sigperp/\sigmaT = 848$, $\sigpar/\sigmaT = 2.46$, and $\sigbar/\sigmaT = 4.31$. These cross section values are radically different from those obtained in the case of Her X-1, which reflects the distinctly different nature of the conditions in the accreting plasma in the two sources. This is further discussed below. Table~2 also includes the scattering optical depth at the top of the thermal mound, $\tau_{\rm th} = 0$, the scattering optical depth at the top of the accretion column, $\tau_{\rm top} = 1.75$, the distance between the stellar surface and the top of the thermal mound, $z_{\rm th} = 0$, the magnetic polar cap radius, $r_0 = 966\,$m, the impact velocity at the stellar surface, $\vel_* = -0.19\,c$, the accretion velocity at the top of the thermal mound, $\vel_{\rm th} = -0.19\,$c, and the thermal mound temperature, $T_{\rm th} = 8.04 \times 10^6\,$K, which is used to compute the blackbody seed photon distribution.

\begin{figure}[t]
\begin{center}
\epsfig{file=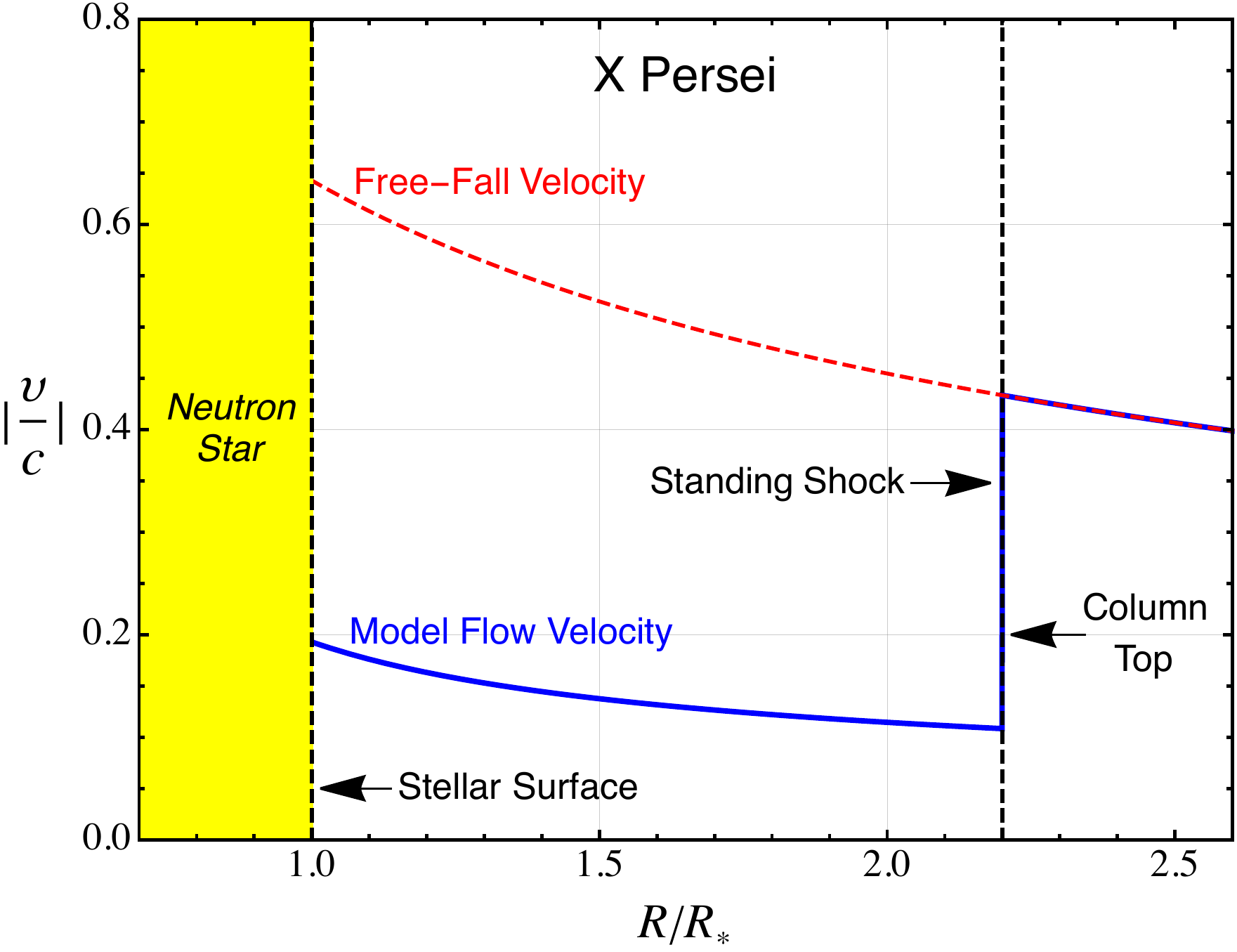,height=9.0cm}
\end{center}
\vskip-0.2truein
\caption{Model velocity profile $\vel$ for X Per ({\it blue line}), computed using Equation~(\ref{eq3.13l}) with $k_0 = 0.3$ and $k_\infty = 0.244$. The gas impacts the stellar surface with speed $0.19\,c$, and the flow satisfies the jump conditions for a strong shock at the top of the accretion column.
\label{fig:xperdyn}}
\end{figure}

We note that in the case of X Per, the thermal mound is located directly on the stellar surface, because there is insufficient absorption optical depth to form an extended region of thermal equilibrium. Hence the thermal mound velocity, $\vel_{\rm th}$, is equal to the impact velocity onto the stellar surface, $\vel_*$. The theoretical spectrum plotted in Figure~\ref{fig:xperspec} also includes interstellar absorption, with a hydrogen column density $N_{\rm H} = 3.0 \times 10^{21}\,{\rm cm}^{-2}$ \citep{BeckerandWolff2005a,BeckerandWolff2005b,LaPalombaraandMereghetti2007}. However, the model for X Per does not include either the cyclotron absorption feature (Equation~(\ref{eq8.3})) or the iron emission line (Equation~(\ref{eq8.5})), since there is no strong observational evidence for either of these features in the X-ray spectrum for this source. The associated velocity profile for the X Per accretion column is obtained by substituting the parameter values $k_0 = 0.3$ and $k_\infty = 0.244$ into Equation~(\ref{eq3.13l}), and is plotted in Figure~\ref{fig:xperdyn}.

Comparison of the various spectral components plotted in Figure~\ref{fig:xperspec} with the {\it RXTE} data for X Per clearly indicates that the X-ray spectrum is dominated by Comptonized blackbody radiation, escaping through the top of the accretion column. The seed blackbody photons for this component are generated in the ``hot spot'' at the base of the accretion column, where the gas collides at high speed with the surface of the star. According to Equations~(\ref{eq3.13l}) and (\ref{eq3.13q}), the surface velocity is determined by the value of the parameter $k_0$, with $k_0 = 0.3$ in the case of X Per. The resulting large impact velocity, $\vel_* = -0.19\,c$, reflects the inability of radiation pressure to decelerate the gas once it passes through the standing, discontinuous shock at the top of the accretion column. Instead, the gas gradually reaccelerates in the post-shock region (see Figure~\ref{fig:xperdyn}). The kinetic energy associated with the large impact velocity powers the blackbody seed emission, making it a much more effective source of seed photons for the Comptonization process than either bremsstrahlung or cyclotron. These behaviors are qualitatively different from those observed in the case of Her X-1, and stem from the fact that the luminosity of X Per is roughly three orders of magnitude lower, which profoundly alters the accretion dynamics and the properties of the accreting plasma.	

The values obtained for the electron scattering cross sections in the application of the model to X Per are listed in Table~2. In contrast to the results obtained in the case of Her X-1, we find that in the case of X Per, the cross sections are very strongly influenced by the effects of vacuum polarization (see Section~\ref{sec:radproc}). This is expected, because the electron number density near the base of the accretion column in X Per is $n_e \sim 10^{18}\,{\rm cm}^{-3}$, which yields for the vacuum energy $\epsilon_{\rm vac} \sim 0.1\,$keV (see Equation~(\ref{eqVacEnergy1})). Since vacuum polarization is important for photons with energy $\epsilon \gtrsim \epsilon _{\rm vac}$, we conclude that the entire X-ray continuum will be strongly modified by this effect in X Per. Further insight into this issue can be obtained by examining the distinction between the scattering cross sections for the ordinary and extraordinary mode photons. According to Figures~\ref{fig:secondelecscatplot} and \ref{fig:secondelecscatplot2}, both the ordinary and extraordinary mode photons with energy $\epsilon \sim \epsilon_c$ in X Per will experience super-Thomson scattering cross sections due to vacuum polarization. However, the angular dependence of the cross sections for the two modes is quite different. In particular, the cross section for the extraordinary mode photons {\it increases} as a function of the propagation angle $\theta$, whereas it {\it decreases} with increasing $\theta$ for the ordinary mode. Keeping in mind that photons propagating perpendicular to the magnetic field see the perpendicular scattering cross section, $\sigperp$, we conclude that {\it if} extraordinary mode photons dominate in X Per, then we would expect to see super-Thomson cross sections, with $\sigperp \gg \sigpar \gtrsim \sigmaT$, which is indeed the behavior observed in our model (see Table~2).

This naturally leads to the question of whether there is any physical reason that extraordinary mode photons would actually dominate in the X Per accretion column. We argue in Section~\ref{sec:modselfcon} that ``mode pumping'' of blackbody seed photons from the ordinary mode to the extraordinary mode will tend to reinforce the dominance of extraordinary mode photons in the X Per accretion column, which supports the super-Thomson cross section results we obtain when applying our model to that source. We also note the surprising similarity between the values obtained for $\sigpar$ and $\sigbar$ in the case of X Per, and suggest that this effect is due to a strong anisotropy in the radiation field inside the accretion column. The anisotropy would result in the beaming of most of the X-rays along the column axis, which is consistent with the dominance of the column-top emission over the column-wall emission in X Per.

\section{MODEL SELF-CONSISTENCY}
\label{sec:modselfcon}

The fundamental physical parameters in the theoretical model developed here are the stellar mass, set using $M_* = 1.4 \, \Msun$, the stellar radius, set to the canonical value $R_* = 10\,$km, the accretion rate, $\dot M$, set using observational estimates of the X-ray luminosity, combined with Equation~(\ref{eq4.8c}), and the source distance, $D$, set using published estimates. The remaining free parameters in the theory are $\alpha$, $\xi$, $\deltapar$, $B$, $\Theta_1$, $\Theta_2$, $k_0$, $k_\infty$, $y_{\rm top}$, and $T_e$. These parameters are varied in order to achieve satisfactory qualitative fits between the theoretical model and the observed X-ray spectra of Her X-1 and X Per, with the results displayed in Figures~\ref{fig:herx1details} and \ref{fig:xperspec}, respectively.

In addition to the need to fit the X-ray data, the theoretical free parameters can also be constrained by considering the dynamical and thermal structure of the accretion column. This is an important issue, since the velocity profile used in our model, given by Equation~(\ref{eq3.13l}), is a mathematical result obtained via application of the separation of variables technique to the solution of the transport equation (Equation~(\ref{eq3.1})). Hence the velocity profile used for each of the sources treated here must be physically validated based on the values selected for the constants $k_0$ and $k_\infty$.
We examine this issue below by performing an analysis of the hydrodynamic and thermodynamic structure of the accretion columns in Her X-1 and X Per obtained using our model.

We must also critically examine the results we have obtained for the electron scattering cross sections $\sigperp$, $\sigpar$, and $\sigbar$ when using our model to analyze the X-ray spectra of Her X-1 and X Per. The results obtained for the cross sections are radically different for the two sources (see Table~2). In the application to Her X-1, we obtain $\sigperp/\sigmaT = 0.537$, $\sigpar/\sigmaT = 6.68 \times 10^{-5}$, and $\sigbar/\sigmaT = 5.90 \times 10^{-4}$, and in the case of X Per, we obtain $\sigperp/\sigmaT = 848$, $\sigpar/\sigmaT = 2.46$, and $\sigbar/\sigmaT = 4.31$. We argue below that these apparently contradictory results can be understood as a consequence of the effect of vacuum polarization in X Per, which strongly alters the entire X-ray continuum in that source, whereas this effect is unimportant in the case of Her X-1.

\subsection{Column Structure}

The dynamical results for the velocity profiles in Her X-1 and X Per, plotted in Figures~\ref{fig:herx1vel} and \ref{fig:xperdyn}, respectively, are based on the substitution of specific values for the parameters $k_0$ and $k_\infty$ into the velocity function given by Equation~(\ref{eq3.13l}). The values for $k_0$ and $k_\infty$ are obtained as part of the spectral fitting process for the two sources. While the resulting spectral fits show good qualitative agreement with the X-ray spectra plotted in Figures~\ref{fig:herx1details} and \ref{fig:xperspec}, it is important to investigate whether the velocity profiles utilized in the models for the two sources satisfy the physical hydrodynamic and thermodynamic conservation equations.

\subsubsection{Thermodynamic Structure}

In order to validate the input dynamical profiles for Her X-1 and X Per, we need to analyze the structure of the accretion columns by solving a set of physical conservation equations. The results of this calculation will be a ``hydrodynamical'' velocity profile for each source that can be compared with the input profiles used in our model, which are computed using Equation~(\ref{eq3.13l}). The thermodynamic and hydrodynamic structures are coupled since the pressure of the radiation field and the gas may each make important contributions. We shall begin by considering the thermodynamic structure of the accretion column by examining the variation of the electron and proton temperatures. In the case of Her X-1, we would expect these two temperatures to track closely, since Coulomb coupling will be very efficient in the dense plasma. However, in the case of X Per, the situation may be quite different. In particular, \citet{LangerandRappaport1982} have shown that the plasma in the accretion column in X Per is so tenuous that a two-temperature plasma may result, in which the proton temperature, $T_p$, is much higher than the electron temperature, $T_e$, due to inefficient Coulomb coupling between the two species.

In our model, the electrons are assumed to form an isothermal distribution, with a temperature, $T_e$, that is regulated mainly by thermal Comptonization \citep{LyubarskiiandSyunyaev1982,SunyaevandTitarchuk1980,West_etal2017a,West_etal2017b}. The isothermal assumption is likely to be approximately satisfied in both the Her X-1 and X Per accretion columns. An analysis of the variation of the proton temperature, $T_p$, requires the solution to an energy equation for the protons, and we can expect to obtain very different results in the cases of Her X-1 and X Per due to the vastly different plasma densities in the two sources. We can write the proton energy equation as
\begin{equation}
\vel \frac{dU_p}{dR} = \gamma_p \, \frac{U_p}{\rho} \, \vel \frac{d\rho}{dR} + \dot U_p
\ ,
\label{ProtonEnergy1}
\end{equation}
where $R$ is the radius from the center of the star, $\rho = m_p\, n_e$ is the mass density, $\gamma_p = 5/3$ denotes the proton adiabatic index, and the proton energy density, $U_p$, is related to the proton temperature, $T_p$, via
\begin{equation}
U_p = \frac{1}{\gamma_p-1} \, \frac{\rho \, k T_p}{m_p}
\ .
\label{ProtonEnergy2}
\end{equation}
The first term on the right-hand side of Equation~(\ref{ProtonEnergy1}) represents adiabatic compression, and the second term, $\dot U_p$, denotes the heating rate for the protons via Coulomb coupling with the electrons. Based on Equation~(C1) from \citet{LangerandRappaport1982}, we can write
\begin{equation}
\dot U_p = - \beta_{e} \, R_0 \, n_e^2 \, \frac{m_e}{m_p} \left(\frac{8}{\pi}\right)^{1/2} \frac{T_p - T_e}{T_{\rm eff}}
\left(\frac{m_e c^2}{k T_{\rm eff}}\right)^{1/2}
\ln \Lambda_{\rm QM}
\ ,
\label{ProtonEnergy3}
\end{equation}
where the effective temperature, $T_{\rm eff}$, is given by
\begin{equation}
T_{\rm eff} \equiv T_e + \frac{m_e}{m_p} \, T_p
\ ,
\label{ProtonEnergy4}
\end{equation}
the constant $R_0$ is defined by
\begin{equation}
R_0 \equiv \frac{3}{4} \sigmaT m_e c^3
\ ,
\label{ProtonEnergy5}
\end{equation}
and $\Lambda_{\rm QM}$ is the Coulomb logarithm, computed using
\begin{equation}
\Lambda_{\rm QM} = 5.41 + \frac{1}{4} \, \ln\left(\frac{k T_{\rm eff}}{20\,{\rm keV}} \, \frac{B}{10^{12}\,{\rm G}}
\, \frac{10^{20}\,{\rm cm^{-3}}}{n_e}\right)
\ .
\label{ProtonEnergy6}
\end{equation}
Note that the heating rate for the protons, $\dot U_p$, is negative if $T_p > T_e$. The quantity $\beta_e$ appearing in Equation~(\ref{ProtonEnergy3}) is an efficiency factor for Coulomb cooling. In a collisional plasma, such as expected in Her X-1, we set $\beta_e = 1$, but in a collisionless plasma, such as expected in X Per, the efficiency of the Coulomb cooling may be reduced significantly below unity, and we therefore set $\beta_e = 0.1$ in this situation. Physically, this reflects the fact that the mean free path for proton-electron collisions is much larger than the system size in a collisionless plasma, and the proton distribution may therefore deviate significantly from an isotropic Maxwellian \citep{Spitzer1962,KirkandGalloway1982}.

\begin{figure}[t]
\begin{center}
\epsfig{file=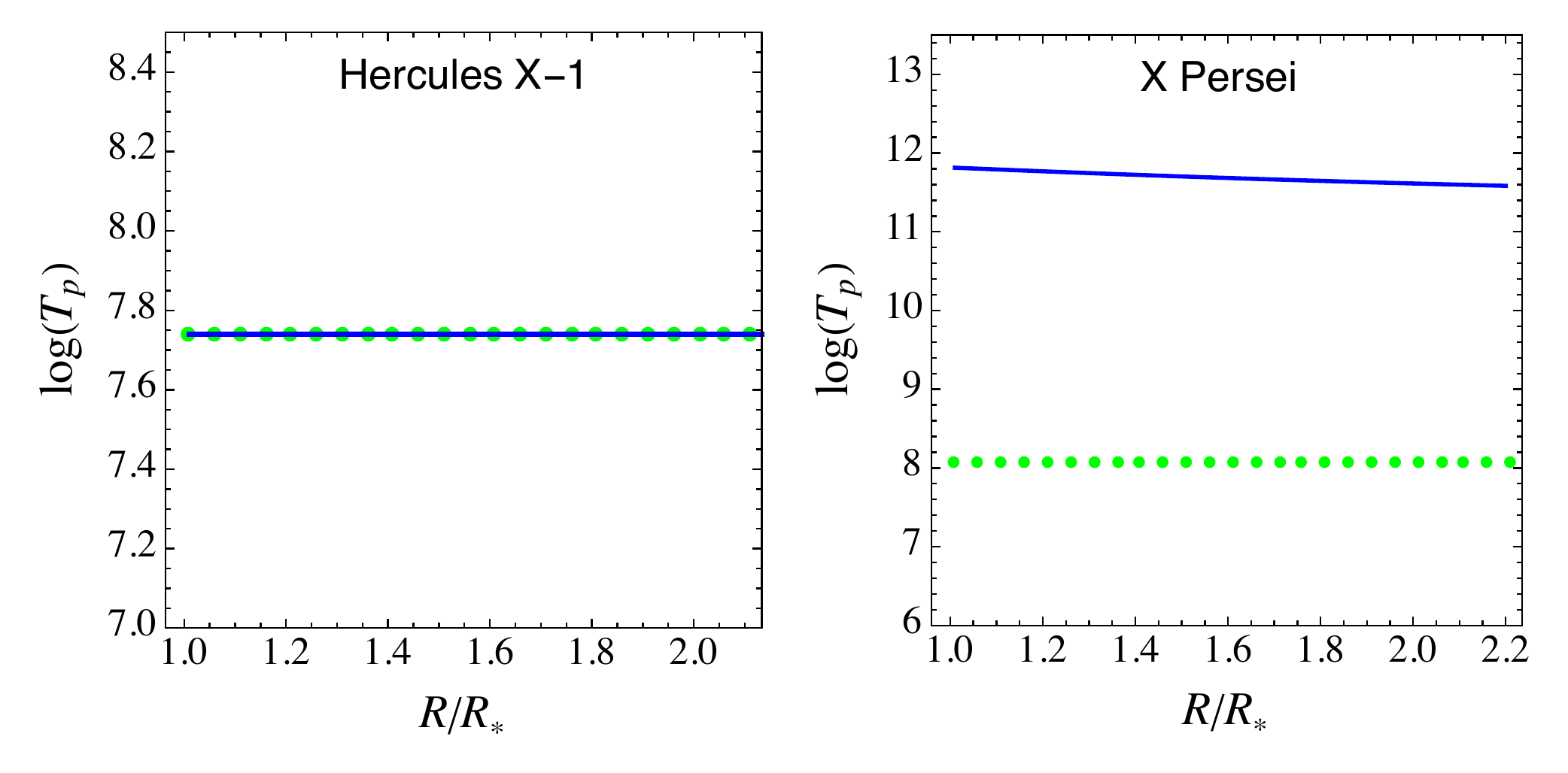,height=7.0cm}
\end{center}
\vskip-0.3truein
\caption{Proton temperature distribution, $T_p$ ({\it blue lines}), plotted as a function of $y=R/R_*$ by numerically integrating Equation~(\ref{ProtonEnergy9}), for Her X-1 (left) and X Per (right). Also plotted for comparison is the isothermal electron temperature, $T_e$ ({\it green dots}).
\label{fig:tplot}}
\end{figure}

We can transform Equation~(\ref{ProtonEnergy1}) into an ordinary differential equation for the proton temperature, $T_p$, by combining it with Equation~(\ref{ProtonEnergy2}), which yields
\begin{equation}
\frac{d \ln T_p}{dR} = -(\gamma_p-1) \frac{d \ln (R^2 \, |\vel|)}{dR} + \frac{1}{\vel} \, \frac{\dot U_p}{U_p}
\ ,
\label{ProtonEnergy7}
\end{equation}
where we have also used Equation~(\ref{eq3.2b}) to relate the mass density $\rho$ to the velocity $\vel$ using
\begin{equation}
\rho = \frac{\dot M}{\Omega R^2 |\vel|}
\ .
\label{ProtonEnergy8}
\end{equation}
By making use of Equation~(\ref{eq3.13l}) for the flow velocity $\vel$, we can simplify Equation~(\ref{ProtonEnergy7}) to obtain the equivalent form
\begin{equation}
\frac{d \ln T_p}{dy} = - \left[\frac{3 (\gamma_p-1) R_g \, c^2 \, k_\infty^2}{R_*}
+ \frac{\dot M}{\Omega R_* m_p} \, \frac{\dot U_p}{n_e^2 \, k T_p} \right] \frac{1}{\vel^2 \, y^2}
\ ,
\label{ProtonEnergy9}
\end{equation}
where we have also transformed the independent variable from $R$ to $y \equiv R/R_*$, and $\dot U_p$ is evaluated using Equation~(\ref{ProtonEnergy3}).

In order to numerically integrate Equation~(\ref{ProtonEnergy9}) to obtain the global solution for the proton temperature, we also need to specify a boundary value for $T_p$. We do this in two different ways for the two sources treated here. In the case of Her X-1, we simply set $T_p = T_e$ at the top of the accretion column. In the case of X Per, we set $T_p$ equal to the post-shock value, $T_{p+}$, obtained on the downstream side of the standing shock located at the top of the column. The standard Rankine-Hugoniot conditions give the downstream proton temperature as
\begin{equation}
k T_{p+} = m_p \vel_{-}^2 \frac{[2+(\gamma_p-1){\cal M}^2][1+\gamma_p(2{\cal M}^2-1)]}{\gamma_p (1+\gamma_p)^2 {\cal M}^4}
\ ,
\label{ProtonEnergy9b}
\end{equation}
where $\vel_{-}$ is the upstream velocity, which is equal to the Newtonian free-fall velocity, and ${\cal M}$ is the shock Mach number. Setting $\gamma_p = 5/3$ and assuming that ${\cal M} \gg 1$ for a strong shock yields
\begin{equation}
k T_{p+} = \frac{3}{16} \, m_p \vel_{-}^2
\ ,
\label{ProtonEnergy9c}
\end{equation}
which agrees with Equation~(16c) from \citet{LangerandRappaport1982}. With the boundary value for $T_p$ specified, numerical integration of Equation~(\ref{ProtonEnergy9}) yields the global solution for $T_p$ as a function of the dimensionless radius $y$.

\subsubsection{Hydrodynamic Structure}

Once the global solution for the proton temperature, $T_p$, is determined via numerical integration of Equation~(\ref{ProtonEnergy9}), we can solve for the proton pressure, $P_p$, using the ideal gas relation
\begin{equation}
P_p = \frac{\rho \, k T_p}{m_p}
= \frac{\dot M}{\Omega R^2 |\vel|} \, \frac{k T_p}{m_p}
\ ,
\label{ProtonEnergy10}
\end{equation}
where we have substituted for the mass density, $\rho$, using Equation~(\ref{ProtonEnergy8}), and $\vel$ is computed using Equation~(\ref{eq3.13l}). We can now use the proton pressure distribution to obtain the solution for the ``hydrodynamical'' flow velocity, denoted by $\vel^*$, which satisfies the hydrodynamical acceleration equation
\begin{equation}
\vel^* \frac{d\vel^*}{dR} = - \frac{G M_*}{R^2} - \frac{1}{\rho} \, \frac{dP_p}{dR}
- \frac{1}{\rho} \, \frac{dP_e}{dR} - \frac{1}{\rho} \, \frac{dP_r}{dR}
\ ,
\label{ProtonEnergy11}
\end{equation}
where $P_r$ denotes the radiation pressure, computed using the solution for the radiation spectrum, and the electron pressure, $P_e$, is given by
\begin{equation}
P_e = n_e k T_e
\ .
\label{ProtonEnergy12}
\end{equation}
Since the electron temperature, $T_e$, is a constant in our model, it follows that the electron pressure, $P_e$, varies in proportion to the electron number density, $n_e$. By combining results, we can numerically integrate Equation~(\ref{ProtonEnergy11}) to determine the hydrodynamical flow velocity, $\vel^*$, associated with the various pressure gradients acting on the gas in our model. Ideally, we would find that $\vel^* = \vel$, where $\vel$ is the input velocity for the model computed using Equation~(\ref{eq3.13l}) based on the selected values for the constants $k_0$ and $k_\infty$. In practice, we can't expect perfect agreement between the two velocity profiles, but by making a comparison between the input velocity $\vel$ and the hydrodynamical velocity $\vel^*$, we can evaluate the validity of our models for the accretion columns in Her X-1 and X Per.

\begin{figure}[t]
\begin{center}
\hskip-0.2truein\epsfig{file=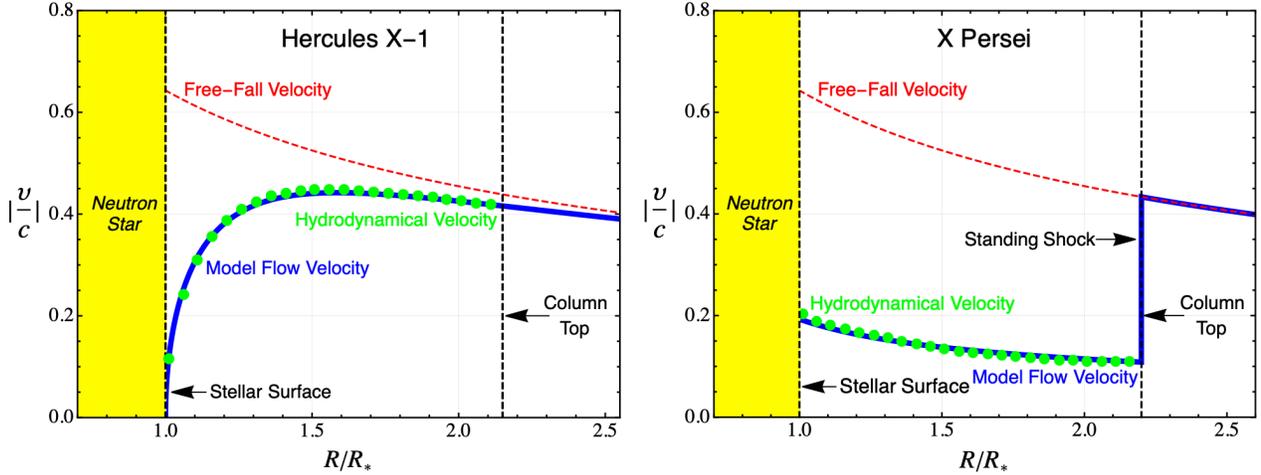,height=17.0cm,angle=-90}
\end{center}
\vskip-1.5truein
\caption{Hydrodynamical flow velocity, $\vel^*$ ({\it green dots}), obtained by numerically integrating Equation~(\ref{ProtonEnergy11}), compared with the model flow velocity, $\vel$ ({\it blue lines}), computed using Equation~(\ref{eq3.13l}) for Her X-1 (left) and X Per (right). The good agreement indicates that the hydrodynamical models for the two accretion columns are physically self-consistent.
\label{fig:vrecomp}}
\end{figure}

In Figure~\ref{fig:tplot} we plot the proton temperature distributions obtained in our models for the Her X-1 and X Per accretion columns. As expected, in the case of Her X-1, the proton temperature tracks the electron temperature closely, due to efficient Coulomb coupling. Since the electrons are assumed to be isothermal, the resulting proton temperature is also constant. On the other hand, in the case of X Per, we note that the proton temperature is significantly higher than the electron temperature, due to the relative inefficiency of the Coulomb coupling between the protons and the electron in the tenuous plasma in this source \citep{LangerandRappaport1982}. In Figure~\ref{fig:vrecomp} we plot the solutions for the hydrodynamical flow velocity, $\vel^*$, obtained for each source by numerically integrating Equation~(\ref{ProtonEnergy11}), and compare them with the corresponding model flow velocity profiles, $\vel$, computed using Equation~(\ref{eq3.13l}). The reasonably close agreement between $\vel^*$ and $\vel$ indicates that the hydrodynamics of the models for the accretion columns in Her X-1 and X Per are reasonably self-consistent.

\subsection{Electron Scattering Cross Sections}

The results we have obtained for the electron scattering cross sections $\sigperp$, $\sigpar$, and $\sigbar$ in our model are very different in the cases of Her X-1 and X Per, as indicated in Table~2. In the case of Her X-1, the results obtained are $\sigperp/\sigmaT = 0.537$, $\sigpar/\sigmaT = 6.68 \times 10^{-5}$, and $\sigbar/\sigmaT = 5.90 \times 10^{-4}$, and in the case of X Per, the results obtained are $\sigperp/\sigmaT = 848$, $\sigpar/\sigmaT = 2.46$, and $\sigbar/\sigmaT = 4.31$. We explore this issue in detail here, and we will show that the disparity between the cross section values in the two sources stems from the strong effect of vacuum polarization in the accretion column in X Per. The low plasma density near the base of the accretion column in X Per, with $n_e \sim 10^{18}\,{\rm cm}^{-3}$, results in a very low value for the vacuum energy, $\epsilon_{\rm vac} \sim 0.1\,$keV (see Equation~(\ref{eqVacEnergy1})). Conversely, the electron number density near the base of the accretion column in Her X-1 is much larger, $n_e \sim 10^{24}\,{\rm cm}^{-3}$, and this yields for the vacuum energy $\epsilon_{\rm vac} \sim 100\,$keV. Since the effect of vacuum polarization is only important for photons with energy $\epsilon \gtrsim \epsilon_{\rm vac}$, we conclude that vacuum polarization is completely unimportant for the X-ray continuum in Her X-1 due to the much larger plasma density in that source.

\begin{figure}[t]
\begin{center}
\epsfig{file=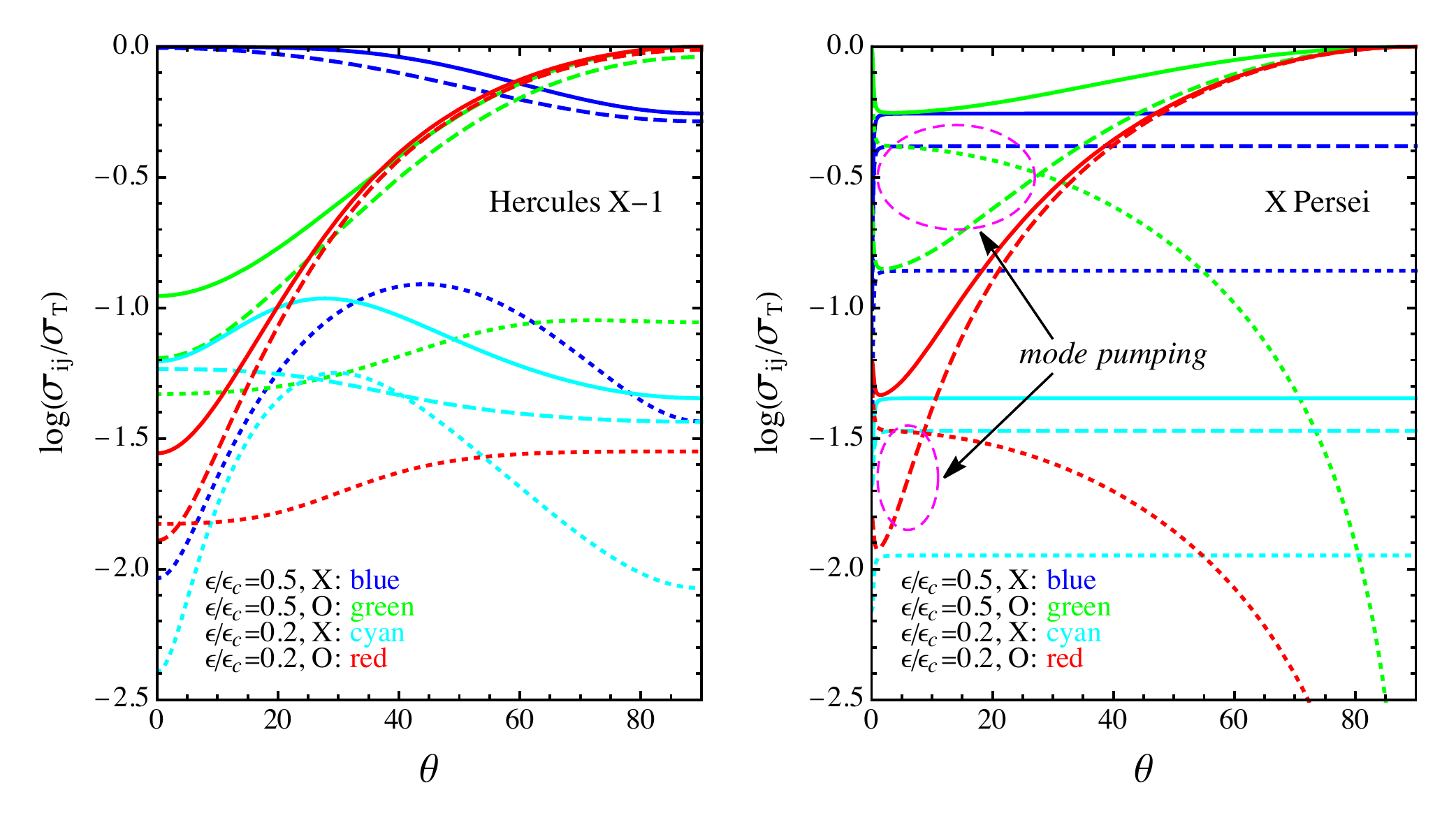,height=9.0cm}
\end{center}
\vskip-0.5truein
\caption{The mode-change cross sections are plotted as functions of the propagation angle $\theta$ using Equation~(29) from \citet{MeszarosandVentura1979} for parameters appropriate for Her X-1 (left) and X Per (right). Results include the total cross sections ({\it solid lines}), the mode-coherent cross sections ({\it dashed lines}), and the mode-changing cross sections ({\it dotted lines}). The energies of the extraordinary (X) and ordinary (O) mode photons are indicated by the color legends.
\label{fig:modechange}}
\end{figure}

Vacuum polarization not only affects the magnitudes of the total scattering cross sections, but it also has a profound effect on the {\it mode-change} cross sections that govern the transition between the ordinary and extraordinary modes that can occur in a single scattering. We plot the energy and angular dependences of the mode-change cross sections in the accretion columns in Her X-1 and X Per using Equation~(29) from \citet{MeszarosandVentura1979} in Figure~\ref{fig:modechange}. The left-hand panel in Figure~\ref{fig:modechange} utilizes parameters appropriate for the surface of the thermal mound in the Her X-1 accretion column, with electron number density $n_e = 1.36 \times 10^{24}\,{\rm cm}^{-3}$ and magnetic field strength $B =  3.26 \times 10^{12}\,$G. The right-hand panel utilizes parameters appropriate for the thermal mound in the X Per accretion column (which is located at the stellar surface), with $n_e = 1.47 \times 10^{18}\,{\rm cm}^{-3}$ and $B =  3.00 \times 10^{12}\,$G. Figure~\ref{fig:modechange} provides a detailed description of the nature of the mode-to-mode transitions that can take place inside the accretion columns in the two sources.

According to the mode-change scattering cross sections for Her X-1 plotted on the left-hand side of Figure~\ref{fig:modechange}, when vacuum polarization is unimportant, as in the case of Her X-1, both the ordinary and extraordinary mode photons will tend to scatter {\it coherently}, meaning that they are unlikely to change mode in a single scattering. We recall that in the case of Her X-1, the observed spectrum is dominated by Comptonized bremsstrahlung emission, which produces ordinary-mode photons because it is a non-resonant process. Since the bremsstrahlung seed photons are emitted in the ordinary mode, and they tend to scatter in a mode-coherent manner, this suggests that the photon population in the Her X-1 accretion column is dominated by ordinary mode photons. In this case, the dependence on the propagation angle $\theta$ for photon energy $\epsilon \ll \epsilon_c$ indicated by the red dashed curve on the left-hand side of Figure~\ref{fig:secondelecscatplot} implies that in the perpendicular direction, photons will scatter with cross section $\sigperp \sim \sigmaT$, and in the parallel direction, they will scatter with cross section $\sigpar \sim 10^{-3}\,\sigmaT$. These cross section estimates agree reasonably well with the cross section values obtained in our model for Her X-1 (see Table~2).

The mode-change scattering cross sections for X Per plotted on the right-hand side of Figure~\ref{fig:modechange} display a qualitatively different behavior than the results for Her X-1. In particular, we note that ordinary mode photons with propagation angle $\theta \lesssim 30^\circ$ and energy $\epsilon \sim 0.5\,\epsilon_c$ (denoted by the green curves) {\it are very likely to mode-change into extraordinary mode photons}. However, the corresponding extraordinary mode photons (denoted by the blue curves) are likely to remain as extraordinary mode photons, regardless of the value of $\theta$. These facts combine to suggest that ``mode pumping'' may occur in the X Per accretion column, which is a process in which the photons tend to become increasingly concentrated in the extraordinary mode. The regions of mode pumping are indicated by the magenta ellipses in Figure~\ref{fig:modechange}, which are the regions where the dotted curves (mode-change cross section) lie above the dashed curves (mode-coherent cross section). This process is physically significant, because the buildup of photons in the extraordinary mode leads to super-Thomson values for the total scattering cross section for photons with energy $\epsilon \sim \epsilon_c$, with a dependence on $\theta$ that yields values for the perpendicular cross section on the order of $\sigperp \sim 10^3\,\sigmaT$ and values for the parallel cross section on the order of $\sigpar \sim \sigmaT$, according to the solid blue curve on the right-hand side of Figure~\ref{fig:secondelecscatplot}. As indicated in Table~2, these cross section values agree reasonably well with the results obtained in our model for X Per. Note that super-Thomson cross section values for X Per are apparent in Figure~\ref{fig:secondelecscatplot}, but not in Figure~\ref{fig:modechange}, because the photon energies considered in Figure~\ref{fig:secondelecscatplot} are closer to $\epsilon_c$.

We must also consider the implications of the variation of the dipole magnetic field for the cyclotron energy, $\epsilon_c$, and how this may cause the resonance in the cross section to sweep through the X-ray continuum for each of the two sources considered here. In connection with this, we point out that the observed X-ray spectrum in X Per is dominated by the column-top component. Hence most of the photons experience a large transition in radius, from $R \sim R_*$ at the stellar surface, up to the column top located at radius $R \sim 2\,R_*$. For a dipole magnetic field, the increase in radius by a factor of 2 from the bottom to the top of the accretion column implies a drop in the magnetic field strength, $B$, (and hence the cyclotron energy $\epsilon_c$) by a factor of 8, which will sweep the cyclotron resonance into the center of the X-ray continuum for X Per. This naturally leads to the super-Thomson scattering cross sections for this source indicated in Table~2. On the other hand, in the case of Her X-1, the escaping radiation is dominated by the column-wall component, and most of the photons escape near the base of the column. It follows that there is not much variation in $B$ or $\epsilon_c$ throughout the emission region in this source. Hence, in the case of Her X-1, the cyclotron resonance will not sweep through the continuum, and therefore most of the X-ray photons have energy $\epsilon \ll \epsilon_c$, which is consistent with the sub-Thomson values for the scattering cross sections listed in Table~2.

\section{DISCUSSION AND CONCLUSION}
\label{sec:concl}

We have developed a new model for the radiative transfer occurring in accretion-powered X-ray pulsars based on a conical geometry for the accretion column, that also incorporates a number of additional enhancements relative to the previous model presented by the authors (BW07). In particular, the new model utilizes a very flexible form for the flow velocity profile (Equation~(\ref{eq3.13l})) that includes two free parameters, $k_0$ and $k_\infty$, where $k_0$ represents the ratio of the flow velocity divided by the Newtonian free-fall velocity at the stellar surface, and $k_\infty$ is the same ratio but evaluated as $R \to \infty$ (see Equations~(\ref{eq3.13q}) and (\ref{eq3.13w})). Many different scenarios can be modeled using this approach. For example, if we set $k_0 = 0$ and $k_\infty = 1$, then the velocity equals zero at the stellar surface, and merges smoothly with the Newtonian free-fall profile far from the star. This velocity profile, depicted in Figure~\ref{fig:herx1vel}, is probably an appropriate choice for luminous sources such as Her X-1, in which radiation pressure decelerates the matter to rest at the base of the accretion column.
On the other hand, in low-luminosity sources such as X Per, radiation pressure is probably insufficient to decelerate the material to rest at the stellar surface, and the gas may impact onto the star with a substantial residual velocity, after passing through a standing, gas-mediated shock at the top of the column.
We use this scenario to model the X Per accretion column, depicted in Figure~\ref{fig:xperdyn}, using the parameter values $k_0 \sim 0.3$ and $k_\infty \sim 0.25$. In this case, the surface impact velocity is $\vel_* = - 0.19\,c$ and the velocity profile at the top of the column is equal to $1/4$ the free-fall value, in accordance with the strong-shock jump condition.

The central result of the paper is the closed-form analytical solution for the Green's function describing the spectrum of the radiation inside the accretion column as a function of the scattering optical depth, $\tau$, and the photon energy, $\epsilon$, given by Equation~(\ref{eq4.26}). This solution represents the response to the continual injection of monoenergetic seed photons with energy $\epsilon_0$ from a source located at optical depth $\tau_0$ inside a conical accretion column with velocity profile given by Equation~(\ref{eq3.13l}). Based on this result, we proceed to derive the corresponding Green's functions for the column-integrated spectrum escaping through the column walls (Equation~(\ref{eq5.1f})), and also the Green's function for the spectrum escaping through the column top (Equation~(\ref{eq4.26xxx})). Since the fundamental transport equation (Equation~(\ref{eq3.1})) includes an escape term to describe the diffusion of photons through the column walls, and the upper surface (the column top) is treated as a free-streaming surface, it follows that our results for the two escaping radiation components are computed self-consistently.

The model includes the implementation of source photon distributions corresponding to blackbody, cyclotron, and bremsstrahlung emission. We confirm that the number of photons escaping from the column per unit time is exactly equal to the number injected, for each radiation mechanism, as required in the steady-state scenario considered here. The contributions to the observed X-ray spectrum due to emission from the walls and top of the accretion column are computed using Equations~(\ref{eq8.1}) and (\ref{eq8.1b}), respectively, and the approximate phase-averaged spectrum is computed using Equation~(\ref{eq8.1c}).
In Figures~\ref{fig:herx1details} and \ref{fig:xperspec}, we display the qualitative fits to the phase-averaged X-ray spectra of Her X-1 and X Per, respectively, obtained using the new model, and we note that the agreement between the theory and the data is reasonably close for both sources. We find that in the case of Her X-1, the observed phase-averaged X-ray spectrum is dominated by Comptonized bremsstrahlung emission escaping through the side walls of the accretion column, and in the case of X Per, it is dominated by Comptonized blackbody emission escaping primarily through the top of the column. The associated velocity profiles for the two sources are presented in Figures~\ref{fig:herx1vel} and \ref{fig:xperdyn}, respectively, and these profiles are validated via comparison with detailed hydrodynamical calculations in Figure~\ref{fig:vrecomp}. The set of results presented here represents the first time that a single overarching model for spectral formation in X-ray pulsars has been able to successfully reproduce the observed X-ray spectra for two sources spanning about three orders of magnitude in X-ray luminosity, from $L_X \sim 10^{34}\,{\rm ergs\,s}^{-1}$ for X Per to $L_X \sim 10^{37}\,{\rm ergs\,s}^{-1}$ for Her X-1.

In the case of Her X-1, the velocity profile plotted in Figure~\ref{fig:herx1vel} begins with a Newtonian free-fall shape far from the star, and smoothly transitions into a radiative, radiation-dominated shock as the gas approaches the base of the accretion column. The radiation-dominated shock decelerates the gas to rest at the stellar surface, and the kinetic energy of the flow is radiated away primarily through the walls of the accretion column. On the other hand, the velocity profile for X Per (Figure~\ref{fig:xperdyn}) exhibits a very different shape. In this source, the flow passes through a standing, gas-mediated discontinuous shock at the top of the accretion column, where the velocity drops by a factor of four, after which it proceeds to reaccelerate as the gas approaches the stellar surface. The gas collides with the neutron star with a residual velocity $\sim 0.19\,c$ (see Table~2).

\subsection{Model Parameters}

The process of obtaining the qualitative fits to the X-ray spectra displayed in Figures~\ref{fig:herx1details} and \ref{fig:xperspec} begins with the selection of values for the stellar mass, $M_*$, the stellar radius, $R_*$, the accretion rate, $\dot M$, and the source distance, $D$. We use canonical values for the stellar mass and radius, with $M_* = 1.4\,\msun$ and $R_* = 10\,$km, and the values for the distance $D$ and the accretion rate $\dot M$ are taken from published estimates. After these parameters are set, the spectral fits are obtained by varying the fundamental free parameters $\alpha$, $\xi$, $\deltapar$, $B$, $k_0$, $k_\infty$, $\Theta_1$, $\Theta_2$, $T_e$, and $y_{\rm top}$. The process for determining the values for the free parameters includes a comparison with the observed X-ray spectrum, as well as a consideration of the thermodynamic and hydrodynamic structure of the accretion column, as discussed in Section~\ref{sec:modselfcon}. Once satisfactory qualitative spectral fits are obtained, and the structure of the column is deemed sufficiently self-consistent, then we can compute the associated values for the scattering cross sections $\sigpar$, $\sigperp$, and $\sigbar$ using Equation~(\ref{eqAlphaNew4}), (\ref{eqXiParNew2}), and (\ref{eqDeltaPar4}), respectively. The results obtained for the fundamental free parameters for Her X-1 and X Per are listed in Table~1, and the corresponding values obtained for the scattering cross sections are reported in Table~2.

The values for the scattering cross sections $\sigpar$, $\sigperp$, and $\sigbar$ obtained by applying our model to Her X-1 and X Per are very different, and this has motivated further investigation. We argue that the difference in the cross sections stems from the effect of vacuum polarization, which is negligible in the case of Her X-1, but very important in the application to X Per, due to the different values for the electron number density, $n_e$, in the two sources.
In X Per, the density at the thermal mound surface is $n_e \sim 10^{18}\,{\rm cm}^{-3}$, whereas in the case of Her X-1, we find that $n_e \sim 10^{24}\,{\rm cm}^{-3}$. According to Equation~(\ref{eqVacEnergy1}), the corresponding results obtained for the vacuum energy, $\epsilon_{\rm vac}$, in the two sources are $\epsilon_{\rm vac} \sim 0.1\,$keV for X Per and $\epsilon_{\rm vac} \sim 100\,$keV for Her X-1. Since vacuum polarization profoundly influences the electron scattering cross sections for photons with energy $\epsilon \gtrsim \epsilon _{\rm vac}$, it is clear the this process plays a crucial role in determining the shape of the X-ray continuum in X Per. Conversely, in the case of Her X-1, vacuum polarization is unimportant.

The detailed implications of vacuum polarization are explored in Section~\ref{sec:modselfcon}, where we plot the mode-change cross sections for extraordinary and ordinary mode photons in Figure~\ref{fig:modechange}, using parameters appropriate for X Per and Her X-1. The plots clearly indicate that in the case of Her X-1, most of the photons will remain in the ordinary mode, because bremsstrahlung emission (which dominates the seed photon production in Her X-1) primarily produces ordinary mode photons, and there is little propensity for these photons to switch to the extraordinary mode, as indicated on the left-hand side of Figure~\ref{fig:modechange}. On the other hand, in the case of X Per, the situation is quite different. In particular, in this source, there is a strong likelihood for ordinary mode photons to switch into the extraordinary mode, and this will lead to ``mode pumping,'' in which photons tend to accumulate in the extraordinary mode, as indicated on the right-hand side of Figure~\ref{fig:modechange}.

The concentration of photons in the extraordinary polarization mode in X Per, combined with the fact that the cyclotron resonance will sweep through the continuum in this source due to the escape of most of the photons through the top of the column (at radius $R_{\rm top} \sim 2\,R_*$), will tend to increase the cross sections, yielding the super-Thomson values listed in Table~2 for this source. Hence in the case of X Per, we find that $\sigperp \gg \sigbar \gtrsim \sigpar \gtrsim \sigmaT$. On the other hand, in the case of Her X-1, the cyclotron resonance will not sweep through the continuum as strongly, because most of the radiation leaks out through the column walls near the base of the flow. The resulting cross sections in this source are generally sub-Thomson, and we find that $\sigmapar \ll \sigbar \ll \sigperp \sim \sigmaT$ for Her X-1. The similarity between the values for $\sigpar$ and $\sigbar$ obtained in the application to X Per, along with the much larger value obtained for $\sigperp$, suggests a strong anisotropy in the radiation field, with most of the X-rays beamed along the column axis. We note that this behavior is consistent with the dominance of the column-top emission versus the wall emission that we observe in our model for X Per (see Figure~\ref{fig:xperspec}).

\subsection{Dependence of Spectrum on Luminosity}

The cyclotron resonance absorption feature in the X-ray spectrum of Her X-1 exhibits an apparent dependence on the X-ray luminosity which has been studied by several authors \citep[e.g.,][]{Staubert_etal2019,Staubert_etal2020}. In general, the observations indicate that the power-law index of the continuum emission reduces, and the spectrum becomes harder, as the luminosity increases. The same behavior is observed in both the long-term variability of the source, as well as on pulse-to-pulse timescales \citep{Klochkov_etal2011}. In the context of the model developed here, this type of spectral variability with changing luminosity is a direct consequence of changes in the radiation pressure, which ultimately controls the impact velocity of the gas onto the surface of the neutron star. In particular, an increase in the luminosity will amplify the radiation pressure, leading to a decrease in the impact velocity at the stellar surface. The decreased surface velocity leads to an increase in the gas density, and the resulting enhanced compression of the gas acts as a ``piston,'' which powers an increase in the mechanical $P$d$V$ work done on the radiation field by the gas via bulk Comptonization. The increased work performed on the radiation field naturally leads to a flatter, harder spectrum in the high-luminosity sources.

The scenario described above is appropriate for supercritical sources, such as Her X-1, but it must be modified in the case of subcritical sources, such as X Per, in which radiation pressure is probably insufficient to accomplish complete deceleration of the gas to rest at the stellar surface. Instead, the gas may ram into the star with a significant residual velocity, perhaps as large as $\sim 0.25\,c$. In this situation, the gas will do significantly less $P$d$V$ work on the radiation field, due to the smaller amount of compression relative to what would occur if the gas were decelerated completely to rest at the stellar surface. The decreased amount of compression implies a decreased amount of bulk Comptonization, which leads to a steeper spectral slope. In this scenario, the final deceleration occurs in the last few cm above the stellar surface as a result of Coulomb collisions, which heat the material, powering the blackbody hot-spot that dominates the production of seed photons \citep{Sokolova-Lapa_etal2021}. We believe that this simple mechanism may provide a simple physical explanation for the steeper slopes observed in the X-ray continuum for low-luminosity sources such as X Per.

\subsection{Conclusion}

We have demonstrated that bulk and thermal Comptonization naturally leads to emergent spectra in accretion-powered X-ray pulsars that are in good agreement with the observational data for both high-luminosity and low-luminosity sources. Furthermore, we have shown that the spectrum of the high-luminosity source treated here, Her X-1, is dominated by reprocessed bremsstrahlung emission, escaping primarily through the walls of the accretion column. This result agrees with the findings of \citet{West_etal2017a,West_etal2017b}. Conversely, for the low-luminosity source treated here, X Per, we have demonstrated that the observed X-ray spectrum is dominated by Comptonized blackbody emission, escaping through the top of the accretion column, rather than through the walls. In principle, the difference between the dominance of the column-wall emission versus the column-top emission in these two sources could suggest a difference in the pulse profiles and the phase-resolved spectra. We plan to investigate this question in future work, using a general relativistic framework such as the one developed by \citet{RiffertandMeszaros1988}.

In recent years, there has been a significant increase in the development of space-based X-ray polarimetry observatories designed to study the polarization properties of the high-energy emission observed from compact sources, including accretion-powered X-ray pulsars. These missions include NASA's {\it IXPE} which launched in 2021, and the Indian mission {\it XPoSat}, with a planned launch in 2022. The availability of these new observatories complements recent advances in theoretical work, which suggest that there may be a complex, phase-dependent signature of polarization contained in the radiation detected from X-ray pulsars \citep[e.g.,][]{CaiazzoandHeyl2021a,CaiazzoandHeyl2021b}. In fact, a possible detection of polarization in the X-ray spectrum of Her X-1 has already been reported by \citet{Doroshenko_etal2022}. We note that it may be possible to develop new predictions for X-ray polarization as a function of pulse phase and photon energy using the new model developed here, although such a calculation is beyond the scope of this paper. However, we can make a straightforward prediction regarding the differences in the overall polarization properties of the spectra emitted by Her X-1 and X Per, since our model implies that the emission from Her X-1 is dominated by ordinary mode photons, and the emission from X Per is dominated by extraordinary mode photons, as viewed in the frame of the star. However, in the frame of the Earth, the situation is complicated by the spin of the star, which will tend to homogenize the polarization properties between the extraordinary and ordinary modes \citep{Doroshenko_etal2022}.

We also note the possible existence of an additional high-energy component in the spectrum of X Per, which manifests as a broad hump centered at photon energy $\sim 70\,$keV \citep{Doroshenko_etal2012}. This has been interpreted as a possible consequence of the emission of cyclotron photons produced via the excitation of electrons into the $n = 1$ Landau state due to collisions with protons in the Coulomb-stopping layer just above the stellar surface. However, no-self consistent model for this process has been developed yet \citep{Mushtukov_etal2022}. In future work, we intend to investigate this scenario, as well as the alternative possibility that the excess high-energy emission is directly connected with the presence of the standing shock at the top of the accretion column \citep{LangerandRappaport1982}.

The authors are grateful for useful discussions with Ekaterina Sokolova-Lapa,
Joern Wilms, Katja Pottschmidt, Kent Wood, Jason Wong, Brent West, and Ken Wolfram.
This work was supported in part by NASA through the \nustar{} Guest Observer Program
and the Astrophysics Explorers Program. The authors would also like to acknowledge
helpful comments from the anonymous referee.
This research has made use of data and/or software provided by the 
High Energy Astrophysics Science Archive Research Center (HEASARC), 
which is a service of the Astrophysics Science Division at 
NASA/GSFC and the High Energy Astrophysics Division of the Smithsonian 
Astrophysical Observatory.  
We acknowledge use of NASA's Astrophysics Data System (ADS) bibliographic 
services and the ArXiv.

\clearpage

\bibliographystyle{aasjournal}
\bibliography{bwbiblio}

% table 1 follows

\clearpage

%\hskip-2truein
\begin{deluxetable}{clccccccccccccc}
\tabletypesize{\scriptsize}
\tablecaption{Input Model Parameters\label{tbl-1}}
\tablewidth{0pt}
\tablehead{
\colhead{Model\!\!\!\!\!}
& \colhead{Object}
& \colhead{$\alpha$}
& \colhead{$\xi$}
& \colhead{$\deltapar$}
& \colhead{$k_0$}
& \colhead{$k_\infty$}
& \colhead{$\Theta_1$}
& \colhead{$\Theta_2$}
& \colhead{$y_{\rm top}$}
& \colhead{$\!\!\!\dot M$ (g~s$^{-1}$)}
& \colhead{\!\!\!$T_e$ (K)}
& \colhead{\!\!\!$B$ (G)}
& \colhead{\!\!\!$D$ (kpc)}
}
\startdata
1
&Her X-1
&0.35
&1.14
&1.42
&0.0
&1.0
&0.315
&0
&2.15
&$1.90 \times 10^{17}$
&$5.50 \times 10^{7}$
&$3.26 \times 10^{12}$
&6.60
\\
2
&X Per
&0.08
&1.20
&0.75
&0.3
&0.244
&5.536
&0
&$2.20$
&$4.16 \times 10^{14}$
&$1.20 \times 10^8$
&$3.00 \times 10^{12}$
&0.725
\\
\enddata

%% Text for table notes should follow after the \enddata but before
%% the \end{deluxetable}. Make sure there is at least one \tablenotemark
%% in the table for each \tablenotetext.

\end{deluxetable}

% table 2 follows

%\clearpage

\begin{deluxetable}{ccccccccccc}
\tabletypesize{\scriptsize}
\tablecaption{Computed Model Parameters
\label{tbl-2}}
\tablewidth{0pt}
\tablehead{
\colhead{Model}
& \colhead{$\sigperp/\sig$}
& \colhead{$\sigpar/\sig$}
& \colhead{$\sigbar/\sig$}
& \colhead{$r_0\,$(m)}
& \colhead{$z_{\rm th}\,$(cm)}
& \colhead{$\Tmound\,$(K)}
& \colhead{$\vel_* / c$}
& \colhead{$\vmound / c$}
& \colhead{$\taumound$}
& \colhead{$\tau_{\rm top}$}
}
\startdata
1
&$0.537$
&$6.68 \times 10^{-5}$
&$5.90 \times 10^{-4}$
&$55.0$
&692
&$6.07 \times 10^7$
&0.0
&-0.03
&0.08
&2.91
\\
2
&$848.0$
&$2.46$
&$4.31$
&$966.0$
&0.0
&$8.04 \times 10^6$
&-0.19
&-0.19
&0.0
&1.75
\\
\enddata

%% Text for table notes should follow after the \enddata but before
%% the \end{deluxetable}. Make sure there is at least one \tablenotemark
%% in the table for each \tablenotetext.

\end{deluxetable}

\end{document}